%% file: main.tex
\documentclass[onecolumn]{aastex631}
\usepackage[utf8]{inputenc}
\usepackage{graphicx}
\usepackage{natbib}
\usepackage{amsmath}
\usepackage{hyperref}
\usepackage{listings}
\usepackage[all]{nowidow}

\newcommand{\esf}{\epsilon_{\rm SF}}
\newcommand{\lbol}{L_{\rm bol}}
\newcommand{\ltot}{L_{\rm tot}}
\newcommand{\menv}{M_{\rm env}}
\newcommand{\msun}{M_\odot}
\newcommand{\smad}{\sigma_{\rm MAD}}
\newcommand{\tbol}{T_{\rm bol}}
\newcommand{\lumrat}{L_{\rm smm}/\lbol}

\shorttitle{Modeling YSO evolution}
\shortauthors{Richardson et al.}
\begin{document}

\title{A framework for modeling the evolution of young stellar objects}
\author[0009-0001-8880-6951]{Theo Richardson}
\email{terichard57@gmail.com}
\affiliation{Department of Astronomy, University of Florida, P.O. Box 112055, Gainesville, FL 32611}

\author[0000-0001-6431-9633]{Adam Ginsburg}
\affiliation{Department of Astronomy, University of Florida, P.O. Box 112055, Gainesville, FL 32611}

\author[0000-0002-5204-2259]{Erik Rosolowsky}
\affiliation{Dept. of Physics, University of Alberta, Edmonton, Alberta, Canada, T6G 2E1}

\author[0000-0002-5937-9778]{Joshua Peltonen}
\affiliation{Dept. of Physics, University of Alberta, Edmonton, Alberta, Canada, T6G 2E1}

\author[0000-0002-4663-6827]{R\'{e}my Indebetouw}
\affiliation{National Radio Astronomy Observatory, 520 Edgemont Road Charlottesville, VA 22903; rindebet@nrao.edu}
\affiliation{Department of Astronomy, University of Virginia, P.O. Box 3818, Charlottesville, VA 22903-0818; remy@virginia.edu}

\begin{abstract}
Measuring properties of young stellar objects (YSOs) is necessary for probing the pre-main-sequence evolution of stars. 
As YSOs exhibit complex geometry, measurement generally entails comparing observed radiation to template populations of radiative-transfer model YSO spectral energy distributions (SEDs). Due to uncertainty on the precise mechanics of star formation, the properties inferred for YSOs using these models often depend strongly on the assumed accretion history. 
We develop a framework for predicting observable properties of YSOs that is agnostic to the underlying accretion history, enabling comparison between theories. This framework links a set of radiative-transfer SEDs with protostellar evolutionary tracks to create models of evolving YSOs. Unlike previous works, we directly relate evolution models to observables through theoretical physical parameters rather than through intermediate, observationally derived analogues. We make flux predictions for YSOs corresponding to stars with birth masses from 0.2 to 50 $\msun$ during their accretion phase following isothermal-sphere, turbulent-core, and competitive accretion histories, showing that these histories may be observationally distinguished by examining the 100-$\mu$m and 3-mm fluxes of a YSO. We discuss the impact of dust models and parameter ranges on the output of radiative transfer simulations through a comparison to another SED model grid. We quantify the degree of confusion between YSO Stages and Classes across a wide range of physical scenarios; for each, we calculate confusion matrices that enable inference of the number of objects of a given Stage from an observed population. Finally, we critically examine the physical significance of various literature Stage and Class definitions.
\end{abstract}

\section{Introduction}\label{sec:1}
Stars play a key role across many areas in astrophysics, and understanding their evolution is necessary to model and interpret the behavior and spectra of galaxies, star clusters, and planetary systems. Much work has therefore been devoted to building appropriate models of stellar evolution and spectra, and a good deal of information has been amassed regarding what happens to a star once it arrives on the main sequence. Less agreed on, however, is its journey to that point from an initial cloud of dust and gas (e.g. how stars gain mass, or where in the prestellar environment that mass comes from). Modeling this phase of a star's life is made more difficult by the relative dearth of information bounding the problem. Knowledge about this stage of stellar evolution is limited by the relatively small number of pre-main-sequence (PMS) stars when compared to existing main-sequence (MS) stars, and observations of these forming stars that could constrain the available theory space are complicated by the high extinction present in the dense environments where stars are born. As such, there are many competing prescriptions for the processes of accretion and protostellar evolution, particularly for higher-mass stars \citep[e.g.][]{shu1977,bontemps1996,bonnell1997,bonnell2001,mckee2002,mckee2003,bally2005}.

Many previous approaches to modeling young stellar objects (YSOs) rely on the assumption of particular protostellar evolutionary tracks or accretion histories \citep[e.g.][]{robitaille2006,robitaille2007,zhang2018}.
These works produce grids of self-consistent YSO evolutionary tracks and spectral energy distribution (SED) models; however, their capacity to measure YSO properties when used as templates for SED fitting is limited by the narrow range of parameters allowed by their evolutionary tracks. That narrowness generally limits the extent of the theory space that the resulting SED models are able to cover. These model grids are therefore generally unable to distinguish between different models of star formation (e.g. isothermal vs. turbulent-core initial conditions). 
Other YSO model grids that do not assume any particular evolutionary theory exist, as in the cases of \citet{furlan2016} or \citet{haworth2018}. However, these grids similarly cover a relatively narrow range of protostellar parameter space and often focus on modeling a single YSO morphology (i.e. star + disk systems, star + disk + envelope systems, etc.).

In this paper, we create a method to model YSO evolution that can be used more broadly without sacrificing the quality of existing model grids. Instead of prefacing the creation of YSO models with assumptions,
we link YSO SED models with no foundational evolutionary history and spanning a wide range of parameters and morphologies with separately generated protostellar evolutionary tracks, enabling prediction over a wider theory space.

We base our work on the model set outlined in \citet[][R24]{richardson2024}, which is an updated version of \citet[][R17]{robitaille2017}. This set is a collection of 3D YSO radiative-transfer models (RTMs) that provides templates for SED fitting. The free parameters in R24 are sampled randomly from a parameter space made from properties of stars, cores, and disks. This approach results in roughly even coverage in every dimension. While this treatment produces a number of models that are unphysical in the sense that they are not compatible with any self-consistent accretion history, and it does not include information about the masses or ages of its stars, the grid does not privilege any theory of accretion or evolution over any other by construction. This approach to YSO modeling has precedent in \citet{nandakumar2018}, which also used R17 to constrain the stellar mass of a set of observed YSOs by determining the proximity of the illuminating sources in well-fitting models to protostellar evolutionary tracks from \citet{bernasconi1996}.

Our method for modeling the evolution of YSOs allows us to generate YSOs according to multiple proposed mechanisms of accretion with variation in quantities such as star formation efficiency over a wide range of stellar masses. This freedom in modeling enables us to probe theory space to a previously unattempted extent, which in turn allows us to gain greater insight into the physical mechanisms taking place in regions of star formation. We present this method in Section \ref{sec:2}, present results from our modeled YSOs in Section \ref{sec:3}, show some further uses of our framework in Section \ref{sec:4}, and conclude in Section \ref{sec:5}.

\section{Framework}\label{sec:2}
\subsection{The R24 models}\label{sec:2.1}
This work makes heavy use of the set of YSO SED models from R24. Here, we provide a brief overview of the aspects of the model set relevant to this paper. A full accounting of the properties and construction of the models can be found in R24 and R17, which R24 extends. To avoid confusion with other instances of ``models" or ``modeling" in this work, we will refer to constituent models from R24 as ``RTMs" or ``R24 models" going forward. (The other main instance is in reference to models of protostellar evolutionary tracks, which will be introduced in Section \ref{sec:2.2}.)

R24 models are divided into subsets based on the presence or absence of certain circumstellar density structures. We refer to these subsets as ``geometries" through the remainder of this work. While all RTMs possess a central luminosity source, they may exhibit combinations of circumstellar envelopes, circumstellar disks, bipolar cavities, and an ambient medium. There are some constraints placed on these combinations: geometries with cavities must also have envelopes, and geometries with envelopes must also have an ambient medium, which allows a cutoff point for an envelope to be defined. Each of these features has its own set of associated parameters that determine its shape and density profile. The number of RTMs in each geometry is influenced by its complexity (i.e. number of free parameters); the two most complex geometries make up a plurality of the total set of models. Every geometry is axisymmetric about the $z$-axis (i.e. no $\phi$ dependence) and reflectionally symmetric across the $x-y$ plane, but many features introduce a $\theta$ dependence in the dust density profile. Geometries are identified by a series of seven characters, e.g. \texttt{spubhmi}, indicating which features are present or absent; see Table 2 of R17 for more detail.

Each set of parameters has an accompanying SED modeling the dust continuum emission, created through the use of the Monte Carlo radiative transfer code \texttt{Hyperion} \citep{robitaille2011}. The SEDs are given as flux densities (i.e. $S_\nu$) in units of mJy. They are calculated over the wavelength range of 0.01-5000 $\mu$m within a series of circular apertures that have radii evenly log-spaced between 10$^2$ and 10$^6$ au. RTMs with a $\theta$ dependence have SEDs with nine lines of sight randomly sampled from ten-degree bins from $0^{\circ}$ (face-on) to $90^{\circ}$ (edge-on). RTMs with no $\theta$ dependence have only one SED, since they are spherically symmetric and look the same along every sightline. All dust in R24 takes its properties from the model of \citet[][D03]{draine2003a,draine2003b} with the \citet{weingartner2001} Milky Way grain size distribution A for $R_{\rm V}$ = 5.5 and carbon abundance C/H renormalized to 42.6 ppm.

As of R24, each RTM is also associated with several properties that emerge from its shape parameters and temperature profile. The quantity most relevant to this work is the circumstellar mass, which is calculated around each source in spherical regions with the same radii as the SEDs. This ``sphere mass" tracks the amount of dust and gas around the central source out to the radius at which the envelope (along with any disk or cavities present) blends into the ambient medium. Since the native density profiles only track dust density, these masses assume a gas-to-dust ratio (GDR) of 100 to arrive at a total mass.

\subsection{YSO composition}\label{sec:2.2}
In this section, we lay out the steps we follow to create a model of an evolving YSO within our framework, as well as the assumptions we make in the course of modeling. For the sake of clarity, we adopt the linguistic convention that a ``YSO'' refers to the combined system of a central luminosity source (star, PMS star) and surrounding density structures (envelope, disk, etc.) while a ``protostar'' refers only to the source. This is consistent with the usage of these terms in the work that we build upon, but may differ from the parlance and working definitions of other entries in the literature.

We begin by generating evolutionary tracks for protostars using the \citet[][K12]{klassen2012} code for modeling protostellar evolution. 
Each of the tracks created by the K12 code predicts mass, radius, luminosity, and other intrinsic properties of a protostar from the initiation of gravitational collapse to arrival on the main sequence, given its final (zero-age main-sequence) mass as input.
As published, the code implements an isothermal-sphere \citep[IS,][]{shu1977} accretion history following the the \citet{offner2009} implementation of the protostellar evolution model of \citet{nakano1995}, as extended by \citet{nakano2000} and \citet{tan2004}. We have modified this code to also generate both turbulent-core \citep[TC,][]{mckee2002,mckee2003} and competitive \citep[CA,][]{bonnell1997,bonnell2001} accretion histories according to prescriptions for accretion rates laid out in \citet[][M10]{mckee2010}. We will refer to these evolutionary tracks as ``protostellar evolutionary models"--abbreviated ``PEMs"--throughout the rest of the work. 

\begin{figure}
    \centering
    \includegraphics[width=0.7\linewidth]{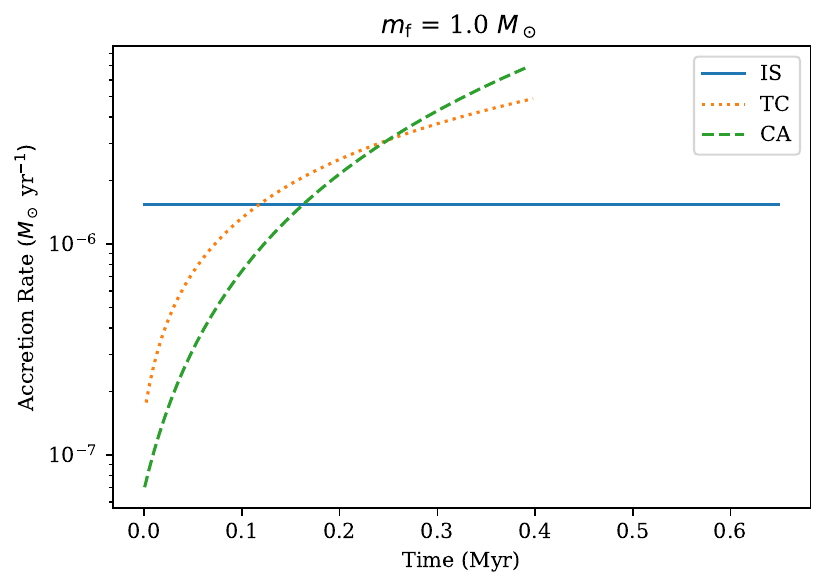}
    \caption{Accretion rates from PEMs generated from our modified K12 code, following prescriptions from M10. The lines show isothermal-sphere (IS), turbulent-core (TC), and competitive (CA) accretion rates as a function of time for a star with a final stellar mass of 1 $\msun$, attained at the end of each line.}
    \label{fig:ratecomp}
\end{figure}

M10 accretion history models all follow the the form
\begin{equation}
    \dot{m}=\dot{m}_1\left(\frac{m}{m_{\rm f}}\right)^j m_{\rm f}^{j_{\rm f}}
    \label{eq:mcmodel}
\end{equation}
where $m_{\rm f}$ is the final stellar mass, and $j$ and $j_{\rm f}$ are real-valued exponents that vary with accretion history. $\dot{m}_1$ is the final accretion rate for a star of unit mass and is set by a scaling parameter, which also varies with history. For IS accretion, this scaling parameter is the gas temperature $T$, for which we adopt a value of 10 K. For TC accretion, the scaling parameter is the gas clump surface density $\Sigma_{\rm cl}$, which we take to be 0.1 g cm$^{-2}$. For CA, given the hierarchical nature of the theory, the scaling parameter is the average number density of hydrogen atoms across a cloud $\bar{n}_{\rm H}$, which we take to be 10$^4$ cm$^{-3}$. (These are the fiducial values by which the respective accretion rates are scaled in M10.) Figure \ref{fig:ratecomp} shows a sample comparison of accretion rates prescribed by these histories.
Varying these scaling parameters changes the accretion rate of a protostar, which will in turn affect protostellar properties sensitive to the accretion rate and the timescale of accretion. This means that the predictions we make using these PEMs will also be affected by these scaling parameters; however, a full analysis of their impact is out of scope for this work.

We adopt different time steps for the K12 code depending on the assumed accretion history and zero-age mass to ensure that evolution is tracked on appropriate timescales. IS accretion in M10 is a constant rate, meaning that it can be well captured with time steps that scale with zero-age stellar mass. We set the time step to be 0.1\% of the total accretion time. The M10 prescriptions for TC and CA depend both on instantaneous and final stellar mass, resulting in accretion that increases with time. Per \citet{bonnell2001}, the characteristic time for star formation in CA is roughly the initial freefall time of the parent cloud of a stellar population, $t_{\rm ff}=0.435\,\bar{n}_{\rm H, 4}^{-1/2}$ Myr, where $\bar{n}_{\rm H,4}\equiv\bar{n}_{\rm H}/(10^4 \;{\rm cm}^{-3})$. Since CA in M10 is tuned to produce stars that accrete in roughly this characteristic time regardless of final mass, we adopt a time step of 0.1\% of this timescale, as with IS. For TC, which has no such characteristic time, we model evolution for 2 Myr and adopt a fixed time step of 2 $\times$ $10^{3}$ yr. These choices provide a time span that is long enough for the most massive protostars to accrete and sampled well enough to capture the increase in the accretion rate.

Once we generate our PEMs, we use the time evolution of the mass for a given protostar to track the time evolution of the corresponding mass contained in its circumstellar material (i.e. the core mass). For the purpose of this work, we assume that each modeled protostar accretes via monolithic collapse, a paradigm where one core forms one star.
(This assumption is at odds with the theory behind CA, where an entire stellar population competes for the material in a single mass reservoir, i.e. hierarchical collapse. However, the nature of the theory makes it difficult to model CA accretion rates without making some choices for the sake of implementation. M10's parameterization of CA is intended to preserve the dependence of the accretion rate on tidal effects and a formation time roughly equivalent to the freefall timescale, which are important aspects of the theory. Since these broad strokes of CA are factored into this treatment, we use the protostellar mass to track the mass of material around a protostar in the same way as for IS and TC models.)
We further assume a core-to-star mass accretion efficiency $\esf$ such that $M_{\rm\star,\,final}=\esf\times M_{\rm core,\,initial}$; as a result, the mass of the core evolves as $M_{\rm core}(t) = M_{\rm core,\,initial} - M_\star(t)/\esf$. For the sake of internal consistency, we adopt the commonly used value of 1/3 for $\esf$ \citep[e.g.][]{motte1998,alves2007,nutter2007}. We use this value for our results in Section \ref{sec:3}, but implement $\esf$ as a variable parameter and allow variation in Section \ref{sec:4.2}.

The accretion models underlying our PEMs come with some caveats. Firstly, predictions for accretion rates essentially track the ``main" accretion phase of a protostar where the majority of its mass is assembled through accretion from an envelope (in our paradigm). However, even once this envelope is depleted, observations indicate that circumstellar disks may persist for several Myr \citep[e.g.][]{haisch2001,hernandez2007,spezzi2008}. During this time, protostars will continue to accrete at a very low level; \citet{hartmann2016} compile a number of measurements placing this level at $\sim10^{-12}-10^{-6}$ $\msun$ yr$^{-1}$ for stars between $\sim0.01$ and $10\,\msun$. Our PEMs, as currently formulated, do not model this phase. 
Secondly, we note that all of M10's accretion models prescribe steady, nonepisodic accretion. Observations of YSOs indicate that accretion is generally variable over a wide range of timescales and is often episodic or stochastic \citep{fischer2023}. If implemented, accretion variability would likely cause modeled protostellar behavior to differ from the steady state, particularly in the case of YSOs in outburst.
Finally, our current treatment of YSO construction, in concert with our assumed PEMs, paints a physical picture in which a single protostar forms from a finite mass reservoir with either a constant or an accelerating accretion rate. Since mass from the natal material of a star is ejected during formation, decreasing the total available mass and lengthening the freefall time, a nondecreasing infall rate cannot be supported without replenishment of the protostellar envelope, i.e. a protostar drawing from a finite reservoir should have a decreasing accretion rate. Such decreasing rates are also better able to accommodate longer-lived star+disk systems.
We acknowledge that the IS/TC/CA models utilized here are therefore not, as implemented, fully consistent with the modern understanding of star formation. We elect to employ simplified models of accretion to broadly examine the observational consequences of distinct physical models for protostellar growth; the evaluation of accretion histories that incorporate additional physics (e.g. tapered or episodic accretion) is deferred to a future work.

Once we perform the star-to-core mapping, we can translate our PEMs into a parameter space that enables comparison to the RTMs. We construct this space using the source temperature $T_\star$, source luminosity $L_\star$, and circumstellar mass $M_{\rm core}$ of each R24 model. $L_\star$ is the total luminosity of the protostar, meaning that it includes both the intrinsic luminosity of the source and luminosity from accretion. The choice between tracking intrinsic or total luminosity impacts flux predictions for wavelengths sensitive to $L_\star$; we use the total luminosity to ensure we fully capture the radiation emitted from the central source, regardless of origin. More discussion about the impact of source luminosity on downstream results is in Appendix \ref{ap:lum}. Each RTM has a calculated mass within each aperture where the SEDs are defined; we adopt the mass contained within the 11th aperture, which has a radius $\sim$10,000 au  (notated as $M_{\rm 10k \, au}$) as a proxy for $M_{\rm core}$. This size encompasses the majority of the mass contained in the envelopes of most of the RTMs without including background dust.

Since we are primarily interested in modeling the phase of a protostar's evolution where it is actively accreting, we limit this parameter space to RTMs from the geometries that have circumstellar envelopes. This makes our set capable of modeling all the evolutionary stages of a YSO in which the central protostar has not yet depleted its mass reservoir (more discussion about these stages and how they intersect with the R24 models can be found in Section \ref{sec:4.2}). These geometries represent about 75\% of the total set of RTMs.

Finally, we model the SED of a YSO by associating the R24 models with our PEMs. To arrive at an evolving SED, we perform a nearest-neighbor search within the $T_\star-L_\star-M_{\rm core}$ space to match the K12 output to R24 models at each time step. In order to reduce the noise in the predicted SED, we identify the 10 nearest RTMs across all geometries in the set and average between them by taking the median of their SEDs. The majority of RTMs have SEDs defined at nine inclinations, which are randomly sampled from within nine evenly spaced inclination bins between $0$ and $90^{\circ}$ (see \S\ref{sec:2.1}). We preserve this inclination dependence in our predicted SEDs. Keeping the same bins, we ensure that each selected RTM contributes the appropriate SED to each inclination bin and average within each bin independently, producing nine SEDs per PEM time step. RTMs from spherically symmetric geometries, which do not have inclination-dependent SEDs, contribute the same SED to each inclination bin for consistency with other geometries.

\begin{figure*}
    \centering
    \includegraphics[width=0.32\textwidth]{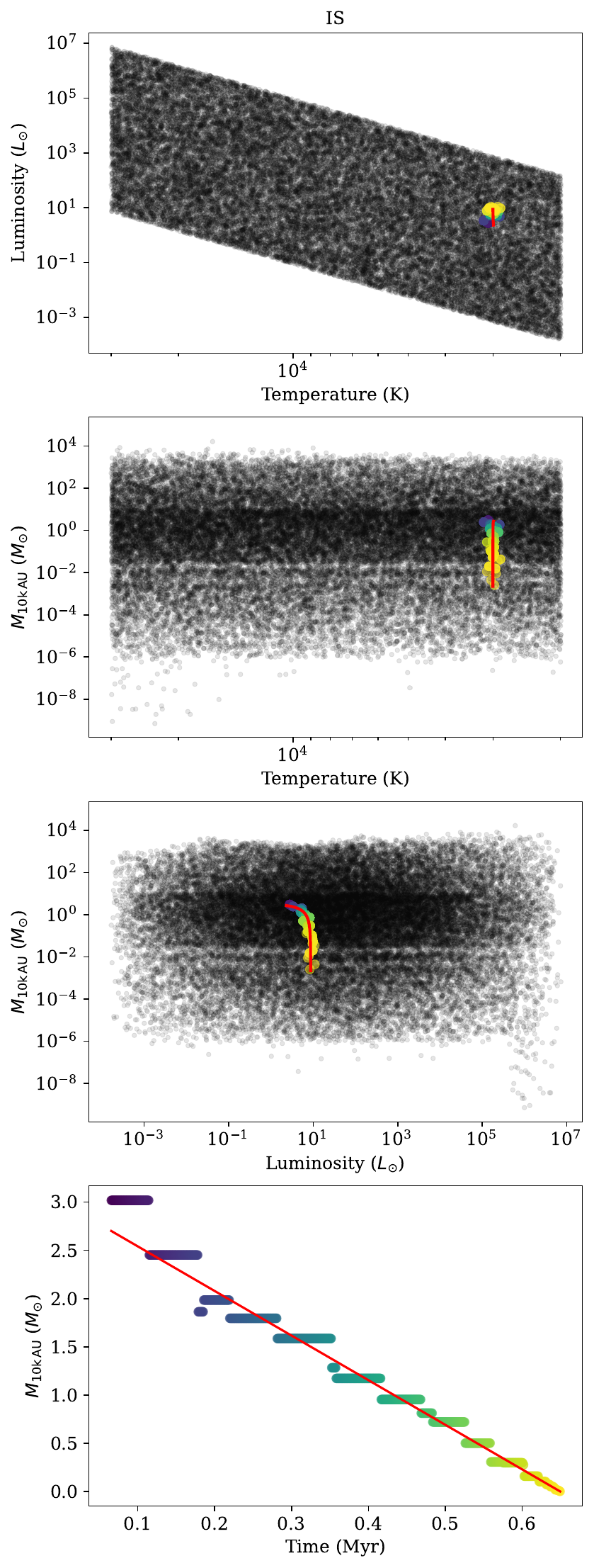}
    \includegraphics[width=0.32\textwidth]{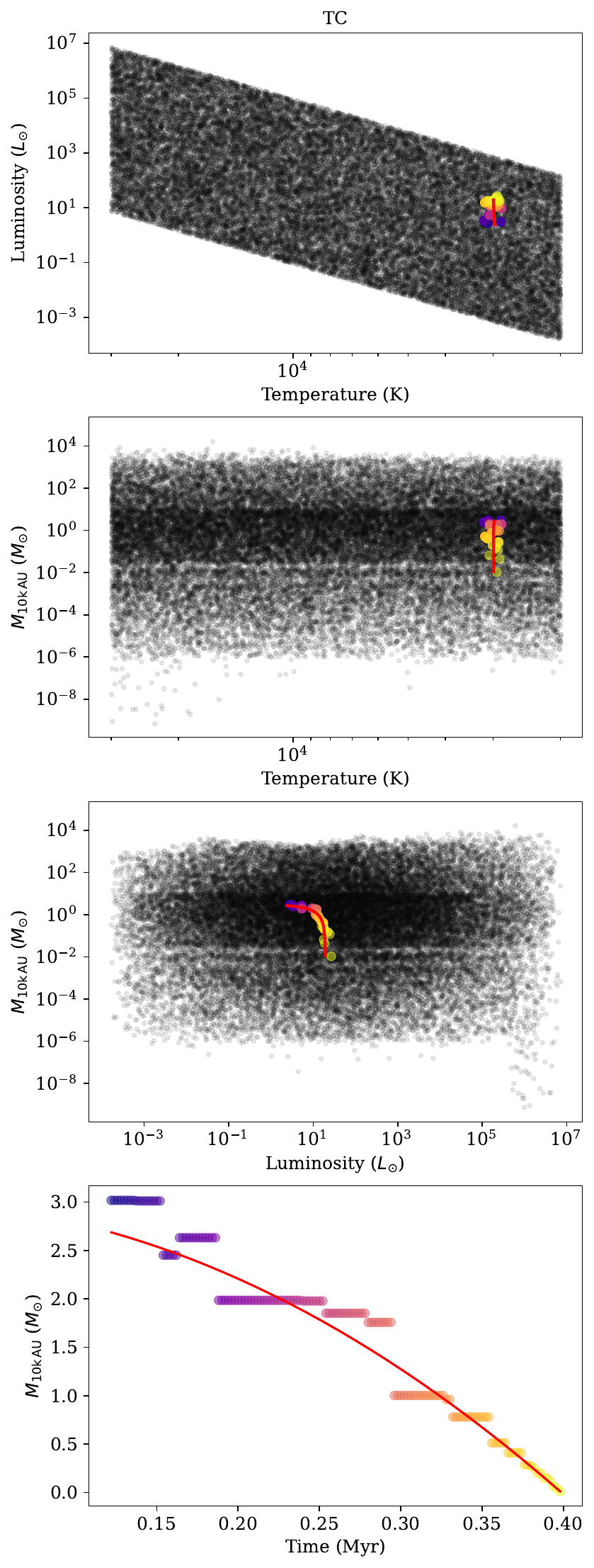}
    \includegraphics[width=0.326\textwidth]{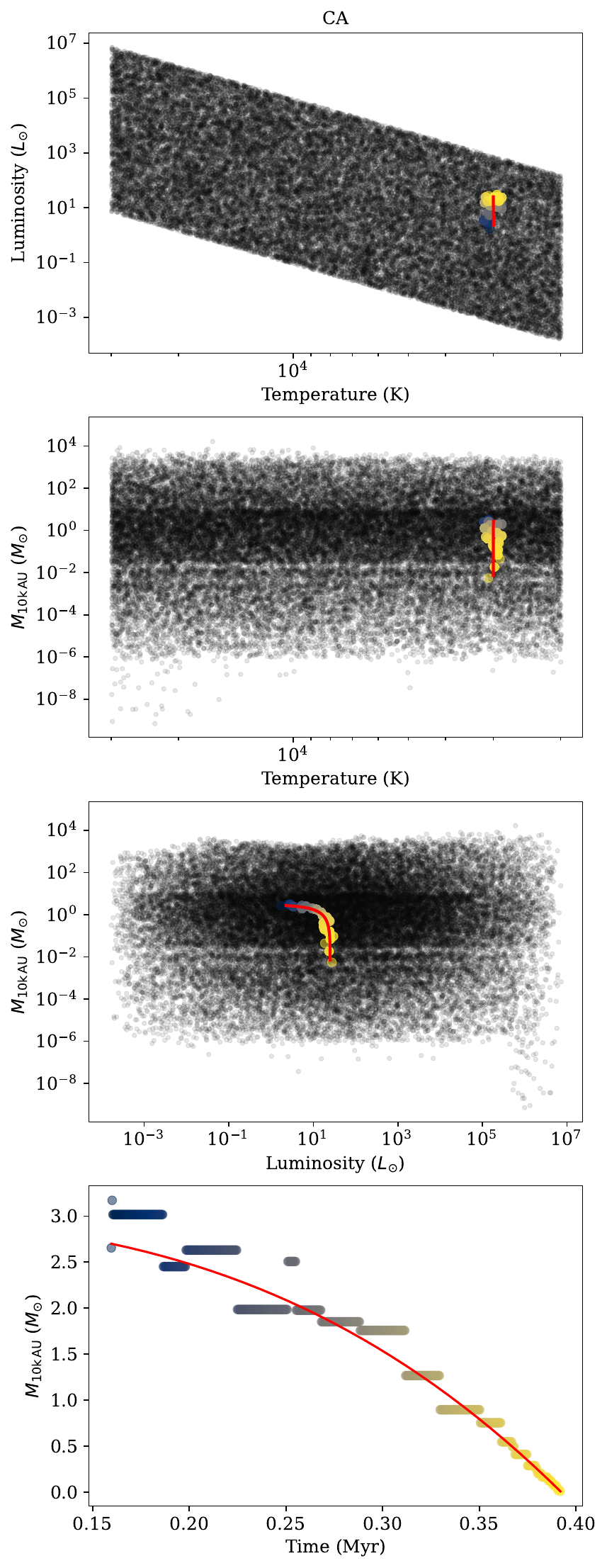}
    \caption{2D projections of the 3D parameter space constructed from stellar temperature, stellar luminosity, and surrounding core mass of all models in one of the R24 geometries. Evolutionary tracks for $1M_{\odot}$ stars generated by our modified K12 code are traced in red. We show IS (\textit{left}), TC (\textit{middle}), and CA (\textit{right}) histories. The nearest-neighbor RTM (per \S\ref{sec:2.3}) to the track at each time step is highlighted. Coloration is determined by time.}
    \label{fig:modeltrack}
\end{figure*}

This step-by-step search allows us to tie any modeled protostellar evolutionary theory to a series of proximal YSO RTMs in our set. We demonstrate nearest-neighbor selection in our parameter space for IS, TC, and CA accretion histories in Figure \ref{fig:modeltrack}. This figure shows the single nearest neighbor to each point on the tracks and restricts itself to our second-most-populous model geometry (see \S\ref{sec:2.1} for more detail) for the sake of visual clarity. In practice, we base our modeled SED on a larger set of RTMs and average over more neighbors to reduce the noise.

\subsection{Proximity}\label{sec:2.3}
In the previous section, we laid out our procedure for creating models of evolving YSOs: tracking the nearest RTMs to PEMs. The question of how to define ``nearest" in our parameter space, however, is a substantive one. Working with different definitions results in different RTMs being identified as ``nearest", which has repercussions for the predicted flux values and uncertainties obtained by averaging over multiple nearest neighbors.

The most common way to define distance in 3D space is by the Cartesian distance metric, $ds^2\,=\,dx^2+dy^2+dz^2$. This is a definition that we can easily adopt, but is not necessarily optimal. The range of values in each of our parameters spans at least two orders of magnitude, meaning that employing the Cartesian metric will not capture distance evenly at different magnitudes. Beyond this numerical consideration, a Cartesian metric is built on the assumption that each included dimension has the same underlying physical significance. For our parameters, this is not the case. Each corresponds to a physical quantity that means something different for the YSO it describes, meaning that a metric that does not allow for independent handling of each dimension may not be desirable.

We adopt an alternate method for characterizing distance that addresses these concerns. Starting from our initial data set of the source temperatures/luminosities and core masses of every YSO with a circumstellar envelope, we use this data to construct a quantile transformer. This transformer maps each parameter to a uniform distribution between 0 and 1 independently by estimating the cumulative distribution function (CDF) of the parameter values. Once this transformer is constructed, we apply it to the evolutionary track and find the nearest neighbors by minimizing the Cartesian distance in this transformed space. Proximity to the evolutionary track is therefore characterized using the following equation:
\begin{align}
\begin{split}
    D_{\rm quant}^2 = \left({\rm QT}_T\{T_{\rm RTM}\}-{\rm QT}_T\{T_{\rm PEM}\}\right)^2+\\
    \left({\rm QT}_L\{L_{\rm RTM}\}-{\rm QT}_L\{L_{\rm PEM}\}\right)^2+\\
    \left({\rm QT}_M\{M_{\rm RTM}\}-{\rm QT}_M\{M_{\rm PEM}\}\right)^2
    \label{eq:quantdist}
\end{split}
\end{align}
for the quantile transform of a quantity $x$ QT$_x$. This scheme for determining distance adequately addresses the issues of straightforward Cartesian distance. Transforming independently based on the CDF of each parameter effectively allows the distance in each parameter to be considered within its own physical context, while mapping to a uniform distribution aligns the value ranges of the disparate dimensions.

We find that this quantile-transform approach produces good results with low uncertainty (see \S\ref{sec:3.2}) and can be easily applied to any set of parameters, making it a very attractive general-purpose definition for distance. The results we present throughout this paper therefore use this definition. In the course of research, we have devised numerous alternate conceptions of ``distance"; Appendix \ref{ap:comp} discusses these alternates, but the quantile-transform approach generally matches or outperforms them.

\section{Results}\label{sec:3}
In this section, we provide an overview of the results that can be obtained from YSO models generated with our framework.

\subsection{Flux predictions}\label{sec:3.1}
\begin{figure*}
    \centering
    \includegraphics[width=0.49\textwidth]{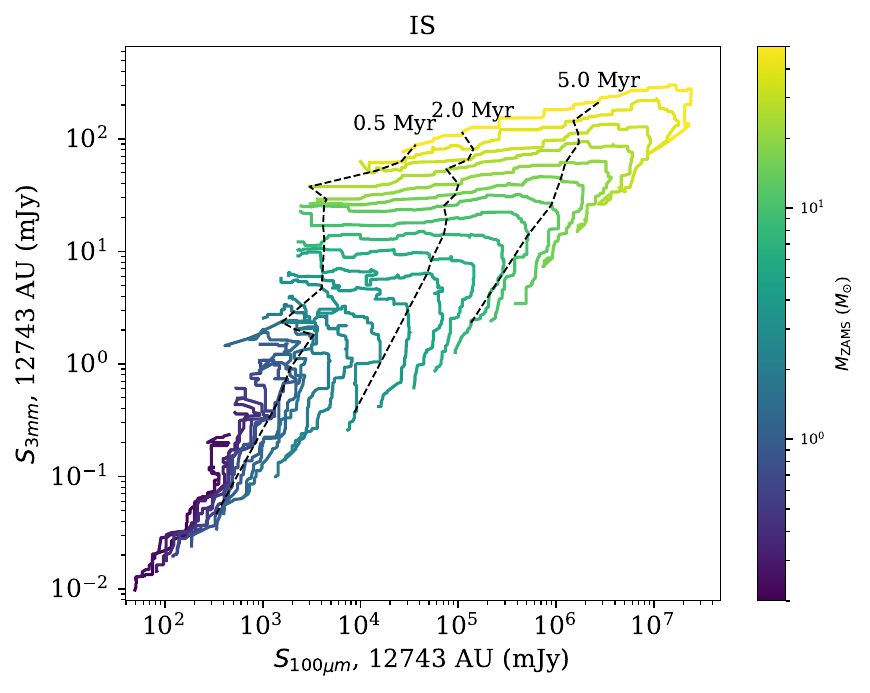}
    \includegraphics[width=0.49\textwidth]{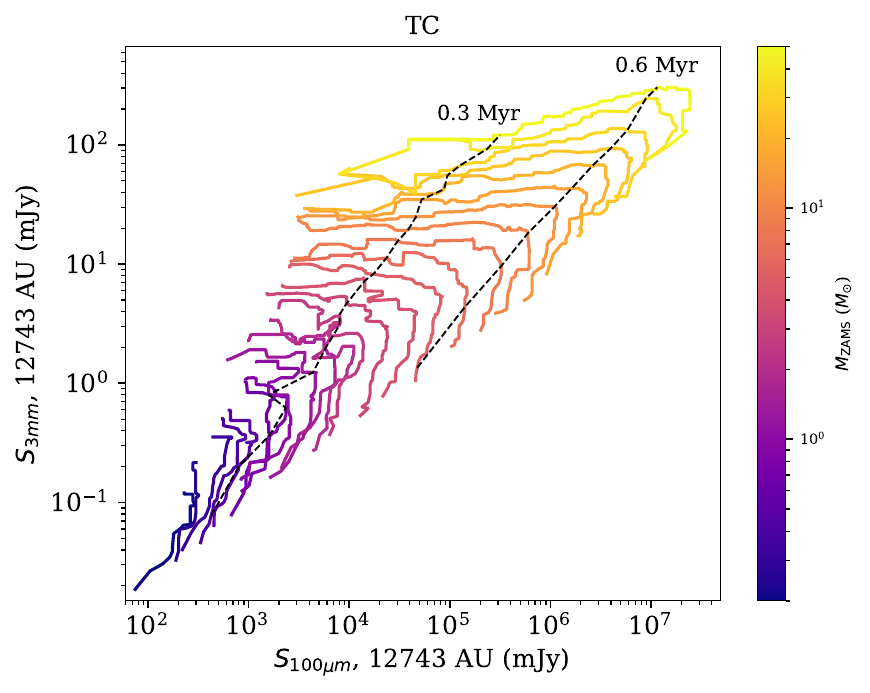}
    \includegraphics[width=0.49\textwidth]{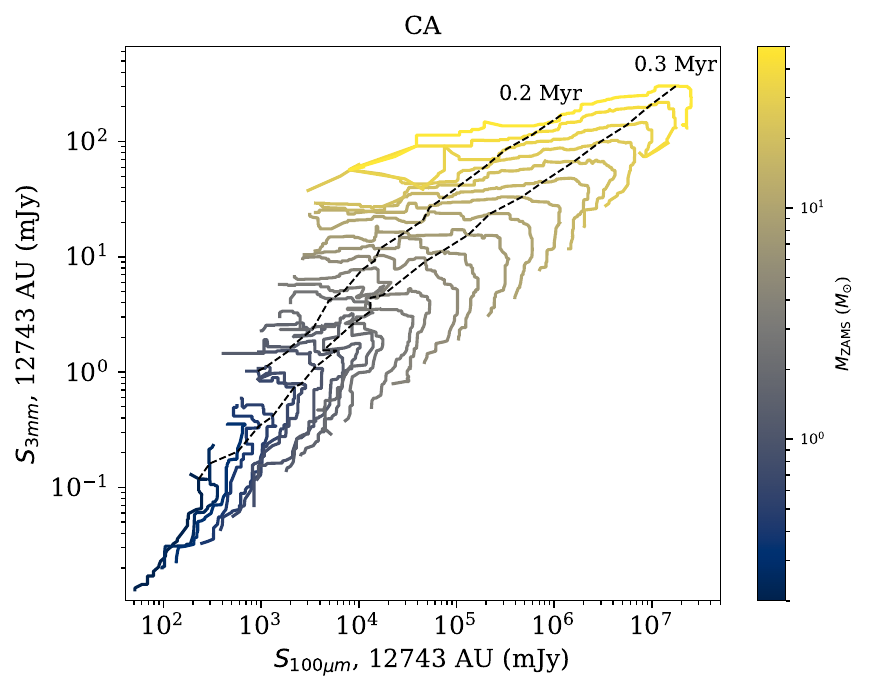}
    \includegraphics[width=0.49\textwidth]{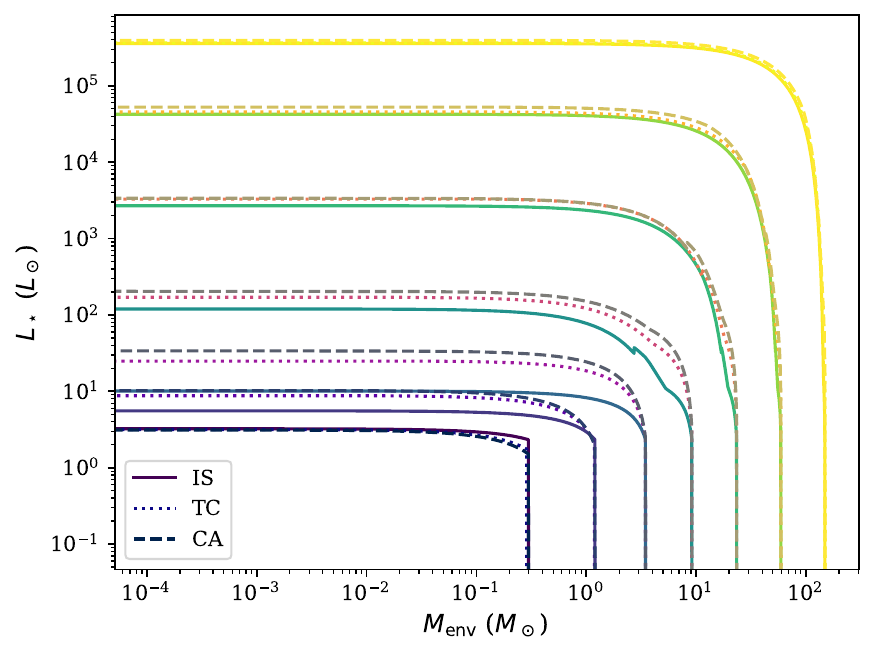}
    \caption{The 3-mm vs. 100-$\mu$m flux of evolving YSOs, constructed through our selection procedure (\S\ref{sec:2.2}). We show IS (\textit{top left}), TC (\textit{top right}), and CA (\textit{bottom left}) histories. Tracks correspond to zero-age stellar masses evenly log-spaced between $0.2-50$ $M_{\odot}$. Each line spans the ignition of a source to depletion of the surrounding mass reservoir, roughly from left to right. Coloration is determined by final stellar mass ($M_{\rm ZAMS}$).
    The flux values plotted here have been smoothed by taking a rolling median of the predicted flux values at each time step; the number of time steps included in the median depends on the length of the track, with tracks corresponding to more massive stars including more time steps.
    We plot isochrones (\textit{black, dashed}) to track the passage of time. There is a clear distinction between the movement of models produced by each history within this flux space. In the bottom right, we plot the evolution of the envelope mass and total bolometric source luminosity corresponding to a subset of evolutionary tracks from all three histories; time travels in the direction of decreasing envelope mass. The plotted tracks correspond to the same set of final stellar masses for each accretion history; color bars remain the same.}
    \label{fig:protohr}
\end{figure*}

The primary functionality of our framework is predicting the flux exhibited by YSOs across time for a given theory. An example is Figure \ref{fig:protohr}, which shows the time evolution of YSO flux at multiple wavelengths for IS, TC, and CA histories. In this figure, we track the flux at 100 $\mu$m and 3 mm. These wavelengths are often used to trace the luminosity and mass of dusty sources, respectively. It is generally assumed that the bolometric temperature of dust peaks around 30 K such that the 100-$\mu$m flux is a good proxy for the total luminosity. Dust is also generally assumed to be optically thin at longer wavelengths, therefore allowing its mass to be well traced by 3-mm emission.

These results can be used to set expectations for how a YSO will appear to an observer over the course of its lifetime. Starting as a clump of gas and dust at stellar densities, a protostar grows by consuming mass from the core and heats its surroundings. For low-mass sources, this growth manifests primarily in a decrease of both 3-mm and 100-$\mu$m intensity. Since low-mass protostars also have low temperatures, the heating from the source is negligible. Most of the dust, and by extension gas, is at the floor temperature of 10 K imposed by the RTM parameters. As such, evolution in flux space for low-mass YSOs is mostly witnessing gas being consumed by the star. High-mass protostars, on the other hand, are able to heat their surrounding material enough to outpace the loss of material. We see a decrease in flux at both wavelengths only once the dust is almost entirely consumed. The resulting ``knee" feature in the high-mass tracks is akin to the transition between the ``accelerating envelope" and ``clean-up'' phases hypothesized in \citet{molinari2008} and \citet{elia2010}, representing the point where the envelope begins to dissipate due to the high energy output of the central source. (Since these phases are conceived using the TC model of accretion, we do not claim exact congruence with this scenario, but note that the general picture of YSO evolution painted by these predictions is qualitatively similar.)

The three accretion histories we model (see \S\ref{sec:2.2} for details) lay out distinct visions of protostellar evolution. Our results allow us to quantify the impact these histories have on the expected flux from our modeled YSOs. For a YSO with some final stellar mass, the assumed history does not significantly change the shape of the path it cuts through the flux space. The change instead occurs primarily in the timescale of accretion and evolution, which is highly dependent on assumed history.

Accretion rates for both CA and TC histories exhibit some dependence on both final and instantaneous stellar mass, and IS accretion depends on neither. This causes the accretion following TC or CA histories to be slowed down relative to IS accretion--to varying degrees--for low-mass stars. As an example: in our PEMs, a 0.2 $\msun$ star takes roughly 0.13 Myr to accrete in an IS scenario, approximately 0.25 Myr to accrete for TC, and about 0.36 Myr to accrete for CA. Low-mass stars will therefore deplete their dust on different timescales, and will consequently be visible at long wavelengths for varying times (putting aside that sources near the substellar boundary are difficult to observe in general). 

Conversely, the mass dependences of TC and CA histories accelerate the accretion of high-mass stars relative to an IS scenario to varying degrees. A 5 $\msun$ star takes about 0.4 Myr to accrete following our CA PEM, approximately 0.6 Myr to accrete for TC, and over 3 Myr to accrete for IS. Since these higher-mass protostars are the main driver of 100-$\mu$m flux, the far-IR luminosity of these YSOs will peak on drastically different timescales, and will linger around the peak for different amounts of time as well.

Given the difference in time scales, it is theoretically possible to distinguish the mechanism of accretion at play through observation. CA YSOs should be expected to reach peak 100-$\mu$m and 3-mm flux around the same time, while low-mass YSOs should peak sooner than high-mass YSOs following TC or IS histories (and much more so for IS histories than TC). For a given 3-mm flux, then, populations following different accretion histories should exhibit markedly different 100-$\mu$m fluxes at the same time.
It is possible that YSOs within different mass regimes may operate in different modes of accretion. High-mass stars have been hypothesized to behave more competitively than low-mass stars, which are generally thought to form from the collapse of isolated mass reservoirs \citep{kennicutt2012}. Such a mixed-mode population would consequently exhibit different time behavior in this flux space than any of our modeled PEMs, and therefore likely appear distinct from other accretion histories.

Aside from the long-wavelength emission of a YSO, if the central protostar reaches a high enough mass to begin burning deuterium during the time it is actively accreting, its evolution in temperature and luminosity also varies significantly with the assumed accretion history. This turnover point in our PEMs occurs at about 2 $\msun$ for IS, 4 $\msun$ for TC, and 5 $\msun$ for CA. (Less massive stars still enter a deuterium-burning phase, but only after they have accreted all of their mass according to the PEM.) Changing the accretion history therefore changes the properties of protostellar sources in the selected RTMs. The cumulative effect from variation in source temperature and luminosity is unlikely to have a large influence on the long-wavelength flux of a YSO (see Appendix \ref{ap:comp}) but may have a stronger effect at shorter wavelengths where the stellar SED is more influential. Differences between the evolutionary tracks can be seen in Appendix \ref{ap:plots}.

Our framework opens a new pathway to observationally determining the evolutionary history of a YSO. 
Many previous studies have attempted to compare observations of YSOs to predictions from various theoretical models for protostellar evolution and accretion. 
The majority of these, however, do not link evolution with radiative transfer as this work does. 
Instead, their focus is on ``summary" properties emerging directly from the accretion models, typically the total protostellar luminosity $\ltot$ or envelope mass $\menv$.

Table \ref{tab:papercomp} compares the coverage and predictions made by this work with those of previous entries in the literature. Broadly, these previous entries break down into two main categories: 
those that chiefly compare distributions of observationally derived summary properties to overall model behavior,
and those that focus more on the evolution of individually simulated models. \citet{offner2011}, \citet{duartecabral2013}, \citet{fischer2017}, and \citet{sheehan2022} exemplify the former category. \citet{offner2011} compare the observed distribution of protostellar luminosities to simulated distributions constructed using a wide array of modeled accretion histories; the remainder compare $\ltot$ and $\menv$ from a sample of YSOs against the coverage of protostellar evolutionary tracks in this total-luminosity/mass space. \citet{duartecabral2013} also perform a similar comparison for $\ltot$\footnote{The original paper notates this as $\lbol$, but defines $\lbol$ as the sum of stellar and accretion luminosity, aligning it with the definition of $\ltot$ used by the other works.} and CO momentum flux $F_{\rm CO}$. (The measured luminosities and masses \citet{fischer2017} and \citet{sheehan2022} use for comparison are derived by comparison to RTMs; however, the radiative transfer is not directly connected with an evolutionary history.)

\citet{dunham2010} and \citet{dunham2012} are good examples of works focusing on detailed modeling. These papers, following the lead of \citet{young2005}, derive accretion histories for the components of modeled YSOs, tie these histories to prescriptions for envelope and disk evolution, and create radiative transfer models by using the resulting star/disk/envelope properties as input in a manner similar to model grids such as \citet{zhang2018}. These simulations allow the prediction of bolometric temperature $\tbol$ and luminosity $\lbol$ in addition to total luminosity and envelope mass; however, these papers do not make predictions for fluxes at specific wavelengths. ($\lbol$ is the luminosity derived from integrating over an observed SED. Although $\lbol$ and $\ltot$ are sometimes equated, we distinguish them because $\lbol$ is subject to observational effects including extinction from nonenvelope dust and inclination dependence resulting from spherical asymmetry in YSO geometry.)

The ability to tie modeled accretion histories to fluxes through radiative transfer to produce direct observables across a wide mass range therefore expands the set of tools used to probe YSO evolution. The majority of papers in both categories also focus on modeling the evolution of YSOs over a mass range corresponding to low-mass to intermediate-mass stars, whereas this work is able to extend farther across the stellar mass spectrum.
\begin{deluxetable*}{c c c c c}
    \tabletypesize{\small}
    \tablewidth{\textwidth}
    \tablehead{\colhead{Reference} & \colhead{$M_{\rm \star,\,final}$ ($\msun$)} & \colhead{Accretion Model} & \colhead{Predicted Properties} & \colhead{RTMs Linked?}}
    \tablecaption{Comparison between works matching YSO observations to predicted behavior.}\label{tab:papercomp}
    \startdata
    This work & 0.2-50 & IS/TC/CA &  $S_\nu/\tbol/\lbol/\ltot/\menv$ & Y \\
    \citet{offner2011} & $<$3 ($M_{\rm protostar}$) & IS/TC/CA/2CTC/2CCA\tablenotemark{a}\tablenotemark{b} & PLF\tablenotemark{c} & N \\
    \citet{dunham2012}\tablenotemark{d} & 0.1-3 ($M_{\rm core}$) & MHD driven & $S_\nu/\tbol/\lbol/\ltot/\menv$ & Y \\
    \citet{duartecabral2013} & 0.06-50 & Const./Exp. taper/Taper + bursts & $\ltot$\tablenotemark{e}$/\menv/F_{\rm CO}$\tablenotemark{f} & N \\
    \citet{fischer2017} & 0.12-2.8 & Exp. taper & $\ltot/\menv$ & N \\
    \citet{sheehan2022} & 0.01-5 & Exp. taper & $\ltot/\menv$ & N \\
    \enddata
    \tablenotetext{a}{``2C" refers to two-component accretion models, which blend the base history with IS-like accretion.}
    \tablenotetext{b}{Tapered versions of each are also included.}
    \tablenotetext{c}{Protostellar luminosity function.}
    \tablenotetext{d}{Extends \citet{dunham2010}.}
    \tablenotetext{e}{This paper equates a YSO's observed bolometric luminosity $\lbol$ with its total protostellar luminosity $\ltot$.}
    \tablenotetext{f}{CO momentum flux.}
\end{deluxetable*}

\subsection{Performance}\label{sec:3.2}
We have demonstrated our ability to predict the observed flux for a YSO across multiple wavelengths and according to multiple evolutionary theories using theory-agnostic YSO RTMs. However, the utility of these predictions is limited without an understanding of their accuracy and level of uncertainty. To characterize the performance of our framework, we attempt to recover the SEDs of existing R24 models following our nearest-neighbor averaging approach. Recovery is performed as in Section \ref{sec:2.2}, with the difference that each model is excluded from the set of models used to recover its flux (i.e. we perform a kind of leave-one-out cross validation of the method on the models used for prediction).

\begin{figure*}
    \centering
    \includegraphics[width=0.49\textwidth]{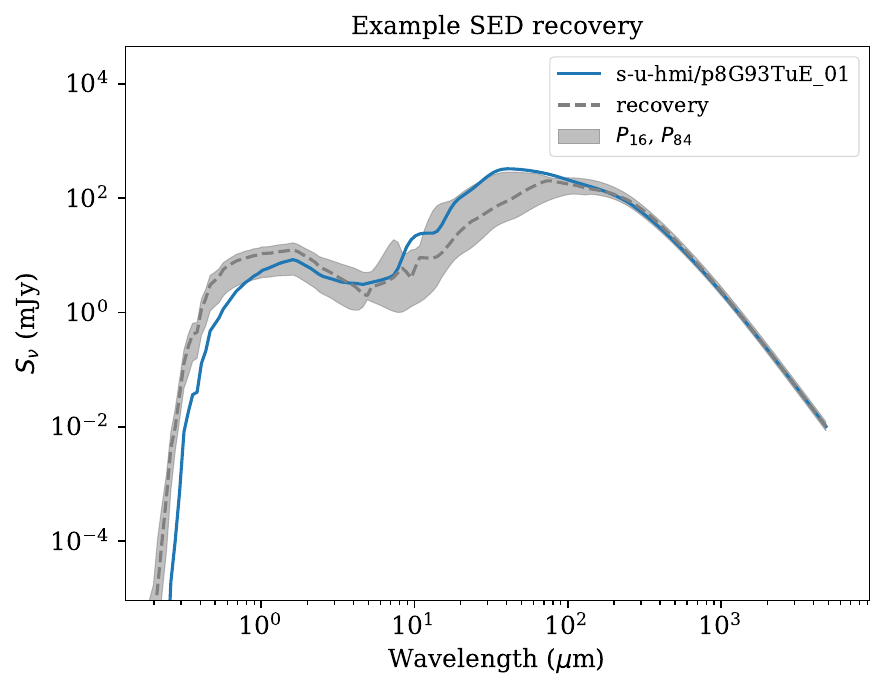}
    \includegraphics[width=0.49\textwidth]{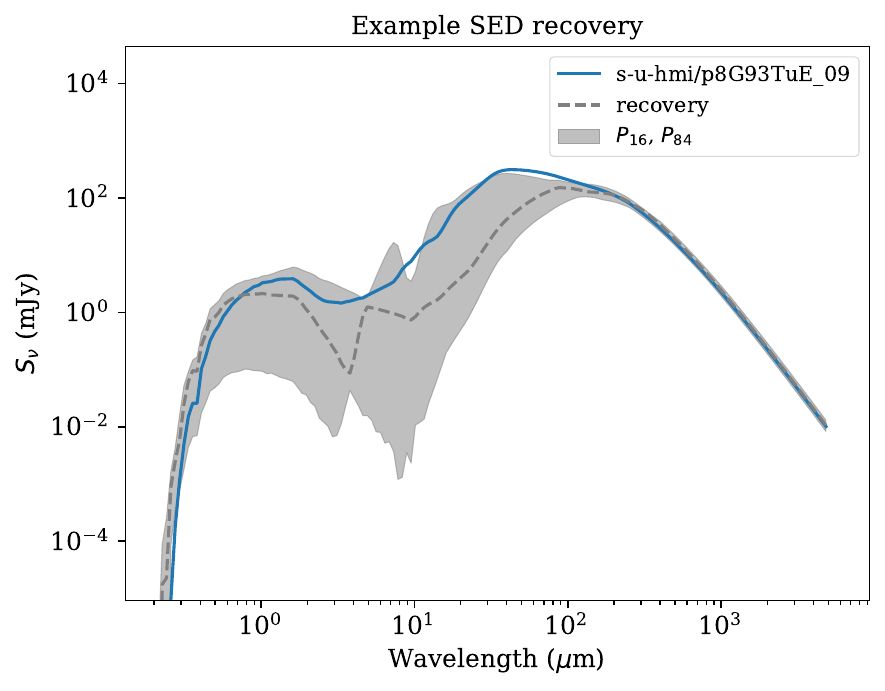}
    \includegraphics[width=0.49\textwidth]{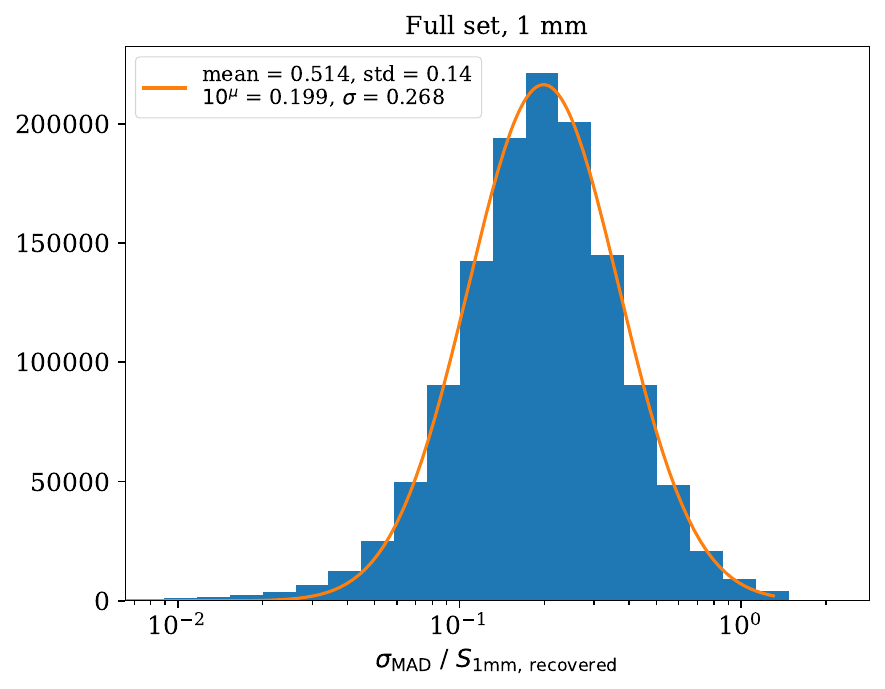}
    \caption{\textbf{Top row:} An SED from the \texttt{s-u-hmi} geometry in the R24 model set (\textit{blue}) plotted against our recovered SED (\textit{gray, dashed}). The shaded region indicates the region between the 16th and 84th percentile of the SEDs used in the reconstruction; percentiles are plotted here to avoid negative values in regions with $\smad$s greater than the associated flux. We show recovery for the same SED at both face-on (\textit{left}) and edge-on (\textit{right}) inclinations. The model identifier for the SED is shown in the legend, and the trailing digits indicate the inclination bin (see \S\ref{sec:2.1}). \textbf{Bottom row:} $\smad$s (median absolute deviation, scaled to standard deviation) for every recovered SED at 1 mm (as a fraction of recovered 1-mm flux). A log-normal distribution is fit to the histogram. We show the arithmetic mean and standard deviation of the log-normal, along with its shape parameters $\mu$ and $\sigma$.}
    \label{fig:recovery}
\end{figure*}

To quantify the uncertainty in our predictions, we adopt $\smad$ (the median absolute deviation--hereafter MAD--of the 10 SEDs, scaled to standard deviation) as a proxy. $\smad$ is defined as follows:
\begin{equation}
    \smad \equiv k_{\rm normal}\times{\rm median}\left(\left|S_{\nu,\,i}-S_{\nu,\,{\rm true}}\right|_{i=1,\,...,\,n}\right)
    \label{eq:smad}
\end{equation}
where $k_{\rm normal}$ is 1.4826, the scale factor applied to the MAD in order to make it an estimator for the standard deviation of normally distributed data. Our uncertainty quantification is based on the MAD to avoid overweighting larger values. To provide a sense of the uncertainty in the fluxes predicted in Section \ref{sec:3.1}, we calculate $\smad$ for every RTM in our set at a wavelength of 1 mm, which we take to be generally representative of behavior at long wavelengths. Given that R24's RTMs exhibit a wide range of flux values, we standardize between models by considering $\smad$ as a fraction of the recovered flux. We show the distribution of this fractional $\smad$ for every recovered SED in Figure \ref{fig:recovery}, along with an example of full SED recovery. In keeping with our treatment of inclination in our YSO modeling procedure (\S\ref{sec:2.2}), we consider each inclination individually, meaning that each model recovery contributes nine data points to the overall sample.

To characterize the overall behavior of the distribution of fractional $\smad$, we fit it with a log-normal distribution and extract the arithmetic mean and standard deviation of the fit log-normal through its shape parameters ($\mu$ and $\sigma$), using the following equations:
\begin{align}
    {\rm mean\,}=&{\,} e^{\mu + \sigma^2/2}\\
    {\rm std\,}={\,}e^{\mu + \sigma^2/2}&\times\sqrt{e^{\sigma^2}-1}.
\end{align}
We choose a log-normal because $\smad$ is strictly positive and is capable of spanning multiple orders of magnitude, rendering a normal distribution unfit for this use case. The fit indicates that $\smad$ is, on average, about half of the recovered 1-mm flux value, with an arithmetic standard deviation of about 14\%.

\begin{figure*}
    \centering
    \includegraphics[width=0.49\textwidth]{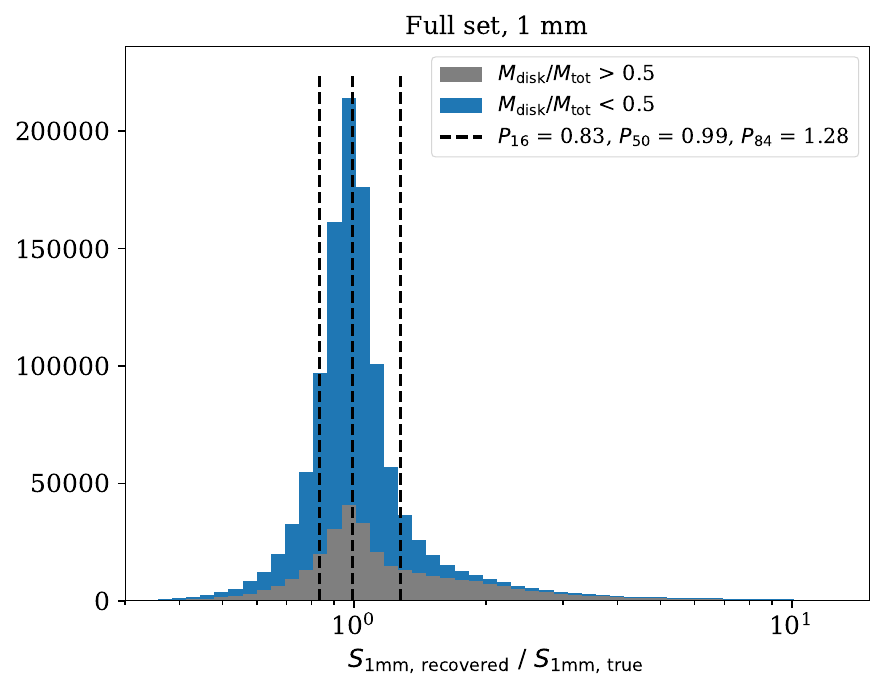}
    \includegraphics[width=0.49\textwidth]{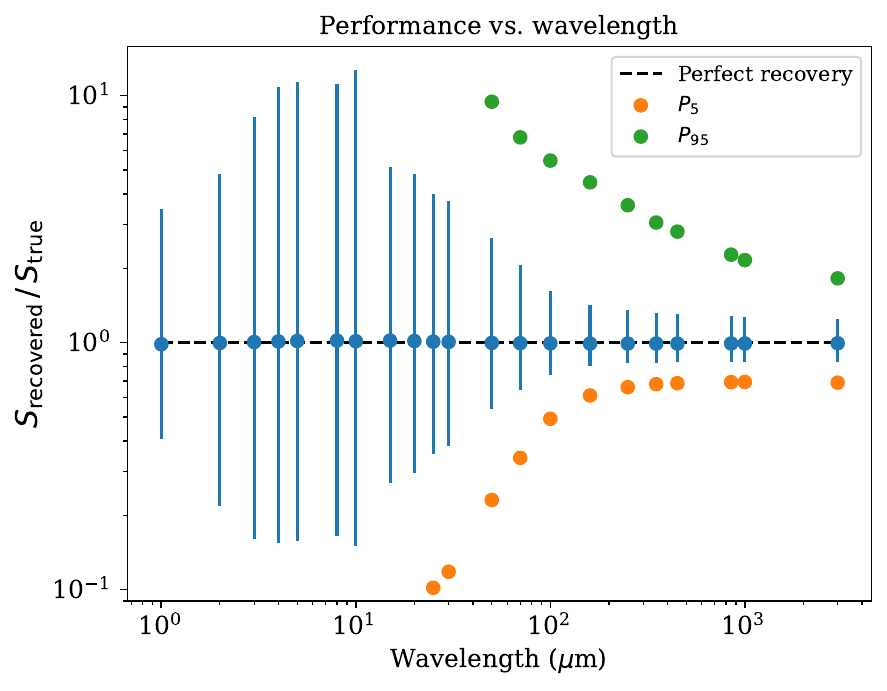}
    \caption{\textbf{Left:} our recovered 1-mm fluxes within an aperture of $\sim$10,000 au as a fraction of true 1-mm flux. Dotted lines indicate the locations of the 16th, 50th, and 84th percentiles of the distribution to indicate its mean and spread. The histogram is broken down into models with over half their total mass contained in a disk (\textit{gray}) and models with less than half their total mass contained in a disk (\textit{blue}). Disk-dominated models exhibit a skew to higher recovered flux that non-disk-dominated models do not. \textbf{Right:} accuracy of our flux recovery at wavelengths commonly used for observation. The error bars show the 16th, 50th, and 84th percentile of the distributions, as in the left panel. The 5th- and 95th-percentile values are also plotted.}
    \label{fig:performance}
\end{figure*}

With the uncertainty of our predicted SEDs characterized, we turn to the accuracy. In Figure \ref{fig:performance}, we evaluate our ability to recover the flux of every RTM in our set across a set of wavelengths commonly used for observations of YSOs, with a particular focus on 1 mm in keeping with our uncertainty quantification. As in Figure \ref{fig:recovery}, each RTM contributes individual data points for each inclination.
Our primary diagnostic tool for evaluating accuracy is the ratio of our recovered flux values to the true fluxes. Since the distributions of these flux ratios are not well fit by a log-normal or other analytical function, we instead characterize them using percentile values, meaning that the 50th percentile serves as the ``mean", and the 16th and 84th percentiles serve as the ``1$\sigma$" bounds.

We now have a comprehensive visualization of the performance of our modeling approach across the spectrum. On the whole, the ``mean" of every distribution is very close to 1, meaning that on average we are able to recover the true flux of models at every considered wavelength. However, the spread in these distributions is uneven across wavelengths. At submillimeter and millimeter wavelengths (spanning the range covered by Figure \ref{fig:protohr}) we are generally able to recover the true value of our RTM SEDs within about 20\%-30\%, meaning that predictions at these wavelengths are likely to exhibit a similar performance. In the infrared, however, the 16th and 84th percentiles of the flux ratio distribution can be anywhere from a factor of two to an order of magnitude away from the 50th percentile. While we recover the correct fluxes at these wavelengths on average, the spread is such that individual predictions are not likely to be accurate.

Figures \ref{fig:recovery} and \ref{fig:performance} show the intrinsic uncertainty in flux predictions made on the basis of $T_\star$, $L_\star$, and $M_{\rm core}$ (which are the quantities from PEMs we track; see \S\ref{sec:2.2}). We are able to reproduce the submillimeter and millimeter flux (starting at $\sim$100 $\mu$m) of the base SED with reasonable accuracy and with an uncertainty (i.e. $\smad$) that is consistently less than the predicted value. At shorter wavelengths, however, our predicted fluxes often diverge significantly from the true value, and the $\smad$ of the predicted SEDs (relative to the predicted value) increases. This disparity in performance between long and short wavelengths is tied to the construction of our framework. Long wavelengths are dominated by dust emission, which is highly dependent on dust mass and source luminosity. In contrast, shorter-wavelength radiation depends much more on dust geometry such as disk inner and outer radii, disk flaring power, or cavity opening angle in the near- and mid-IR \citep[e.g.][]{whitney2004,robitaille2006,furlan2016}.

The reduced IR performance of this framework therefore originates in a limitation of star formation theory: $T_\star$, $L_\star$, and $M_{\rm core}$ only predict the IR flux to the observed level of accuracy. In our modeling procedure, we currently marginalize over disk and cavity properties by drawing from a set of RTMs that includes both asymmetric models, which have disks and/or cavities, and spherically symmetric models, which lack those features. It is possible that, were we to include the disk and cavity, we could improve our flux predictions at shorter wavelengths. However, as it stands, there is little understanding of the degree to which these features are actually predictive of a YSO's IR flux. Likewise, there are currently no good models for how these features are expected to evolve with time, in our PEMs or otherwise. Higher uncertainty at shorter wavelengths is therefore expected and useful to preserve.

Our goal with this work is to develop a way to predict properties of YSOs like mass and luminosity that are visible on a population level. These observables are generally tracked well by longer-wavelength radiation. Consequently, we prioritize good performance at long wavelengths in order to better recover these key properties, and focus on predictions in the submillimeter and millimeter regimes in the remainder of the work. Overall, our approach to modeling YSO SEDs accurately recovers the long-wavelength fluxes of our RTMs, and therefore allows us to make good predictions for observable quantities that rely on dust continuum emission. It is also possible to utilize our framework to make predictions in the IR, with the knowledge that these predictions will come with markedly greater uncertainties. A discussion on the robustness of these results is contained in Appendix \ref{ap:comp}.

\paragraph{Overestimates in the flux recovery distribution are thick disks} There are a nontrivial fraction of cases where our recovered 1-mm fluxes are over a factor of two larger than the true value, and in general, the distributions of flux ratios skew greater than 1 regardless of wavelength (see Figure \ref{fig:performance}). Flux overestimations occur in cases where the mass of the RTM is largely contained within an optically thick disk. Since the mass that we use to find nearest neighbors includes all circumstellar material (i.e. both envelopes and disks), these RTMs are matched with ones that have the same amount of material, but distributed in a way that renders the dust more optically thin. In turn, this leads to a recovered flux that is higher on average due to more dust being visible in the matched RTMs.

\subsection{Caveats}\label{sec:3.3}
In the previous sections, we presented and evaluated the quality of flux predictions made using our modeling framework. Overall, these predictions are reasonably accurate regardless of where they occur in parameter space, and therefore enable good modeling of a wide range of theoretical scenarios. However, there are some aspects of our framework worth keeping in mind when examining and interpreting our predicted SEDs.

Firstly, to reiterate a point from Section \ref{sec:3.2}, the setup of our modeling framework causes better performance in the submillimeter and millimeter than in the near- or mid-IR. We essentially track properties of the central protostar and the overall dust content, while flux in that regime is also sensitive to the shapes of dust density structures (i.e. disk inner radius) that are not tracked by any available PEMs. The uncertainty for predictions we make is therefore markedly higher in that regime compared to longer wavelengths.

Beyond performance over specific wavelength ranges, our procedure for SED modeling causes the quality of results to be fundamentally dependent on the density of RTMs in parameter space. Since the parameters of R24 models are randomly sampled, flux values predicted using our framework may be vulnerable to reduced accuracy due to a decreased model density in the relevant area. While we acknowledge this possibility, we do not anticipate this to be a major source of error in general. Given the large size of our model set, even after down-selection to geometries with envelopes, the degree to which any combination of parameters within coverage will be underpopulated is limited. (It should also be noted that this density issue is not unique to our approach. While we place a greater emphasis on interpolation between models in our parameter space than most literature model grids, sampling every possible combination of relevant parameters is highly infeasible, hence why no currently used grids attempt completeness in this sense even when based within a single theoretical framework.)

We are also somewhat limited in our ability to model some high-mass stars as a consequence of our chosen set of RTMs. The maximum source temperature in R24 (as in R17) is 30,000 K. This places some constraints on our ability to model the evolution of proto-O stars; based on output from the K12 code, temperatures at or over that level are expected for stars with mass $\gtrsim 16\, \msun$ once they have moved onto the main sequence, regardless of the assumed history. Our predictions for the short-wavelength radiation of high-mass YSOs are therefore likely to be underestimates for YSOs with high-mass MS sources due to artificially low stellar temperatures. High-mass stars will also interact with their environments through ionization, the effects of which are currently not modeled within our framework.

Finally, the PEMs underlying the predictions in Section \ref{sec:3.1} are dependent on scaling parameters that we hold invariant in this work (see \S\ref{sec:2.2}). Changes in the values of these parameters should be expected to impact, at minimum, the absolute timescales of the flux evolution we lay out (although applying a different scaling parameter across the mass range of PEMs is not expected to significantly alter the relative behavior of low- and high-mass YSOs we observe). A change in the scaling parameter will also likely impact the specific 100-$\mu$m flux values we predict, which are sensitive to the accretion rate. The impact of scaling parameters on our predictions will be revisited in a future work.

\section{Further uses}\label{sec:4}
We have outlined the functionality of our modeling framework and provided some examples of the base output. In this section, we leverage our infrastructure to pursue open questions in star formation theory.

\subsection{Comparison to a contemporary grid}\label{sec:4.1}
In Section \ref{sec:3.2}, our focus is largely on being able to reproduce known results within our set of RTMs as a way to evaluate the performance of our modeling approach. However, with that performance evaluated, the wider purpose of our framework is to be able to make predictions with the same level of quality across multiple theories of protostellar growth. Ideally, then, we should be able to reproduce not only our own results but also those of contemporary grids of YSO models purposefully built on those theories. We perform a detailed comparison of our results to the RTM grid of \citet[][ZT18]{zhang2018} as a case study of our ability to replicate the results of other model grids.

\subsubsection{Zhang and Tan (2018) SED recovery}\label{sec:4.1.1}
The ZT18 grid is a set of 432 YSO models based on the theory of TC accretion. Every model is assumed to have a central source, a circumstellar envelope, a disk, and bipolar cavities. ZT18 does not include an ambient medium. The grid has three fundamental physical parameters: core mass $M_{\rm core}$, mass surface density of a star-forming clump containing the core $\Sigma_{\rm cl}$, and protostellar mass $m_\star$. Model parameters are sampled from within this space along protostellar evolutionary tracks generated using \citet{hosokawa2009} and \citet{hosokawa2010}. These tracks enable the calculation of other quantities, such as the source temperature and disk mass, at the sample points along the track.

The SED associated with each RTM is calculated using the radiative transfer code \texttt{HOCHUNK3D} \citep{whitney2003a,whitney2013}. Every SED has 20 viewing inclinations, evenly sampled in $\mu\equiv\cos{\theta}$ over the interval of (0.975, 0.025), for a total of 8640 SEDs in the entire grid. The SEDs are entirely produced by the modeled YSOs; no emission from a theoretical parent clump is included.

\begin{figure*}
    \centering
    \includegraphics[width=0.505\textwidth]{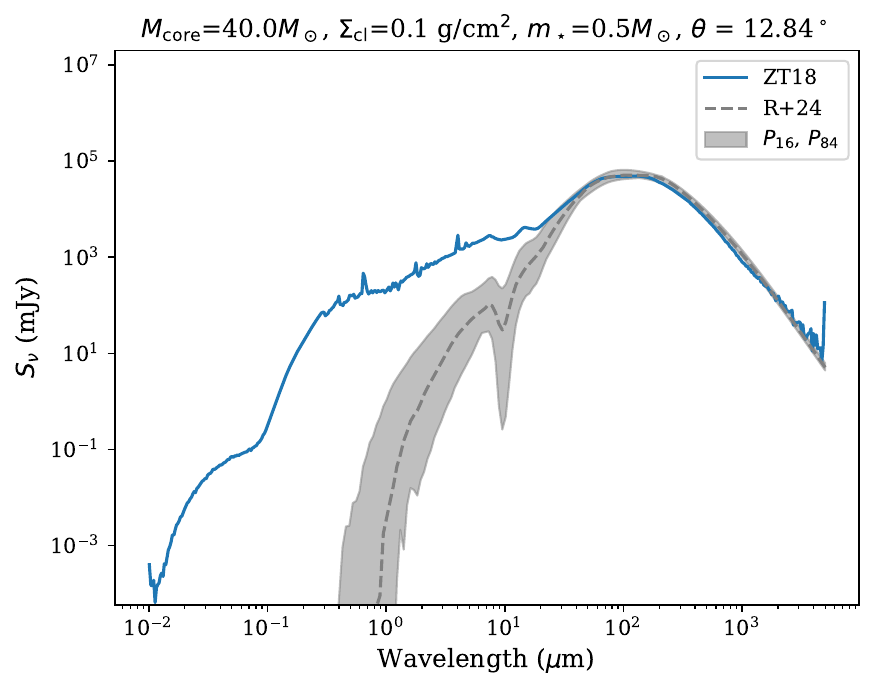}
    \includegraphics[width=0.485\textwidth]{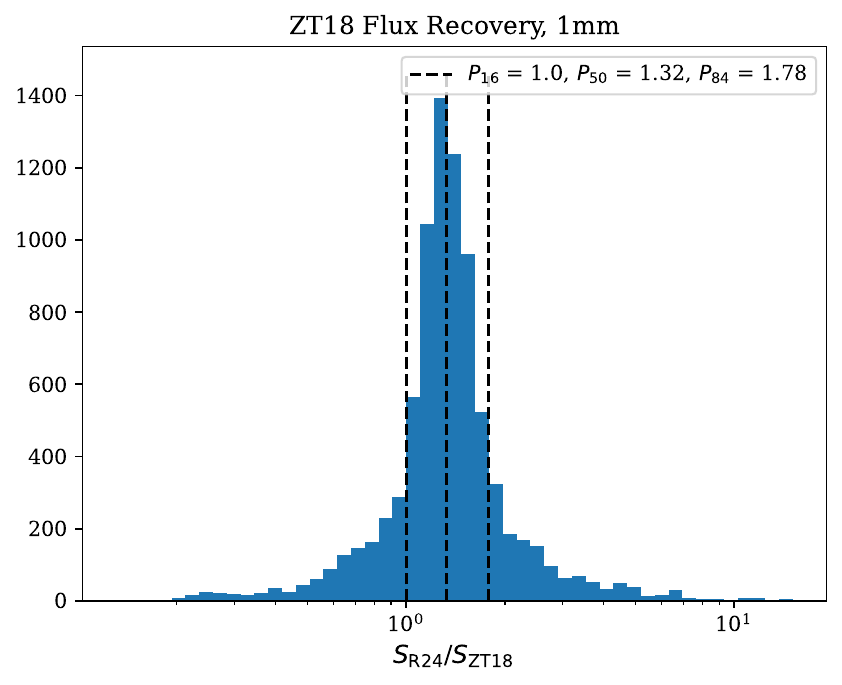}
    \includegraphics[width=0.485\textwidth]{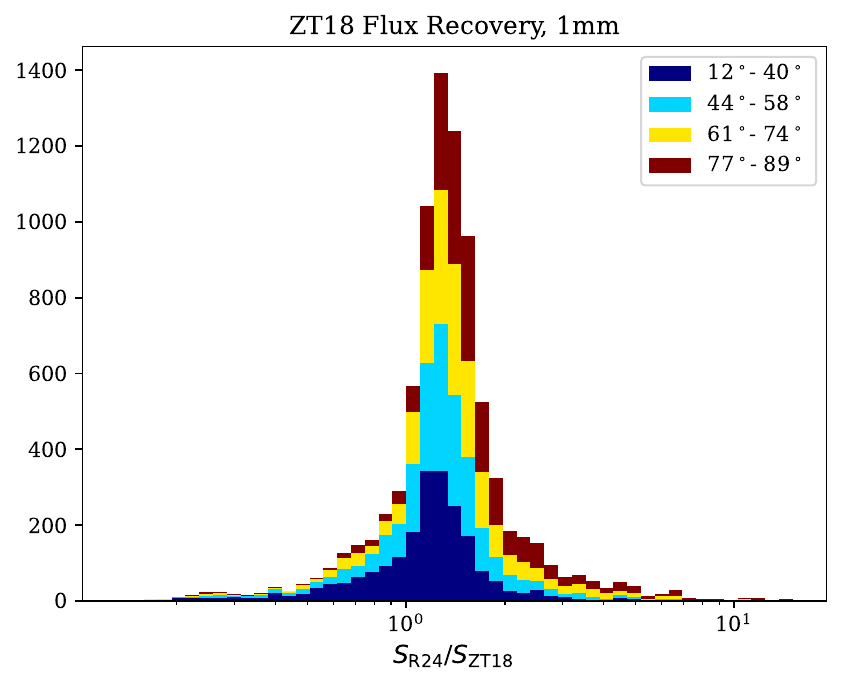}
    \includegraphics[width=0.495\textwidth]{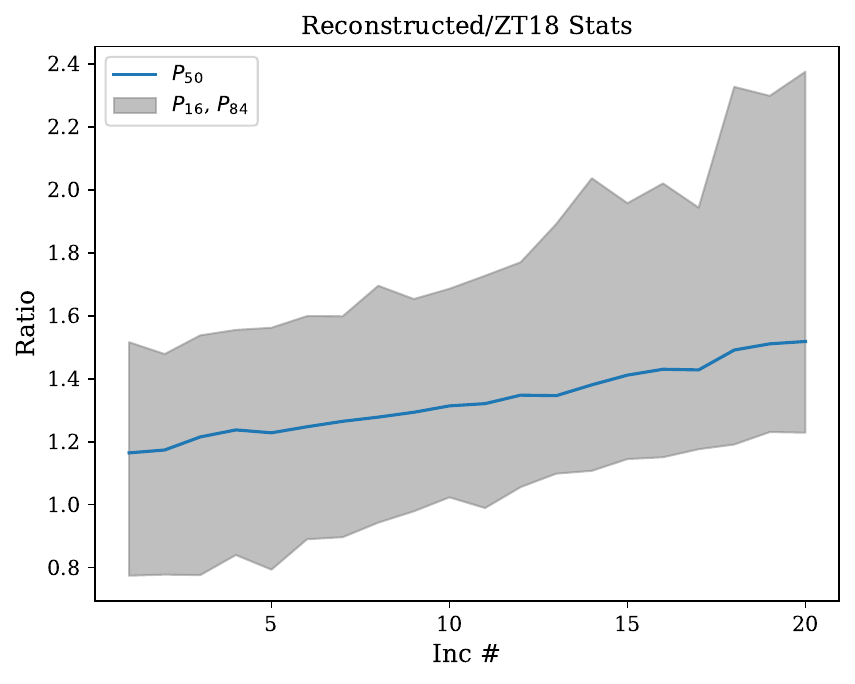}
    \caption{\textbf{Top left:} the SED of a model in ZT18 (\textit{blue}), along with our reproduction (\textit{gray}) and its uncertainty (\textit{shaded}), as in Figure \ref{fig:recovery}. \textbf{Top right:} the left panel of Figure \ref{fig:performance}, but comparing our predicted 1-mm fluxes to the models of ZT18 instead. As there, we indicate the locations of the 16th, 50th, and 84th percentiles as proxies for the mean and 1-$\sigma$ of the distribution. \textbf{Bottom left:} the histogram in the top right panel broken down into four viewing angle bins. The bins contain five inclinations each from the set of inclinations defined in ZT18. \textbf{Bottom right:} the 50th percentile (\textit{blue}) and 16th/84th percentiles (\textit{shaded}) of the distribution of flux ratios at each inclination in ZT18.}
    \label{fig:ztcomp}
\end{figure*}

To test our ability to recreate the results of ZT18, we attempt to recover the 1-mm flux of each RTM in the same way as with our own models (see \S\ref{sec:3.2}, Figure \ref{fig:performance}). Each ZT18 RTM has an instantaneous core mass $\menv$ forward-modeled from the initial core mass, accretion rate, and assumed $\esf$ as well as a disk mass $M_{\rm disk}$ tied to the mass of the central protostar. We use the combined envelope and disk mass for RTM selection in order to remain consistent with our internal treatment. Since the envelopes of ZT18 models have defined outer radii (necessary due to the lack of an ambient medium), we use the masses and fluxes associated with the closest radius in the set of apertures for R24 (see \S\ref{sec:2.1}) to ensure that we capture the ZT18 models appropriately. Since each ZT18 RTM has 20 associated SEDs corresponding to different viewing angles, we compare each ZT18 SED to our recovered SED within the appropriate 10$^\circ$ inclination bin (see \S\ref{sec:2.1}) to match the lines of sight as closely as possible. We show results from this comparison in Figure \ref{fig:ztcomp}.

Overall, as with our own flux values, we are able to recover the long-wavelength fluxes of ZT18's RTMs fairly well. The mean ratio of recovered flux to true flux is approximately 1.32, meaning that on average we overestimate the true flux by $\sim$30\%; however, this offset is systematic and can consequently be compensated for. (We discuss the origin of this offset further in \S\ref{sec:4.1.2}.) With that adjustment, we are generally able to recover the true flux of any ZT18 SED to within approximately 30-45\%, based on the 16th and 84th percentile values; there is a slight bias toward the upside in these results, with the 84th percentile being further away from the median than the 16th. This performance is generally consistent with our results in Section \ref{sec:3.2} in the millimeter regime, although the difference between percentiles is slightly larger. The spread is likely exacerbated relative to our internal recovery by a smaller sample and by the noise in the long-wavelength fluxes of ZT18, which is an artifact from radiative transfer. The bottom panels of Figure \ref{fig:ztcomp} reveal an inclination dependence in the ratio of recovered flux to true flux. We generally overestimate the flux of highly inclined SEDs (i.e. closer to edge-on) by slightly more than that of models closer to face-on, with the mean ratio increasing from 1.16 at an inclination of 12$^\circ$ to 1.52 at an inclination of 89$^\circ$.

As in our own internal recovery, we are not able to recover the shorter-wavelength fluxes of ZT18 to the same level of fidelity as at long wavelengths; flux at shorter wavelengths has a strong dependence on geometric features we marginalize over due to a general lack of models for their evolution (see \S\ref{sec:3.2} for more discussion). ZT18 does contain prescriptions for some of these features (disk outer radius, cavity opening angle); we continue to marginalize over these in order to better capture the uncertainty in our current approach.

\subsubsection{The impact of model construction}\label{sec:4.1.2}
In reconstituting SEDs from ZT18, we have attempted to follow their parameters and structure as closely as possible within the context of our framework. However, there remains a systematic offset between their long-wavelength fluxes and our attempted reproductions, the accuracy of which also exhibits an inclination dependence. Since we are able to reproduce the fluxes of our own models at long wavelengths, these disparities are likely due to the varying ways in which the sets of models are constructed. Where the differences originate is a substantive question, as it indicates ways in which the construction of a set of models may affect the predictions it makes. In turn, this may introduce additional uncertainty into measurements made using the model set. We evaluate the effect of two major differences between R24 and ZT18: the different dust opacities used by the model sets, and the differences in the treatment of disks.

\begin{figure}
    \centering
    \includegraphics[width=0.8\linewidth]{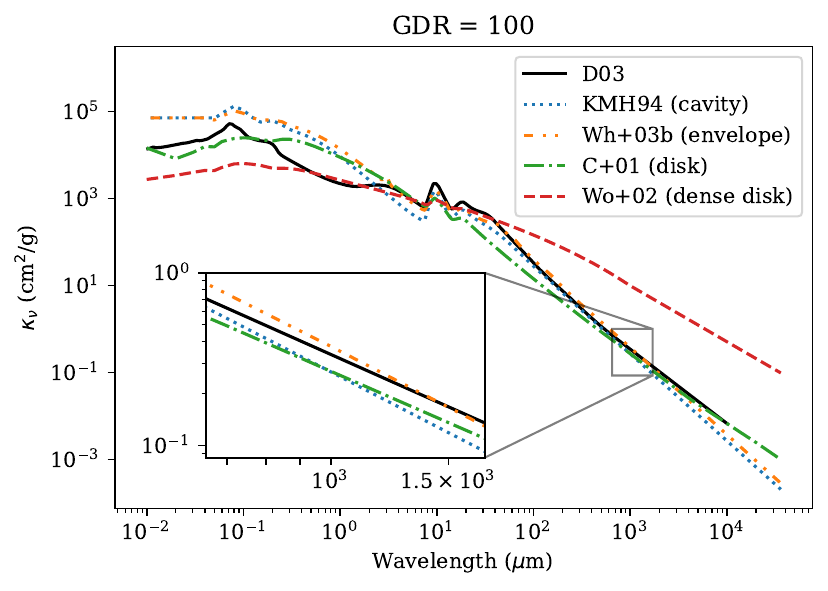}
    \caption{Dust opacities (in cm$^2$ g$^{-1}$) for each dust type used in the RTMs of ZT18, compared to the D03 dust used in R24. We assume a GDR of 100 to place all values in terms of dust opacity, as opposed to total material opacity. The ZT18 opacity models originate from \citet{kim1994}, \citet{whitney2003b}, \citet{cotera2001}, and \citet{wood2002}.}
    \label{fig:dusts}
\end{figure}

As stated in Section \ref{sec:2.1}, R24 employs a single dust type sourced from D03 for every density structure. By contrast, dust in ZT18 follows the configuration of \citet{whitney2003a}, which varies the model used by region within a YSO. Disks in ZT18 include two dust species, separated by a density threshold. Dense regions in the disk are modeled with the large-grain dust from \citet[][W02]{wood2002}; these regions have gaseous hydrogen number density $n_{\rm H} > 2\times10^{10}\;{\rm cm}^{-3}$. Regions in the disk below this density use a model of intermediate grain size from \citet{cotera2001}. Envelopes contain the ice-covered grains from \citet{whitney2003b}. Dust in the cavity uses the small ISM grains of \citet{kim1994}. The opacities of these dust models are illustrated in Figure \ref{fig:dusts}; they are all distinct from that of D03, particularly for the dust in dense regions of ZT18's disks.
While the impact of changing the dust model is difficult to predict solely by comparing the opacities, it is reasonable to expect the use of these differing dust models to alter the resulting emission.

To quantify the impact of dust opacity model on the resulting SED, we rerun a subset of R24 models with ZT18's dust configuration to isolate the effect of the dust opacity. This subset is composed of 500 randomly selected RTMs from the \texttt{spubhmi} and \texttt{spubsmi} geometries (250 each). We choose these geometries to ensure that every RTM being rerun has envelopes, disks, and bipolar cavities, which are the common features of ZT18's RTMs. Since \texttt{Hyperion} is capable of including multiple dust models and handling the properties of envelopes, disk, and cavities independently, we are able to reproduce the setup of dust in ZT18, although with the addition of an ambient medium, which is required when setting up an R24 RTM. As in R24, we set the temperature of the medium at 10 K and its dust density at 10$^{-23}$ g cm$^{-3}$. We assign the dust model of \citet{kim1994} used in the cavities to the ambient medium as well, due to the similarly low density. To determine which parts of the disk are above the density threshold for the larger-grain dust species, we divide the mass density of dust in each disk cell by the mass of hydrogen to arrive at a number density, which is scaled assuming a GDR of 100. For our reruns, we retain the same grid configurations as the original R24 RTMs and postprocess the SEDs in the same way by subtracting the background radiation from the ambient medium, making S/N cuts, and interpolating to a common set of apertures (see \S4.2.4 of R17 for details).

As a caveat, our reruns are not an exact match to the full model setup of ZT18, which also includes gas opacities, adiabatic heating/cooling, and advection for the purpose of outflow modeling and providing corrections to calculated thermal energies \citep{zhang2011,zhang2013}. We do not implement these additions. While they do impact the temperature profiles of YSOs, that impact is primarily felt in very hot regions ($>$10$^4$ K) and at the edges of outflows, which contribute small fractions of the total flux at the wavelength considered herein. The effects on the radiative transfer from including additional physics are therefore expected to be secondary to those from the dust model at long wavelengths.

\begin{figure*}
    \centering
    \includegraphics[width=0.4975\textwidth]{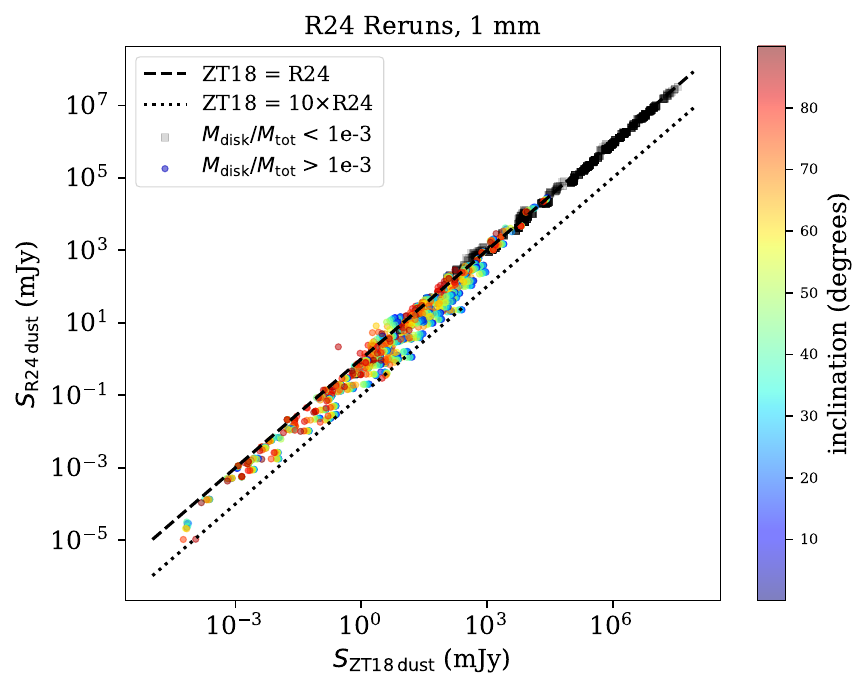}
    \includegraphics[width=0.4825\textwidth]{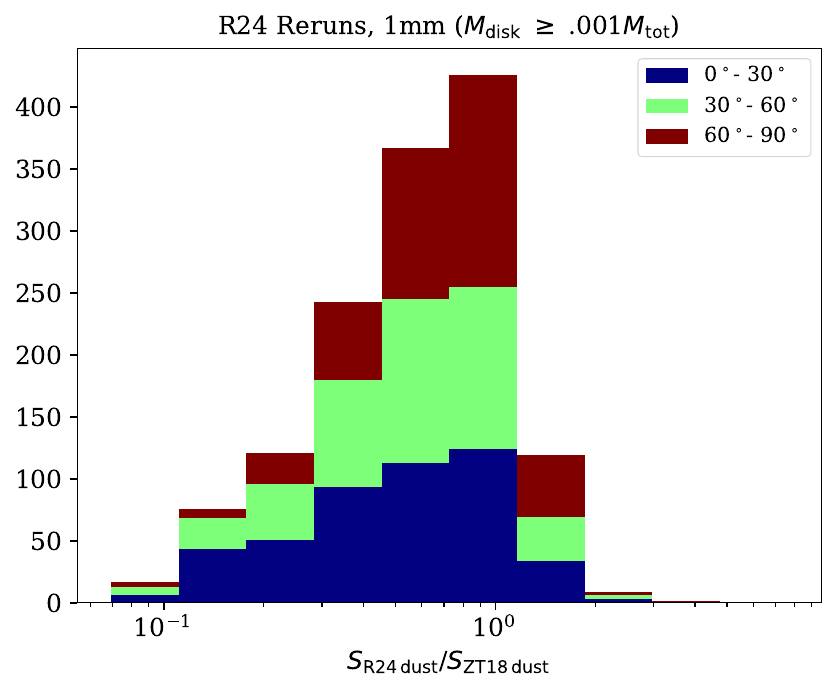}
    \caption{\textbf{Left:} the 1-mm fluxes of R24 models rerun with the ZT18 dust configuration, plotted against the original fluxes. We separate the models into two regimes based on the fraction of total model mass contained in the disk. Models with disks comprising less than 0.1\% of the total mass are represented by black squares, while models with disks greater than 0.1\% of total mass are colored by inclination. \textbf{Right:} the same as the bottom middle plot of Figure \ref{fig:ztcomp}, but comparing the original R24 fluxes to the rerun models, and split into three inclination bins with three viewing angles each to better match the setup of R24. Only models that are in the ``disky" regime in the left panel (i.e. models with $M_{\rm disk}/M_{\rm tot} > 10^{-3}$) are included in this histogram.}
    \label{fig:reruns}
\end{figure*}
In Figure \ref{fig:reruns}, we compare the 1-mm fluxes of our rerun RTMs to the originals. In general, fluxes from the rerun models are higher than their counterparts with R24's dust opacity model. Beyond that larger trend, the reruns generally fall into two categories separated by the ratio of disk mass to envelope mass. The first category is RTMs where the disk is less than $\sim$0.1\% of the total mass in the model. The correlation between the original flux and the flux in the rerun model is very linear; most rerun models in this group are approximately 15\%-20\% brighter than the originals, with very little variation. The majority of these models also exhibit fluxes greater than 10$^3$ mJy. The second category is composed of RTMs where the disk is 0.1\% or more of the total mass. Fluxes of models in this category exhibit more spread around a linear correlation and also an inclination dependence, with the discrepancy between the rerun and original flux growing the closer a viewing angle is to being face-on. The average increase in flux for this group of models is approximately 70\%-80\% of the original, although with a higher variance.

With the information from these rerun RTMs, we return to our recovery of ZT18 RTM fluxes in Section \ref{sec:4.1.1}. The difference in dust model provides a straightforward explanation for the inclination dependence in our recovery, as almost every RTM in ZT18 falls into the ``disky" regime we observe in the rerun R24 RTMs. However, since the net effect of changing the adopted dust opacities is to increase the fluxes relative to the originals, the difference in dust configuration is unlikely to be the sole cause of the systematic offset between ZT18's fluxes and our recoveries for the simple fact that we tend to overrecover the fluxes of ZT18.

Since the difference in dust opacities is not enough to explain the systematic flux offset in and of itself, we turn to the other major disparity between the setup of ZT18 and R24; the properties of their circumstellar disks. Disks in ZT18 are generally more massive (mass between .16 and 53.3 $\msun$, median 2 $\msun$) and more compact (outer radius between 5 and 1354 au, median 70 au) than disks in R24 (mass between $10^{-6}$ and $10$ $\msun$, median $\sim 3\times10^{-3}$ $\msun$, outer radius between 50 and 5000 au, median $\sim$500 au)\footnote{These mass values refer to the total mass in the disk, assuming a GDR of 100 as both ZT18 and R24 do.}, which creates disks that are higher density in ZT18 than in R24 on average. (It should be noted that many of the massive and dense disks in ZT18 would likely be considered Toomre unstable, and are therefore unlikely to contribute meaningfully to the observed radiation from populations of YSOs. As constructed, then, the ZT18 models have components that are potentially physically unrealistic, or at least unlikely to have observable real-world analogues.) Disks in ZT18 also include W02 dust, which has a much higher opacity than D03. While dust opacity is insufficient to explain the nature of the offset alone, we examine the possibility that this very opaque dust model is interacting with the construction of ZT18 disks to affect flux recovery.

\begin{figure*}
    \centering
    \includegraphics[width=0.7\linewidth]{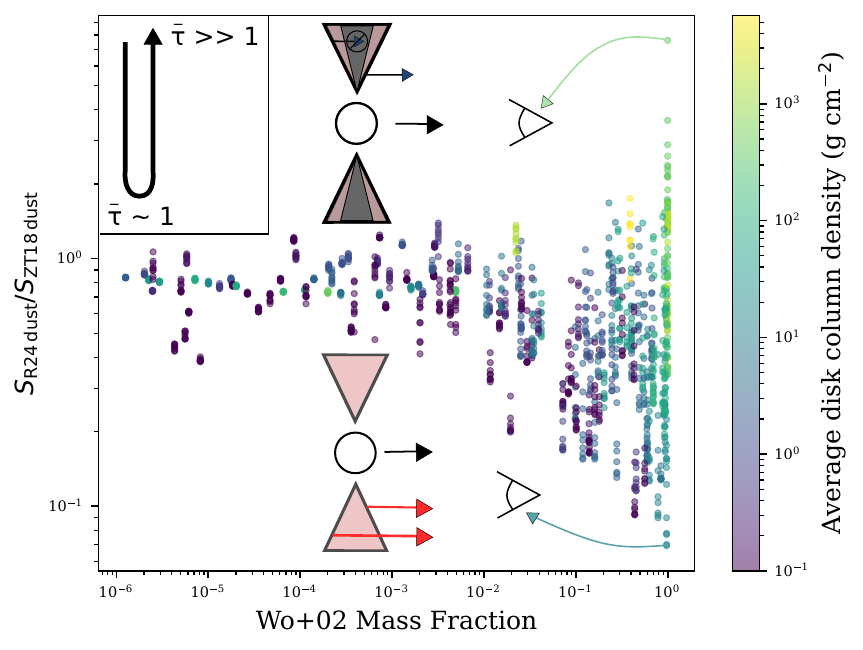}
    \caption{The ratio of 1-mm R24 flux to rerun flux for our set of rerun models, plotted against the fraction of mass in the model in W02 dust. Models are colored by the average disk column density. The $x$-axis is limited to mass fractions of over 10$^{-6}$; some models have lower W02 mass fractions, but those models fall into the ``nondisky" regime in Figure \ref{fig:reruns} and do not exhibit any trends in this space. (Models that have become brighter when rerun have an increased $S_{\rm ZT18\,dust}$, meaning that they are below 1 in this plot. We plot the ratio of $S_{\rm R24\,dust}$ to $S_{\rm ZT18\,dust}$ for consistency with Section \ref{sec:4.1.1}, where ZT18 is the ``true" flux and is therefore the denominator.) We illustrate the reason behind the spread in recovered fluxes using cartoon elements. The opaque W02 dust has higher emissivity, powering additional dust emission for disks with low optical depth. However, for denser disks, the high optical depth resulting from W02's high opacity hides dust emission, decreasing the flux.}
    \label{fig:wood}
\end{figure*}

To determine the impact of disk structure in concert with the effect of W02 dust in our reruns, we examine the relationship for each model between its flux ratio, mass fraction in W02 dust, and the column density of its disk in Figure \ref{fig:wood}. 
Increasing the mass fraction of a model in W02 dust has a clear impact on the agreement between the original and rerun models. In general, the more a model is composed of W02 dust, the higher the rerun flux becomes. Since W02 dust has a higher opacity than D03 at long wavelengths, it also has a higher emissivity, meaning that for the same amount of visible W02 and D03 dust, the W02 dust will be brighter. (Models that are envelope-dominated also exhibit a slight increase in flux in the reruns, which in a similar fashion is due to the increased opacity of \citet{whitney2003b} dust relative to D03.) However, there are also cases where a higher W02 mass fraction results in a lesser increase--or even a decrease--in flux. This is a result of the high column density of these models' disks, which prevents the light emitted by W02 dust from escaping due to the resulting high opacity. The spread in flux ratios is therefore essentially visualizing the transition from low to high disk optical depth.

When considering these results within the context of recovering the fluxes of ZT18 models, the combination of disk structure and dust opacity provides a reasonable explanation for our general tendency to overrecover the fluxes of actual ZT18 models. Every disk in ZT18 has an average column density that is roughly 10$^2$ g cm$^{-2}$ or greater, placing them squarely within the ``dense disk" regime in our reruns. The emission from higher-opacity dust which would otherwise power additional flux is therefore blocked due to high optical depth, leading to a recovered flux that is higher than the original. The results of our attempted recovery of ZT18 fluxes can consequently be fully explained through a combination of the assumed dust opacities and model construction. With sufficient knowledge about these aspects of a model grid, then, we are able to produce predictions with our modeling framework that are consistent with predictions made by current grids.

\subsection{YSO classification}\label{sec:4.2}
Despite the plethora of available PEMs, there is a consensus on a general qualitative picture of YSO evolution. This picture is broken down into several evolutionary Stages, with each successive Stage becoming less envelope dominated and closer to a bare pre-main-sequence star \citep{evans2009b,kennicutt2012}. Insight into the physical state of an observed YSO generally comes from its IR spectral index, $\alpha$, defined as follows:
\begin{equation}
    \alpha = \frac{d\log{\lambda F_\lambda}}{d\log\lambda}.
    \label{eq:alpha}
\end{equation}
This index allows the YSO to be assigned one of several Classes, which are empirically determined ranges of spectral index thought to exist at roughly the same Stage. $\alpha$ is generally calculated within the wavelength range of 2-25 $\mu$m, but the actual wavelengths used may differ by work depending on the availability of data and sensitivity to extinction \citep[e.g.][]{mcclure2010,furlan2016}. In this work, we calculate $\alpha$ between 2 and 25 $\mu$m\footnote{We interpret $\alpha$ as the slope of the line connecting the SED at the endpoints, instead of fitting a power law to data points within that range.}.

In practice, these concepts are often conflated, with the Class of a YSO taken to represent its evolutionary Stage. However, since YSO Class is a fundamentally observational quantity, its measurement may be impacted by observational effects. For example, a YSO observed at Stage II (i.e. when it is disk dominated) may appear to be Class I when observed edge-on through the disk due to extinction. Without direct insight into the orientation and spatial structure of dust in a YSO, the extent to which its Stage and Class may be confused is difficult to ascertain through observation.

In R24, we characterized this potential for confusion by comparing the observational Classes and evolutionary Stages of every RTM in the set. In general, we found that while the Classes and Stages of many R24 models aligned as expected, there were significant fractions of RTMs with mismatched Classes and Stages. However, given R24's lack of a foundational evolutionary theory, we did not attempt to narrow the scope of this comparison based on ``physicality", i.e. whether the models comport with a particular modeled accretion history. The results from that paper therefore indicate the ways in which Class and Stage may be confused and provide a general sense of proportion, but do not attempt to represent YSOs as they occur in nature.

The major advancement presented by this work, then, is the ability to determine whether an RTM is ``physical", i.e., whether it can occur assuming a given PEM. We therefore revisit the comparison between the Classes and Stages of our RTMs with this additional constraint, making the results more directly applicable to observed YSOs. We adopt the following definitions for Class and Stage:
\begin{itemize}
    \item Class:
    \begin{itemize}
        \item 0: $L_{\rm 350+\,\mu m}/\lbol$ $>$ $0.005$, $S_{\rm 24\,\mu m}(1000\,{\rm au},\,1\,{\rm kpc}) < 0.1\,{\rm mJy}$
        \item I: $\alpha \geq 0.3$
        \item Flat: $-0.3 \leq \alpha < 0.3$
        \item II: $-1.6 \leq \alpha < -0.3$
        \item III: $\alpha < -1.6$
    \end{itemize}
    \item Stage:
    \begin{itemize}
        \item 0: $\menv\, >\, 0.1\,\msun$, 
        $M_{\star}\, <\, M_{\rm \star,\,final} / 2$
        \item I: $\menv\, >\, 0.1\,\msun$, 
        $M_{\star}\, >\, M_{\rm \star,\,final} / 2$
        \item II: $\menv\, <\, 0.1\,\msun$, disk present
        \item III: Bare pre-main-sequence star (no envelope, no disk)
    \end{itemize}
\end{itemize}
These definitions are generally the same as in R24; however, the dividing line between Stages 0 and I has been changed. This definition incorporates the Stage 0/I boundary employed by \citet[][F17]{fischer2017}, occurring when half of the final stellar mass has been accreted\footnote{We note that F17 implicitly define the Stage 0/I boundary in two ways. The first of these is when the envelope mass equals the star mass, originating from \citet{andre1993}, while the second is when the star reaches 50\% of its final mass. These definitions agree in the specific case where accretion onto a protostar from an isolated mass reservoir is 100\% efficient, in keeping with the singular IS formation of \citet{shu1977}; however, other scenarios would cause these definitions to diverge. We illustrate this point further in Section \ref{sec:4.2.3}.}.
Additionally, we make a small modification to the R24 definition of Class 0. As published, R24 models are assigned Class 0 if they have a submillimeter luminosity ratio $\lumrat>0.005$ and cannot be assigned a Class by their spectral index, i.e. if \texttt{Hyperion} does not predict a flux at either 2 or 25 $\mu$m due to a lack of photons.
In this work, Class 0 includes models that exhibit the appropriate millimeter luminosity ratio and also have 24-$\mu$m fluxes falling under the $\sim$0.1 mJy sensitivity limits of Spitzer. We take $S_{\rm 24\,\mu m}$ in R24's sixth aperture, which has a radius of $\sim$1000 au, at the models' native 1 kpc distance. This expands R24's definition slightly to include models that may have spectral indices, but would not be feasibly detected in the near- or mid-IR. The Class 0 models of this work are therefore more closely aligned with the constituents of recent IR YSO surveys \citep[e.g.][]{stutz2013}.
Section \ref{sec:4.2.3} contains more discussion on Class and Stage definitions.

To determine which RTMs should be included in the comparison, we repeat our procedure for YSO composition from Section \ref{sec:2.2}. For each modeled accretion history, we consider a set of 50 final stellar masses evenly log-spaced between 0.2-50 $\msun$, similar to Figure \ref{fig:protohr}. To translate our PEMs into RTM parameter space, we have so far assumed a mass accretion efficiency $\esf$ of 1/3. We now allow $\esf$ to vary across the set $\esf\in$ (1/6, 1/4, 1/3, 1/2, 2/3) in order to include scenarios where this efficiency varies (ZT18, for example, allows this quantity to vary between approximately 0.2-0.6 as a function of core mass and clump density, which is now accommodated by our value range). Each RTM that is selected along a PEM track is included in the resulting matrix. In cases where the same model is selected multiple times along the same track, it is included in the comparison each time to ensure that the resulting percentages are representative of the full population of selected RTMs. The SEDs of $\theta$-dependent RTMs (see \S\ref{sec:2.1}) are sampled within nine bins with $10^\circ$ widths, providing nine distinct SEDs to contribute to the matrix. However, for observed YSOs, the occurrence rates of inclinations within these bins will be uneven if the inclinations are randomly distributed as expected \citep[e.g.][]{otter2021}.
For better congruence with the expected distribution of viewing angles, each SED from these inclination-dependent RTMs is contributed $\sin(\theta_i)/\sin(5^\circ)$ times, rounded to the nearest number, where $\theta_i$ is the average inclination in each bin. We normalize by $\sin(5^\circ)$, the average inclination within the bin closest to face-on, to ensure that SEDs within that bin are contributed exactly once.
Spherically symmetric RTMs contribute their single SED 65 times, matching the number of data points provided by an inclination-weighted RTM.

\begin{figure*}
    \centering
    \includegraphics[width=0.49\textwidth]{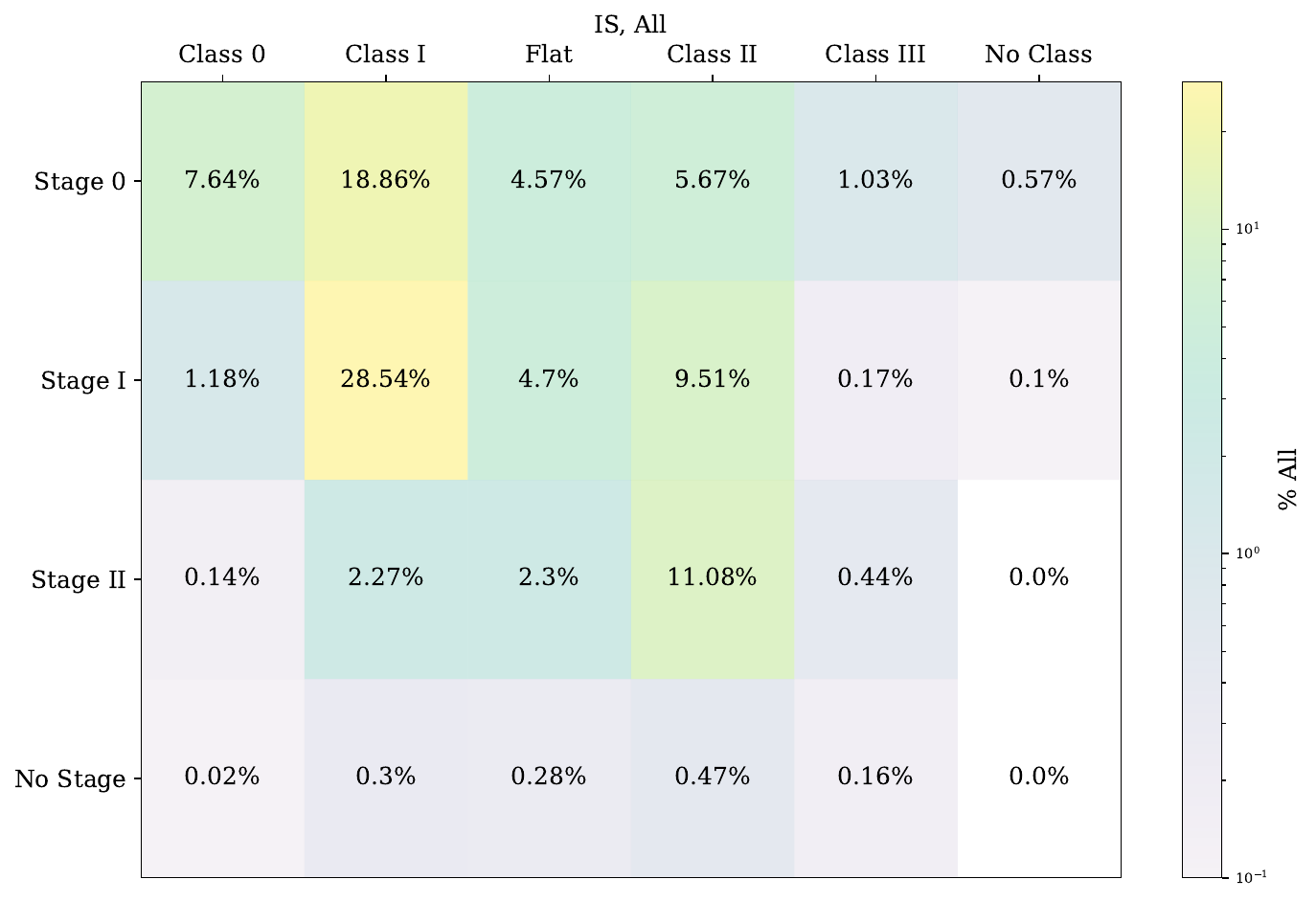}
    \includegraphics[width=0.49\textwidth]{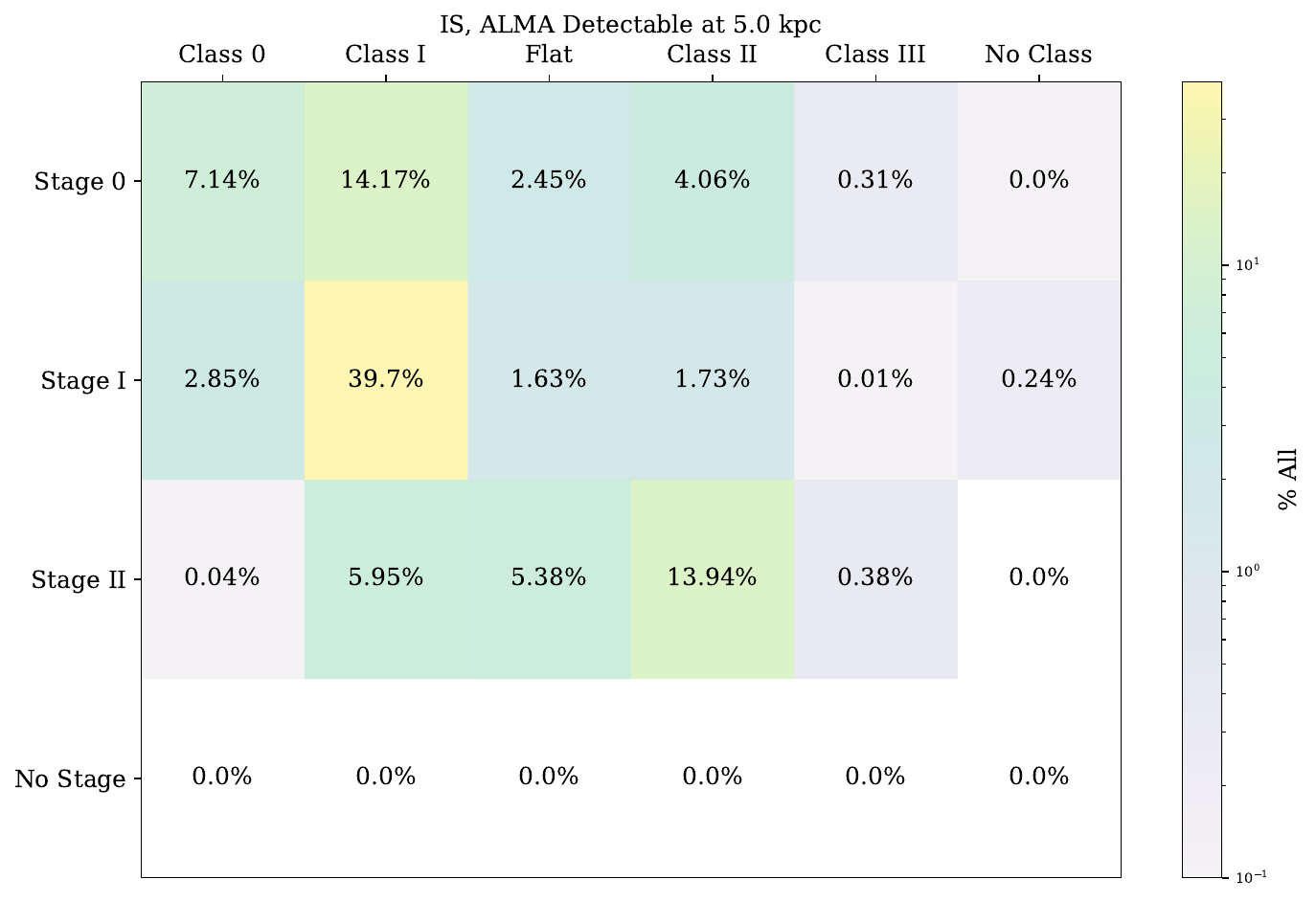}
    \includegraphics[width=0.49\textwidth]{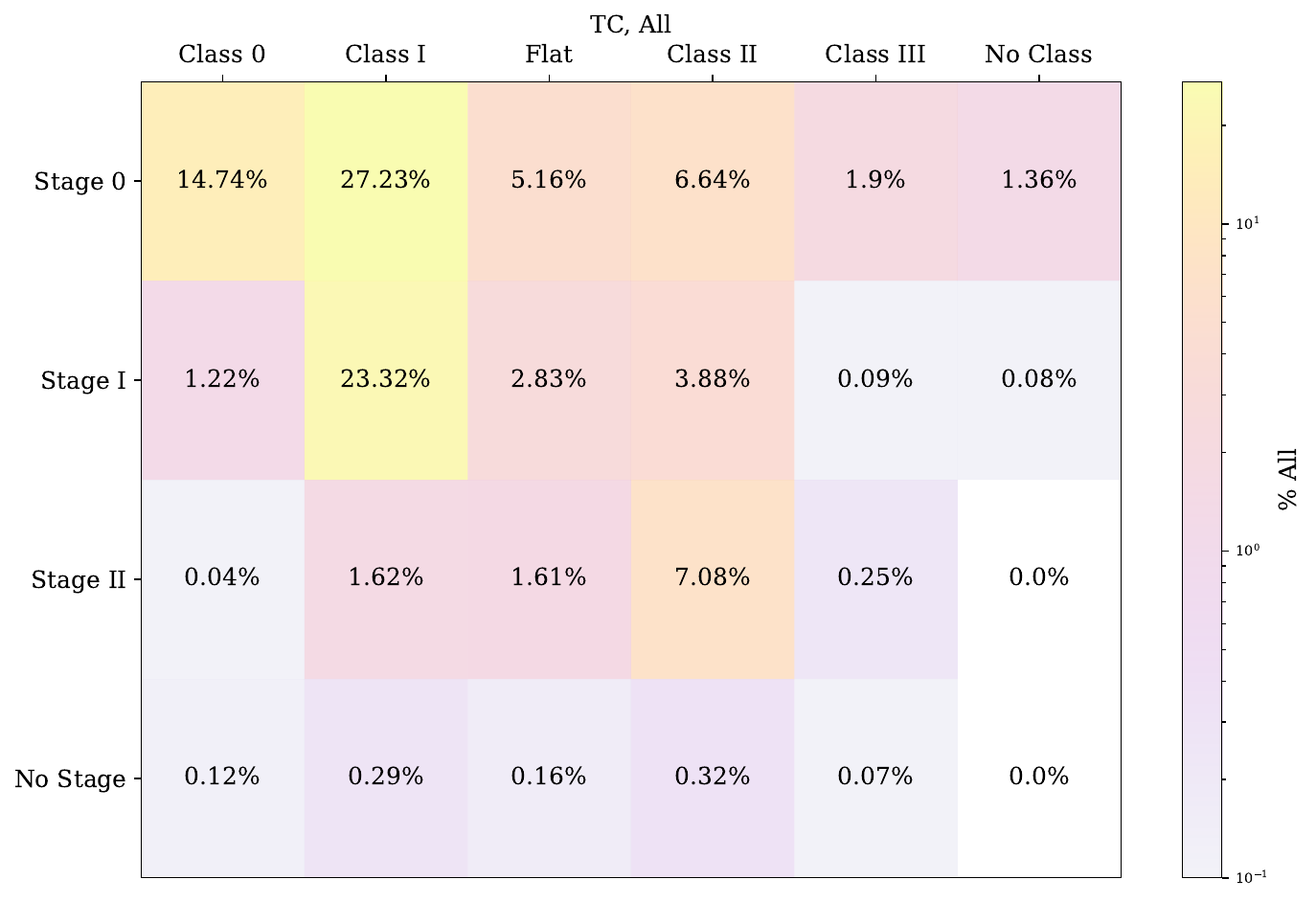}
    \includegraphics[width=0.49\textwidth]{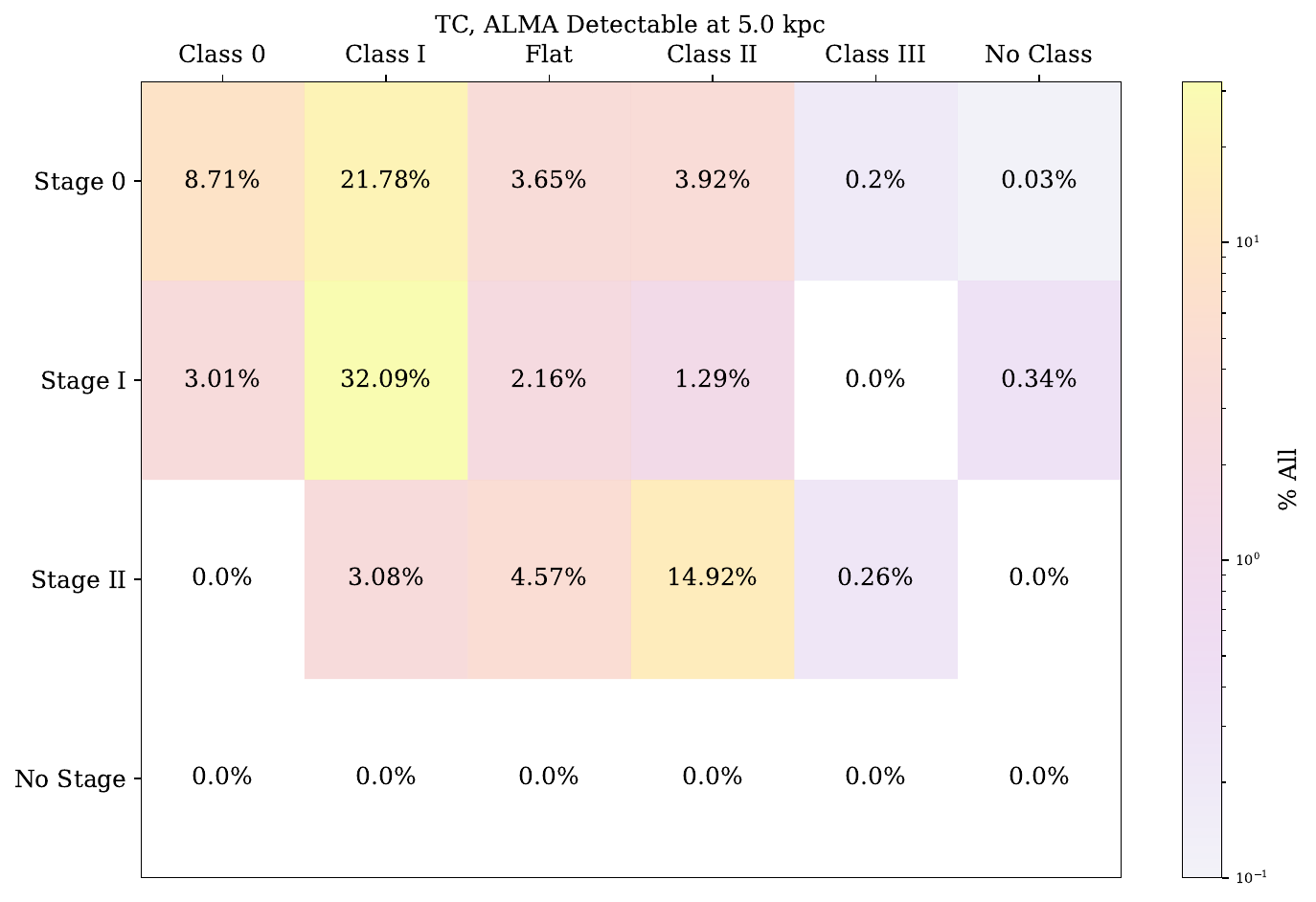}
    \includegraphics[width=0.49\textwidth]{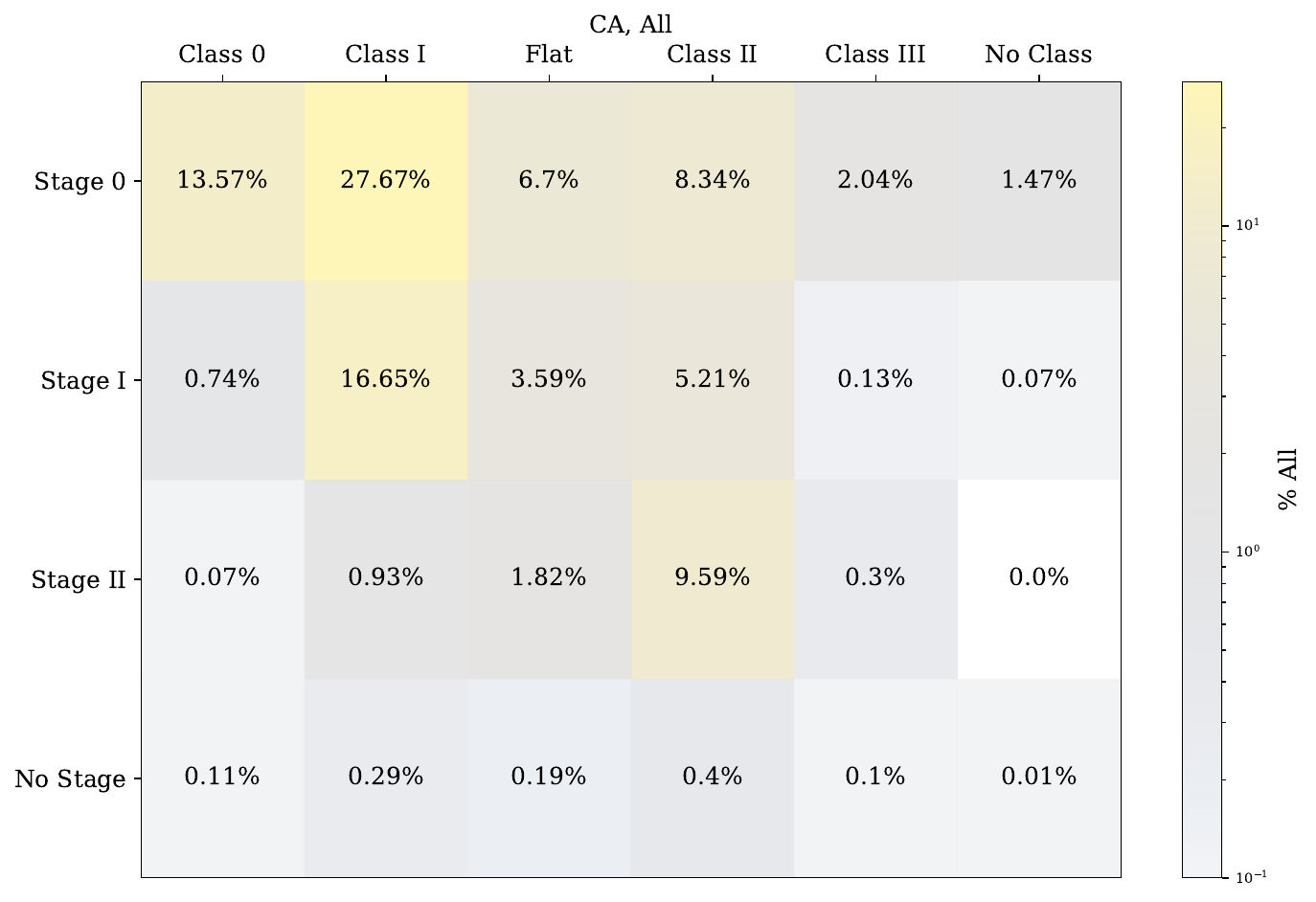}
    \includegraphics[width=0.49\textwidth]{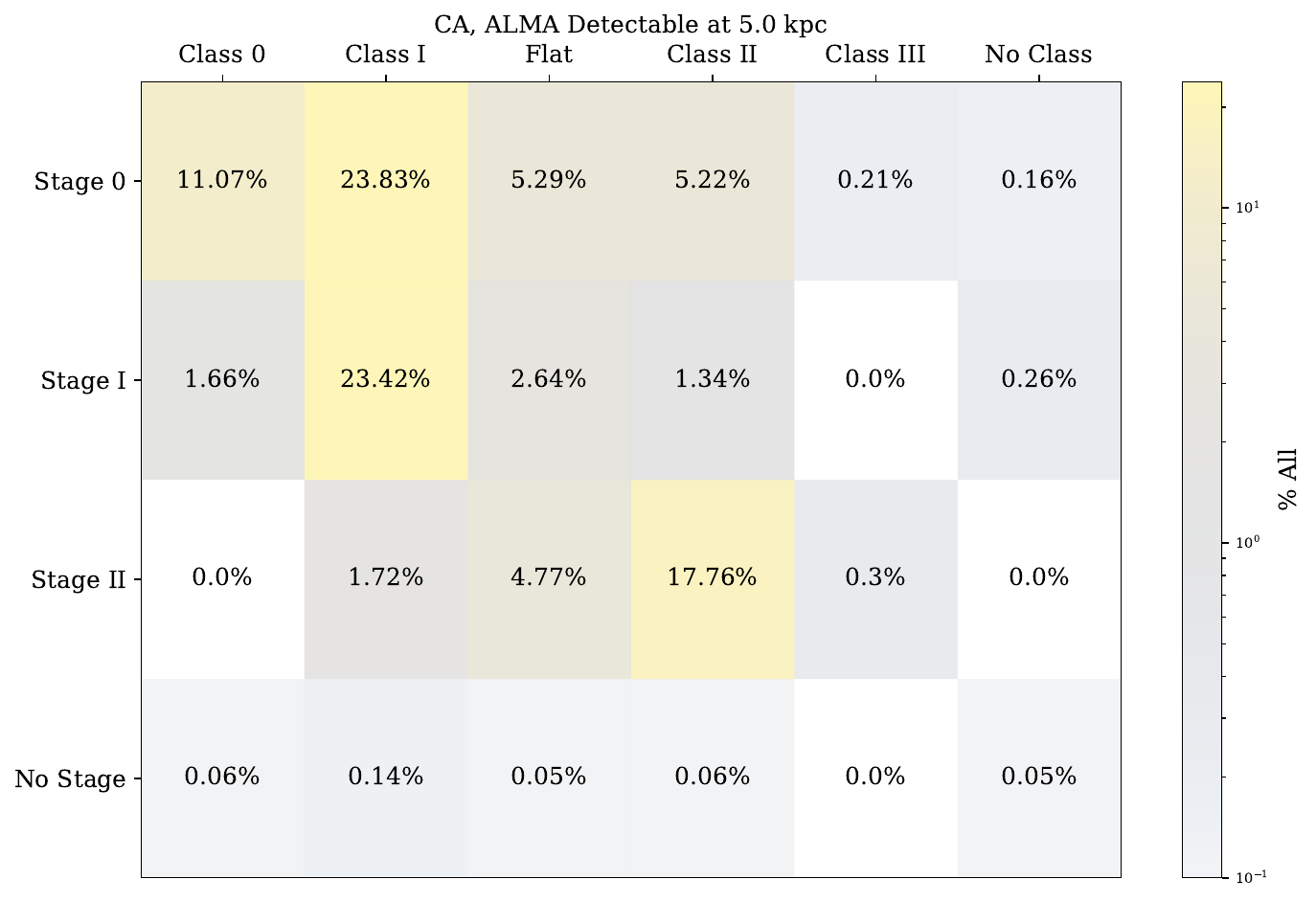}
    \caption{Class/Stage confusion matrices for RTMs consistent with our IS (\textit{top}), TC (\textit{middle}), and CA (\textit{bottom}) PEMs. We show matrices that include all selected models (\textit{left}) as well as ones restricted to models that are plausibly ALMA-detectable at 5 kpc (\textit{right}).}
    \label{fig:matrices}
\end{figure*}
Results from our Class/Stage comparison can be seen in Figure \ref{fig:matrices}. In this figure, we consider two cases for each PEM. 
The first is the ``all-inclusive" case, which includes all RTMs associated with every final mass, age, and $\esf$. The second is the ``detectable" case, which include the RTMs with fluxes detectable by ALMA at a distance of 5 kpc. ``Detectability" is here defined as exhibiting a 1-mm flux of at least 1 mJy within an aperture of physical radius $\sim$2000 au (aperture 7 in the R24 models);
in keeping with our focus on long wavelengths, this definition is targeted at a typical ALMA Band 6 observation of YSOs within the Milky Way \citep[e.g.][]{ginsburg2022,louvet2024,sanchezmonge2025}. These cases are intended to provide a general overview of Class/Stage confusion within the context of particular accretion histories as well as an indication of the extent to which they may be confused in observations. We consider more cases in \S\ref{sec:4.2.2}.

Given that the RTMs underlying this confusion matrix are now more physically motivated, we are able to use the matrix more directly to interpret observations. For YSOs observed in a particular Class, we can map back to Stages by summing the percentages of objects with that Class (i.e. in a particular column)--disregarding any ``no-stage" models\footnote{These are models that do not comport with our Stage definitions. Their envelopes are not massive enough for Stages 0 or I, they have no disks and therefore cannot be Stage II, and they are precluded from being Stage III due to having nonambient circumstellar dust. \S4.2 of R24 discusses these in greater detail.}--and determining what fraction of those total percentages are contained in each Stage. (Since percentages are all in relation to the total number of models in the matrix, they can be summed.)

As an illustration, we consider some simple cases using our all-inclusive matrices. For an IS PEM, the mapping is as follows (we assume 100 of each Class for the sake of simplicity):
\begin{itemize}
    \item Class 0 $\rightarrow$ 85 Stage 0 / 13 Stage I / 2 Stage II
    \item Class I $\rightarrow$ 38 Stage 0 / 57 Stage I / 5 Stage II
    \item Flat $\rightarrow$ 39 Stage 0 / 41 Stage I / 20 Stage II
    \item Class II $\rightarrow$ 22 Stage 0 / 36 Stage I / 42 Stage II
\end{itemize}
If instead we consider a TC PEM, the numbers are as follows:
\begin{itemize}
    \item Class 0 $\rightarrow$ 92 Stage 0 / 8 Stage I
    \item Class I $\rightarrow$ 52 Stage 0 / 45 Stage I / 3 Stage II
    \item Flat $\rightarrow$ 54 Stage 0 / 29 Stage I / 17 Stage II
    \item Class II $\rightarrow$ 38 Stage 0 / 22 Stage I / 40 Stage II
\end{itemize}
and for CA:
\begin{itemize}
    \item Class 0 $\rightarrow$ 95 Stage 0 / 5 Stage I
    \item Class I $\rightarrow$ 61 Stage 0 / 37 Stage I / 2 Stage II
    \item Flat $\rightarrow$ 55 Stage 0 / 30 Stage I / 15 Stage II
    \item Class II $\rightarrow$ 36 Stage 0 / 23 Stage I / 41 Stage II
\end{itemize}
These mappings provide another clear indication of how the interpretation of observations is dependent on the assumed PEM. Inferred Stage counts for TC and CA are shifted to earlier Stages compared to IS; in particular, they are more Stage-0 heavy. 
The variable accretion rates of TC and CA cause YSOs to spend a proportionally greater amount of their accretion time in Stage 0 compared to I, as the bulk of mass assembly occurs toward the end. Since the accretion rates of individual TC and CA protostars are often similar (see Figure \ref{fig:ratecomp}), TC and CA exhibit broadly similar Stage counts.

These matrices provide a better sense of the ways in which confusion between Class and Stage manifests (and may differ) following varying physical scenarios.
As presented, however, they come with some caveats. 
Firstly, as stated in Section \ref{sec:2.2}, the PEMs underlying our modeling framework do not model accretion after the circumstellar envelope is depleted, meaning that they do not cover the full Stage II lifetime of a YSO. Stage II models are therefore systematically underrepresented in these matrices when considering the entire theorized duration of star formation, which is likely to impact the statistics of most Classes. 
Secondly, the models within these specific matrices are selected across a wide range of ages, final masses, and efficiencies, meaning that the resulting matrices are not representative of specific observed populations. We narrow this scope in Section \ref{sec:4.2.2} to allow for more targeted use.

\paragraph{Class III YSOs} Our modeling framework captures the time when a protostar has circumstellar material, i.e. it does not extend to Stage III, leaving that stage out of our matrices. We therefore do not extend our analysis to Class III YSOs. Given the low extinction necessary for a YSO to be Class III, failing to include Stage III models (the only models with no nonambient circumstellar dust) would yield unrealistic results, unlike for classes implying a greater level of extinction where such models are easier to discount.

\subsubsection{Example application}\label{sec:4.2.1}
As a further demonstration of how to apply our matrices to observed populations of YSOs, we consider the data of the Spitzer ``Cores to Disks" (c2d) sample \citep{evans2009a}, a comprehensive survey of YSOs in nearby molecular clouds. We use the Class counts from the fifth row of their Table 6 (24 Class 0, 125 Class I, 223 Class II) which are derived through dereddened bolometric temperature\footnote{Note that only a fraction of Class II YSOs in this sample have accurately calculated $\tbol$s, and more YSOs are assigned Class II by alternate methods. This affects the overall sample statistics, but does not impact the usage of these numbers for our purposes.}.
c2d is estimated to be complete in mass down to the stellar/substellar boundary and has few other constraints on the acceptable theory space from observation. 
For simplicity, we therefore use the all-inclusive matrices from Figure \ref{fig:matrices} to map Class to Stage.
For the IS case:
\begin{itemize}
    \item 24 Class 0 $\rightarrow$ 21 Stage 0 / 3 Stage I
    \item 125 Class I $\rightarrow$ 47 Stage 0 / 72 Stage I / 6 Stage II 
    \item 223 Class II $\rightarrow$ 48 Stage 0 / 81 Stage I / 94 Stage II
\end{itemize}
For the TC case:
\begin{itemize}
    \item 24 Class 0 $\rightarrow$ 22 Stage 0 / 2 Stage I
    \item 125 Class I $\rightarrow$ 65 Stage 0 / 56 Stage I / 4 Stage II
    \item 223 Class II $\rightarrow$ 84 Stage 0 / 49 Stage I / 90 Stage II
\end{itemize}
For the CA case:
\begin{itemize}
    \item 24 Class 0 $\rightarrow$ 23 Stage 0 / 1 Stage I
    \item 125 Class I $\rightarrow$ 76 Stage 0 / 46 Stage I / 3 Stage II
    \item 223 Class II $\rightarrow$ 80 Stage 0 / 50 Stage I / 93 Stage II
\end{itemize}
Summing the Stage counts, the physical interpretation of c2d's Class counts following different PEMs would be as follows:
\begin{itemize}
    \item IS: 116 Stage 0 / 156 Stage I / 100 Stage II
    \item TC: 171 Stage 0 / 107 Stage I / 94 Stage II
    \item CA: 179 Stage 0 / 97 Stage I / 96 Stage II
\end{itemize}

Regardless of the assumed PEM, the c2d sample maps to a population of YSOs much more weighted toward earlier Stages than would be assumed through the Class counts alone. Current estimates of the durations of various phases of star formation are based on these counts, with the Class 0/I lifetime estimates from \citet{evans2009a} made in relation to the estimated Class II lifetime. If taken at face value, interpreting the c2d sample using these matrices would give reason to reexamine the canonical values, likely in the direction of increasing the Stage 0/I lifetime relative to Stage II.
However, we remind the reader that these particular matrices come with caveats that complicate a direct interpretation from the values presented in this section (see the end of \S\ref{sec:4.2}); our purpose is largely to demonstrate application to a real sample.

\subsubsection{Binned confusion matrices}\label{sec:4.2.2}
By tying our RTMs to protostellar evolutionary tracks, we narrow the scope of our Class/Stage comparison to physically motivated scenarios. However, these scenarios still encompass a wide range of final masses, ages, $\esf$s, and levels of detectability. 
In practice, there will likely be additional constraints placed on YSO measurements, whether theoretical (e.g. the mass-varying $\esf$ of ZT18 or the history of star formation within a region) or observational (e.g. observing YSOs at a particular age, only being able to resolve within a particular mass range, etc.).

In addition to the all-inclusive confusion matrices, we generate a set of matrices that captures various slices of the full set of ``physical" RTMs for more direct applicability to observations performed under these additional constraints. As an example, if one were to observe an embedded cluster in which the maximum YSO mass is known and the age is well constrained, the all-inclusive matrix will likely include models predicted to be more massive at that age than any detected YSOs. Instead of using the full matrix that contains these overly massive stars, one should use a downselected set that includes only the range of possibly detected YSOs, as well as restricting the time span covered by the matrix to times consistent with the measured age.

We establish bins in mass and age by constructing histograms of the values associated with the PEM tracks. Our set of mass bins covers the range of 0.2-50 $\msun$, as in Figure \ref{fig:protohr}; it is invariant with accretion history and is roughly even in log-space to promote an even mass coverage in each bin. Age values included in the bins cover the time that every protostar in a set of tracks (i.e. over the entire mass range) is actively accreting; the edges therefore vary with accretion history given the disparate time scales. TC and CA PEMs are given linearly spaced time bins, since the difference between the earliest start and latest end of the accretion within the set of tracks is about an order of magnitude at most (within $\sim$0.1-1 Myr), so linear spacing is able to capture the distribution of ages. IS PEMs are given log-spaced bin edges because the timescale of accretion can span multiple orders of magnitude ($\sim$0.1-10+ Myr).

Like the matrices in Figure \ref{fig:matrices}, we allow the underlying PEMs for this set of cut-down matrices to vary by accretion history. We also allow $\esf$ to vary and add three additional potential values to the set: (1, 2, 3). These are intended to capture scenarios where mass accreted onto a protostar may come from outside the core, similar to the behavior of the YSOs modeled by F17 in their effort to match an exponentially tapered accretion model to the distribution of YSO bolometric luminosities observed by \citet{furlan2016}. The precise mass accretion efficiency implied by their modeling varies with final stellar mass between approximately 150\%-400\%; we consider a broadly similar range of values.
Additionally, for each combination of bins and accretion histories, we compute matrices with cuts for detectability at the following distances: 0.1, 0.5, 1, 5, and 10 kpc. ``Detectability" retains its former definition: exhibiting a flux $>$1 mJy at 1 mm within an aperture of a radius $\sim$2000 au.

\begin{figure*}
    \centering
    \includegraphics[width=0.80\textwidth]{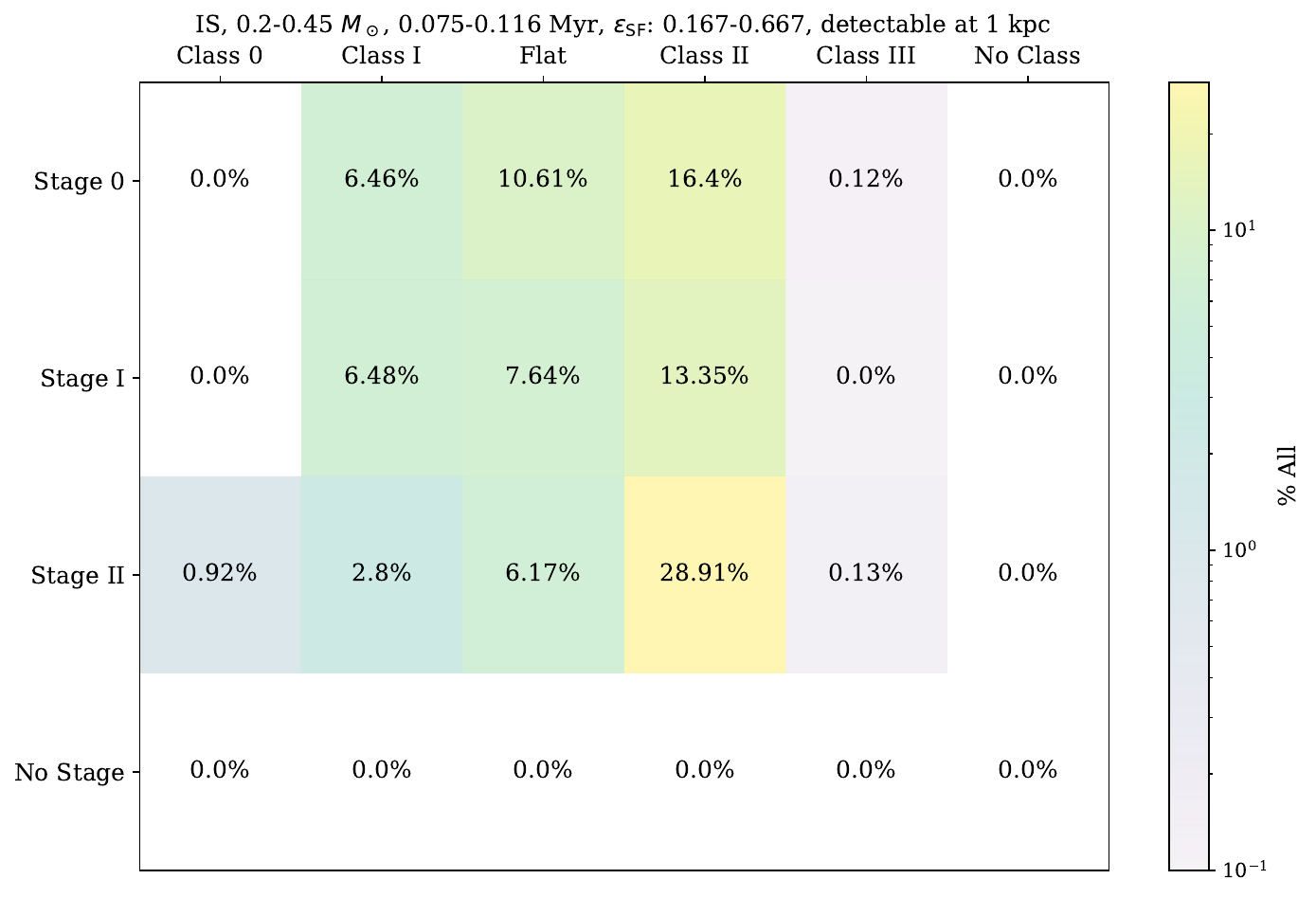}
    \caption{The same as Figure \ref{fig:matrices}, but restricted to a smaller range of final stellar masses and ages.}
    \label{fig:cut_matrix}
\end{figure*}

We generate matrices for a given scenario by including the Classes and Stages of RTMs that are selected along PEMs which follow a particular accretion history, occur within a particular mass and age range (specified by the outer edges of a set of mass/time bins), exhibit some set of $\esf$, and are detectable at a particular threshold. These scenarios cover every unique continuous combination of bins (e.g. IS accretion, mass bins 3-5, time bins 2-4, efficiencies 5-7, detectable at 10 kpc). We do not consider discontinuous combinations (e.g. mass bins 1-2 + 5-6). Figure \ref{fig:cut_matrix} shows an example of a matrix with a narrowed scope.
This matrix includes RTMs that were selected along IS tracks with final stellar masses between 0.2-0.45 $\msun$ and ages between .075-0.116 Myr that have $\esf$ values within the range of 1/6-2/3 and are detectable at a distance of 1 kpc. ($\esf$ is still limited to values in the set.)

If we repeat our number mapping for the matrix shown in Figure \ref{fig:cut_matrix}, we get
\begin{itemize}
    \item Class 0 $\rightarrow$ 100 Stage II
    \item Class I $\rightarrow$ 41 Stage 0 / 41 Stage I / 18 Stage II
    \item Flat $\rightarrow$ 44 Stage 0 / 31 Stage I / 25 Stage II
    \item Class II $\rightarrow$ 28 Stage 0 / 23 Stage I / 49 Stage II
\end{itemize}
which provides a clear demonstration of how constraints beyond just the PEM can change the physical scenario. 
For one, there are very few Class 0 objects here.
In general, the RTMs making up this confusion matrix likely do not have enough dust to reach Class 0 by our definition, which requires a significant contribution from dust to the bolometric luminosity. The Stage II models that are Class 0 are low-luminosity sources viewed edge-on, resulting in YSOs exhibiting high submillimeter luminosity ratios.
This matrix is relatively evenly populated by Stages 0, I, and II with a slight tilt toward Stage 0, in contrast to the all-inclusive IS case (\S\ref{sec:4.2}). This is due to the low masses of the protostars and surrounding envelopes; since the Stage boundaries are very close together at this end of the mass spectrum, the relative time spent in each Stage during accretion is closer to equal than at the high-mass end, where Stage II is barely represented.
The imposition of additional constraints therefore has large ramifications for the Stage counts inferred through the use of a confusion matrix, meaning that care should be taken to use a matrix suitable for a given set of observations.

We have released a data table containing all of the calculated matrices alongside this paper, which may be found in \href{https://doi.org/10.5281/zenodo.13922040}{this Zenodo repository}.

\subsubsection{Class and Stage definitions}\label{sec:4.2.3}
Earlier in this section, we laid out our adopted definitions for YSO Class and Stage. Alternate definitions for these concepts have been proposed and used throughout the literature, raising the question of how these definitions differ in a practical sense, as well as which are the most appropriate or useful. Here, we use our modeling framework to compare the observational and theoretical consequences of various categorization schemes.

\noindent\textbf{Class.} Our adopted Class definitions have largely remained the same from R24 and are consistent with the scheme laid out in \citet{andre1993} and \citet{greene1994}, which forms the basis of many studies of YSOs. However, due to the difficulty in obtaining the submillimeter photometry necessary to identify a Class 0 YSO via this definition, many surveys \citep[e.g.][]{dunham2013,furlan2016,pokhrel2023} instead employ the bolometric temperature $\tbol$ of a YSO to discriminate between Classes 0 and I, with the usual dividing line occurring at 70 K \citep[from][]{chen1995}. Class 0 is intended to correspond to Stage 0, i.e. to identify deeply embedded YSOs where most of the total mass is contained in circumstellar material. As we have assigned Stages to a number of RTMs in R24 through this work, we compare the performance of these definitions at recovering Stage 0 models, calculating $\tbol$ individually for each RTM SED with Equation (1) from \citet{myers1993}:
\begin{equation}
    \tbol = 1.25\times10^{-11}\int_0^\infty \nu S_\nu d\nu \bigg/ \int_0^\infty S_\nu d\nu {\rm \;K\,Hz}^{-1}
    \label{eq:tbol}
\end{equation}

\begin{figure*}
    \centering
    \includegraphics[width=0.49\textwidth]{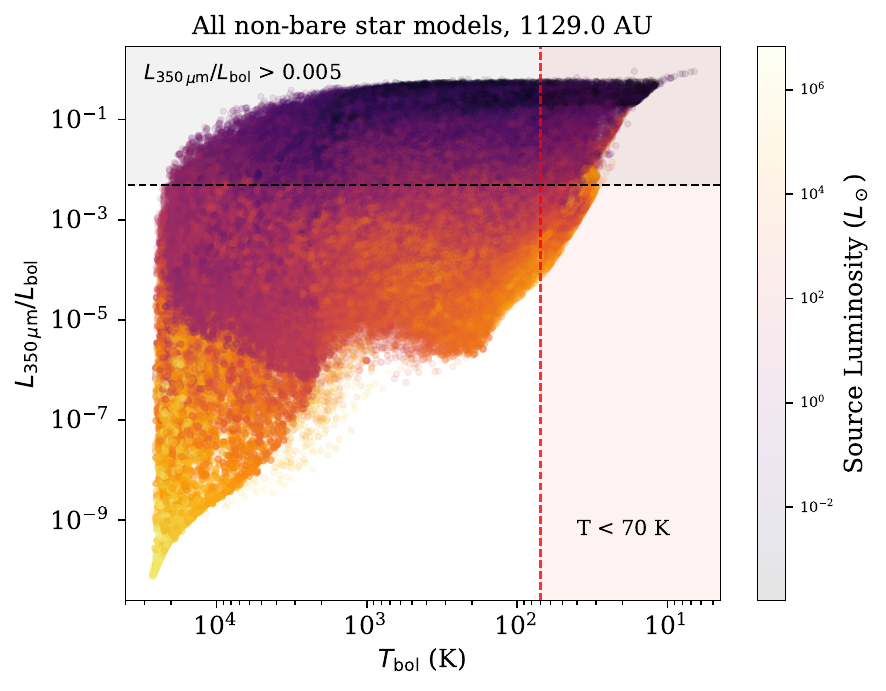}
    \includegraphics[width=0.49\textwidth]{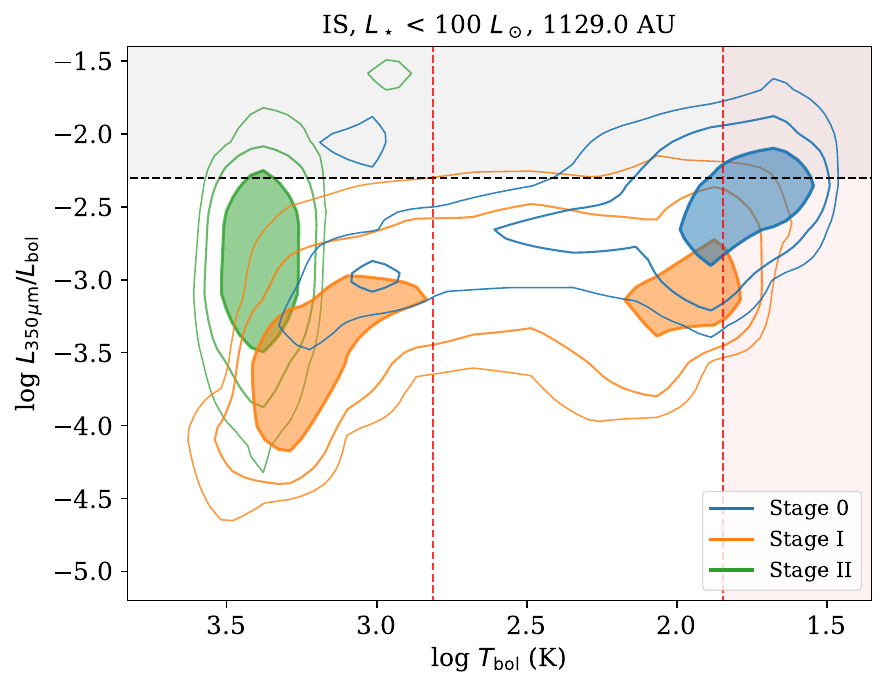}
    \includegraphics[width=0.49\textwidth]{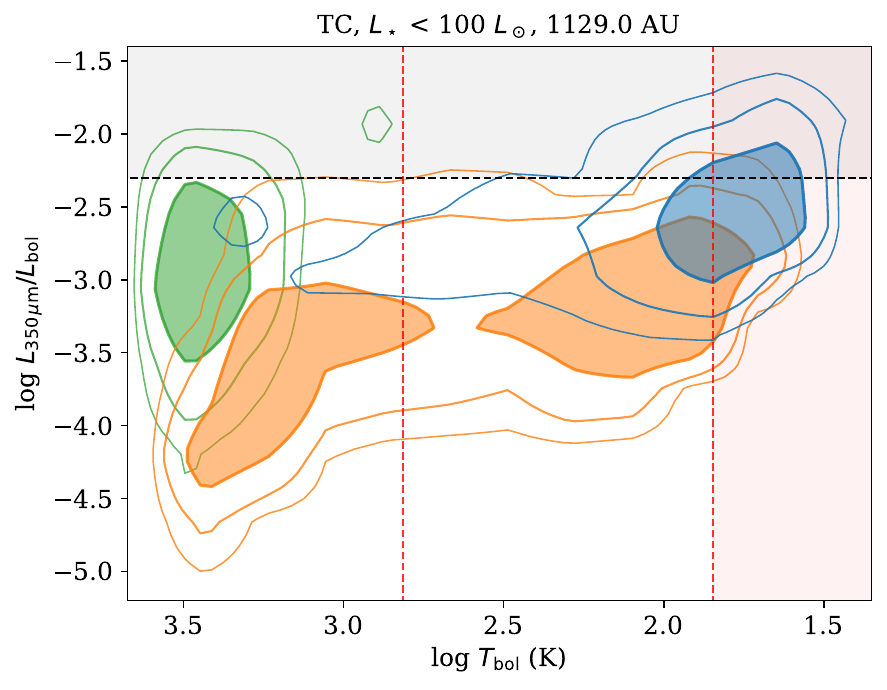}
    \includegraphics[width=0.49\textwidth]{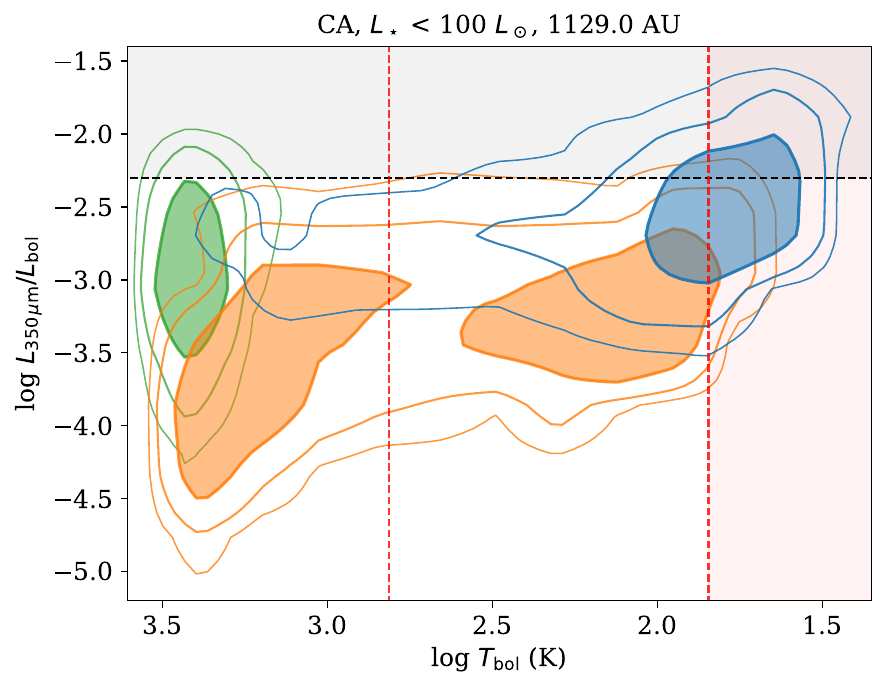}
    \caption{The line-of-sight mass-weighted temperatures and ratios of submillimeter to bolometric luminosity for R24 models. Regions corresponding to different definitions of Class 0 are shaded, and divisions between $\tbol$ Class 0/I/II and $\lumrat$ Class 0/I are plotted as dashed lines. The top left plot shows every RTM in R24 save the bare-star-only geometries (which are necessarily Stage III), colored by the bolometric luminosity of the central source. The remaining panels show contour plots containing samples of RTMs selected alongside IS (\textit{top right}), TC (\textit{bottom left}), and CA (\textit{bottom right}) histories, colored by Stage. Included RTMs have been weighted by inclination as described in \S\ref{sec:4.2}, meaning that more inclined SEDs comprise the majority of the visualized populations. Contour levels are set at 10, 20, and 50\% of the peak value; the region within the 50\% contour is filled.}
    \label{fig:classcomp}
\end{figure*}

We construct samples for each accretion history, starting from the set of RTMs constituting the ``all-inclusive" matrices from Figure \ref{fig:matrices}. In keeping with our confusion matrices, each selected RTM contributes its SEDs weighted by viewing angle (see \S\ref{sec:4.2}). From this inclination-weighted set, we apply a cut for detectability; however, as we are now concerned with the submillimeter luminosity, we consider ``detectability" as exhibiting a 350-$\mu$m flux greater than 1 mJy at the R24 models' native distance of 1 kpc. Further, as low-mass YSOs are the majority of constituents in most surveys of forming stars within the Galaxy, we limit each of our samples to RTMs whose central protostars have a bolometric luminosity of less than 100 $L_\odot$ for better congruence with data. As in \S\ref{sec:4.2}, we calculate all values within an aperture with a physical radius $\sim$1000 au; this is sufficient to capture most circumstellar material at a resolution commensurate with local YSO surveys. 
Figure \ref{fig:classcomp} shows the results of this comparison for each history, as well as the location of every non-bare-star RTM in $\tbol$--$\lumrat$ space.

Comparing $\tbol$ and $\lumrat$ for all of R24's RTMs (except the uniformly Stage III models of bare-star geometries \texttt{s---s-i} and \texttt{s---smi}), it becomes apparent that the majority of models considered Class 0 by virtue of $\tbol$ would also be considered Class 0 by their luminosity ratios. On the other hand, there are many RTMs that have $\tbol$ $>$ 70 K and high luminosity ratios. These results are largely consistent with the comparison between class definitions performed by \citet{dunham2014} using YSOs from various surveys.

Examining the locations of Stages within Figure \ref{fig:classcomp} in greater detail yields a mixed picture of the effectiveness of Class definitions. It is clear that Stage 0 models generally have higher $\lumrat$ and lower $\tbol$ than those in other Stages. However, Stage 0 exhibits an uneven correspondence with both definitions of Class 0, as the area of highest Stage 0 concentration straddles both Class 0/I dividing lines. Consequently, neither definition captures the full range of Stage 0 models; moreover, each comes with its own pros and cons. Class 0 by $\lumrat$ does not capture a majority of Stage 0 models; however, contamination from Stage I appears relatively low. Meanwhile, Class 0 by $\tbol$ contains a higher proportion of total Stage 0 models, but faces more contamination by Stage I; in addition, Stage 0 also exhibits a wide variance in $\tbol$ even after weighting by inclination.
Stage I models have a stronger correspondence with both definitions of Class I in comparison with Class 0, i.e. they mostly have $\tbol$ $>$ 70 K and $\lumrat$ $<$ 0.005 (although a nontrivial amount would be Class 0 by $\tbol$).
At the same time, many Stage I models also have $\tbol$ $>$ 650 K, meaning that they would be considered Class II according to the usual $\tbol$ classification scheme.
Stage II appears to fare the best; these models largely have $\tbol>650$ K and $\lumrat<0.005$, i.e. would or could be considered Class II (although it is worth pointing out that a $\tbol$ divide closer to $\sim$2000 K would yield a cleaner break from models in Stages 0/I). Deviations from Class II are due to a combination of low source luminosities and edge-on viewing. The connection between Class II and Stage II is therefore still subject to observational effects; however, the degree of potential for confusion is considerably lessened, and the remaining degeneracy could be addressable with additional data (e.g. $\alpha$, which is not considered here).

Our results--regardless of definition--do not support an exclusive link between Class 0 and Stage 0, the evolutionary state of a YSO to which it is meant to correspond. 
Overall, Class 0 appears mainly to separate models in Stages 0 and I--which make up the embedded phase of a YSO--from models in Stage II.
Translating this to the observational determination of evolutionary status, an observed Class 0 YSO is unlikely to be an evolved source (Stage II or later) given the high amount of dust extinction necessary to achieve a Class 0 SED.
This is consistent with the general understanding of Class 0 YSOs as inferred from observations \citep[e.g.][]{federman2023}.
However, beyond generally identifying embedded sources, Class 0 appears to be an incomplete indicator of evolution within the embedded phase; at best, Class 0 YSOs have an enhanced likelihood to be Stage 0, particularly if employing the $\lumrat$ criterion.
In light of our findings, we continue to use our modified R24 definition for Class 0 for this work, although with the understanding that its physical import is somewhat limited.

\noindent\textbf{Stage.} The main difference between the Stage definitions of this work and those of R24 \citep[which extends][]{crapsi2008} is the changed divide between Stages 0 and I. R24 distinguished these Stages using the temperature of the source; Stage 0 YSOs had sources with temperatures of $<$ 3000 K, i.e. sources that had not yet entered the Hayashi or Henyey tracks. Part of the motivation for these Stage definitions was the absence of stellar masses and evolutionary histories underlying R24's models. 
The half-mass definition for Stages 0 and I requires knowledge of both the instantaneous and final stellar mass associated with a particular YSO, in turn requiring the assumption of a history. This requirement is now satisfied through the ability to associate R24 models with PEMs, allowing us to implement alternate dividing lines. 

\begin{figure*}
    \centering
    \includegraphics[width=0.49\textwidth]{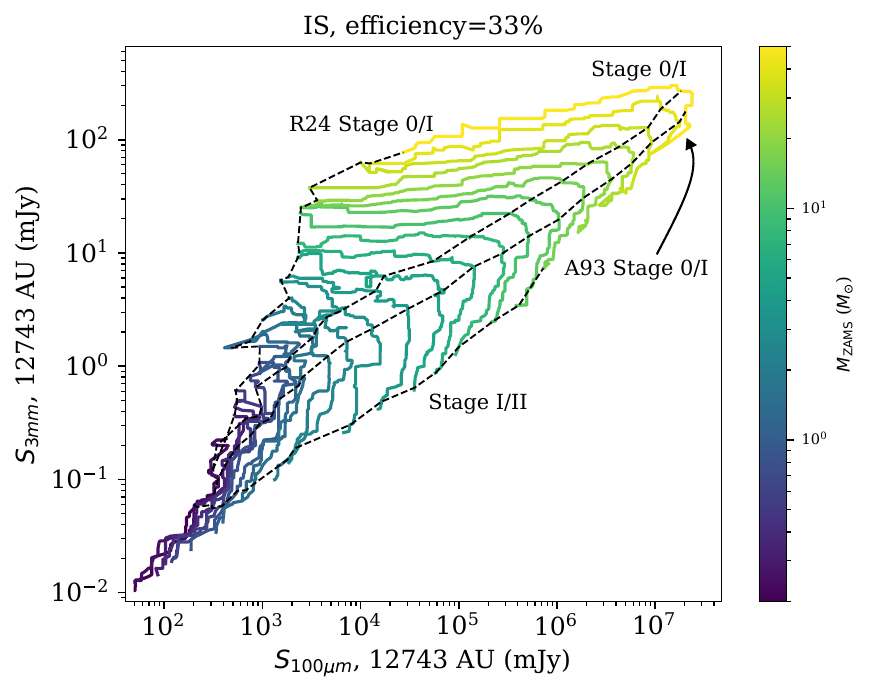}
    \includegraphics[width=0.49\textwidth]{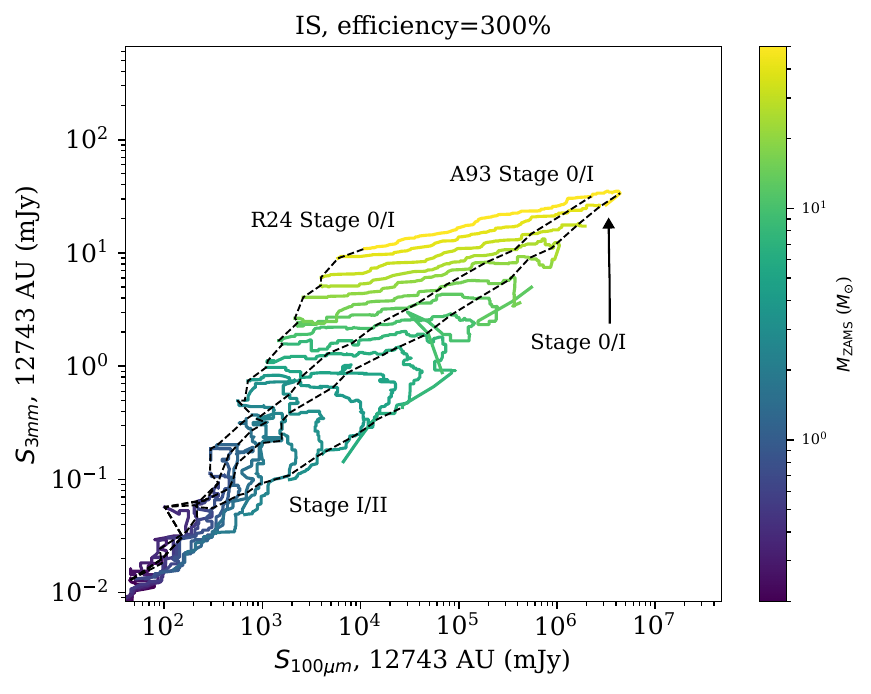}
    \caption{One of the plots from Figure \ref{fig:protohr}, but the dotted lines now indicate the time step corresponding to the dividing line between the Stage definitions of \citet{andre1993}, \citet{richardson2024}, and this work (\S\ref{sec:4.2}). Predictions are shown for scenarios where $\esf=1/3$ (\textit{left}) and $\esf=3$ (\textit{right}); axis limits are the same for both plots.}
    \label{fig:stagehr}
\end{figure*}

To evaluate the impact of this redefinition on our models, we compare the consequences of adopting the R24 and half-mass Stage criteria within the context of our flux predictions (as in Figure \ref{fig:protohr}) in Figure \ref{fig:stagehr}. 
We also include the traditional definition of ``Stage 0" from \citet[][A93]{andre1993}, that being where $M_\star < \menv$, taking the converse of that inequality to be their Stage I\footnote{Another set of definitions that sees use is that of \citet{robitaille2006}, which distinguishes stages based on a YSO's instantaneous protostellar mass $M_\star$, envelope infall rate $\dot{\menv}$, and disk mass $M_{\rm disk}$. Since only $M_\star$ is explicitly included in our PEMs and $M_{\rm disk}$ is not tracked at all, we do not evaluate this definition here.}.
We consider $\esf$s of both 1/3 and 3 in order to locate Stage dividing lines for both isolated collapse with mass ejection and collapse with external infall.

R24's Stage definitions generally bookend our predicted flux tracks, except for very low-mass ($\lesssim\msun$) stars, which spend a nontrivial fraction of their accretion time in Stage II (in other words, with an envelope mass $<$ 0.1 $\msun$). The transition from R24's Stage 0 to I consistently corresponds to the beginning of the tracks, regardless of mass. This close correspondence between the R24 Stage dividing line and our ability to make predictions for the flux of a YSO indicates that the R24 definitions for Stages 0/I correspond to a meaningful physical divide in YSO evolution. However, these definitions also limit Stage 0 to a brief and mostly unobservable early period of a YSO's life.

The half-mass-assembly Stage 0/I dividing line, in contrast with that of R24, exhibits a general correspondence with the ``knee" feature in our predicted flux tracks. These Stage definitions essentially create two distinct observational phases: an increase in far-IR flux at roughly constant millimeter flux and a decrease in millimeter flux at roughly constant far-IR flux.
Consequently--with a sufficiently good measurement of its mass--it should be possible to determine which phase a YSO is in from its position in this flux space, and therefore to directly identify and distinguish Stage 0 and I YSOs through photometric observations at multiple wavelengths\footnote{Figure \ref{fig:stagehr} compares flux at 100 $\mu$m and 3 mm; however, sources for new submillimeter data are limited when compared to the near- and mid-IR or millimeter regimes. Identification of Stages 0 and I by comparison of flux across wavelength regimes is in principle not limited to these particular wavelengths, but use of shorter-wavelength data will likely be complicated by the wider variance in IR flux predictions, as discussed in Section \ref{sec:3.2}.}. Our adoption of the half-mass-assembly divide between Stages 0 and I over R24's is in response to this generally higher level of observational significance, although the definitions are more closely aligned for lower-mass YSOs.

Comparing the half-mass-assembly criterion with the A93 mass-ratio divide, both behave similarly. However, the location of the A93 divide within the predicted tracks exhibits a dependence on the efficiency of mass accretion, occurring after half mass assembly in sub-100\% (isolated-collapse) scenarios and before half mass assembly in super-100\% (external-infall) scenarios. This mirrors the expected order of events within the corresponding evolutionary tracks (see Figure 14 of F17 for an illustration of star+envelope evolution with external infall). As a result, the A93 definition also does not correspond as closely as the half-mass criterion to the transition between behavioral phases in either modeled scenario. We observe the same behavior across all of our modeled accretion histories.
Within the context of our approach to YSO modeling, then, the time at which half of a star's final mass is assembled appears to be a reasonable discriminator between predicted behavioral phases that is consistent across modeled accretion histories and more robust to variation in mass accretion efficiency than the traditional $M_\star/\menv$ criterion.

In addition to examining the stage definitions within the context of the predicted fluxes, Figure \ref{fig:stagehr} also allows us to evaluate the impact of varying efficiency on the predictions themselves. While the shape of the tracks generally remains the same, YSOs with lower mass accretion efficiency are consistently brighter than their high-efficiency counterparts, with a difference in flux of roughly an order of magnitude at both 100 $\mu$m and 3 mm for the highest-mass YSOs at the efficiencies we model. This disparity implies that star formation via localized collapse should appear brighter in the far-IR/mm regimes than formation fed via external infall. Overall, this behavior is consistent with expectations; the mass of the envelope component for a YSO corresponding to the same eventual stellar mass will be systematically less in a scenario where most of the stellar mass originates outside the YSO than one in which stellar growth is entirely fed by the birth mass reservoir, therefore leading to less emission from the dust in the envelope.

As a general caveat to these predictions, the Stage I fluxes calculated for our more massive superefficient YSOs are subject to increased noise and lack of predicted model flux. These YSOs are composed of highly luminous sources with minimal nonambient circumstellar dust, meaning that they are likely to fall below the long-wavelength S/N threshold employed by R17 due to widespread emission from the heated ambient medium (see \S4.2.4 of that paper for more detail on SED postprocessing). Consequently, many models selected by our procedure within the Stage I phase do not exhibit defined fluxes within most apertures.

\section{Conclusion}\label{sec:5}
We have developed a new approach to modeling the evolution of young stellar objects. By associating existing radiative-transfer SEDs from a well populated and formation-agnostic set of YSO models with independent models of protostellar evolution, we create a modeling framework that is capable of predicting the properties and flux of a YSO regardless of the assumed pathway of star formation.

We used this framework to predict the flux emitted by YSOs with initial stellar masses ranging from 0.2-50 $\msun$ across the time they are actively accreting following isothermal-sphere, turbulent-core, and competitive accretion histories. By comparing the 100-$\mu$m and 3-mm flux of these modeled YSOs, we found that the different rates of protostellar growth projected by these histories translate into observable differences in the long-wavelength flux of YSOs. This is particularly true for more massive stars, where the timescale of accretion differs the most. Our modeling approach links the evolutionary history of YSOs to direct observables, advancing the practice of comparing theory to observation, which is often done through the use of intermediate quantities that require additional inferences (e.g. bolometric temperature or luminosity).

We characterized the uncertainty in these predictions by attempting to reproduce the 1-mm flux of SEDs from our set with our framework, finding that we are generally able to recover the expected flux to within approximately 20\%, although with a slight bias to the upside. This good performance is not uniform across the spectrum; our approach does not recover shorter-wavelength radiation to the same degree, largely due to the dependence of this radiation on quantities that are not well modeled. In particular, we highlight the near- and mid-IR as areas suffering from this issue; given their frequent usage in studies of star formation, better modeling of the evolution of disks and outflow cavities is needed to maximize the impact of the current- and next-generation data in this regime.

As a test of the flexibility of our framework, we have applied it to the YSO models of \citet{zhang2018}, a contemporary grid based on the turbulent-core theory of protostellar growth. As with our own models, we are able to recover long-wavelength fluxes that are reasonably consistent with ZT18's predictions. Furthermore, we investigated the extent to which the assumed dust opacity models are responsible for differences between our SEDs and their ZT18 counterparts by recalculating the SEDs of a subset of R24 models with the dust configuration of ZT18. We found that alterations of the dust model are capable of reproducing the kinds of discrepancies we observe between ZT18's fluxes and our recoveries, illustrating the potential uncertainty introduced into radiative-transfer modeling by the choice of dust. We developed an explanation for our overrecovery of ZT18's fluxes based on its higher disk dust opacities and densities.

Leveraging our ability to associate R24 models with evolutionary histories, we have revisited the concept of confusion between the observational Class and evolutionary Stage of YSOs. We created confusion matrices quantifying the relationship between Class and Stage for every available slice of a parameter space composed of accretion history model, final stellar mass, protostellar age, mass accretion efficiency, and level of detectability at long wavelengths. In doing so, we provide a tool to infer the physical reality of observed protostellar populations from observed Class counts that is applicable across a wide range of theoretical scenarios. These confusion matrices are released to the public at \href{https://doi.org/10.5281/zenodo.13922040}{10.5281/zenodo.13922040}.
Further, we have investigated the theoretical and observational significance of various commonly used definitions for Class and Stage within sets of RTMs identified as being consistent with our PEMs. Our results do not evidence an exclusive correspondence between the Classes and Stages of YSOs following any PEM by any set of definitions, with embedded-phase (i.e. Stage 0/I) objects appearing to be particularly prone to confusion.
However, we also find that a Stage 0/I dividing line set when a star accretes half of its final mass creates distinct behavioral phases in the fluxes of our modeled YSOs, potentially facilitating the identification of YSOs in the earliest phases of their evolution through an observable other than Class.
%pretty good, right? told you
\\

%\begin{acknowledgments}
We thank Yichen Zhang for helpful communication regarding the ZT18 model grid and the provision of its dust opacities, as well as the anonymous referee for substantive comments and useful suggestions. AG acknowledges support from the NSF under grants AST 2008101 and CAREER 2142300. ER acknowledges the support of the Natural Sciences and Engineering Research Council of Canada (NSERC), funding reference number RGPIN-2022-03499 and from the Faculty of Science at the University of Alberta. The authors acknowledge \href{http://www.rc.ufl.edu}{University of Florida Research Computing} for providing computational resources and support that have contributed to the research results reported in this publication.
%\end{acknowledgments}

\software{Astropy \citep{astropy:2013, astropy:2018, astropy:2022}, Dask \citep{dask}, Matplotlib \citep{matplotlib}, NumPy \citep{numpy}, Scikit-learn \citep{scikit}, SciPy \citep{scipy}.}

\bibliography{ref.bib}
\bibliographystyle{aasjournal.bst}

\appendix
\section{The impact of accretion luminosity}\label{ap:lum}
\input{lum_appendix}

\section{More on YSO composition}\label{ap:comp}
\input{comp_appendix}

\section{Additional plots}\label{ap:plots}
\input{plot_appendix}

\end{document}

%% file: lum_appendix.tex
In Section \ref{sec:2.2}, we outlined our procedure for making models of evolving YSOs. In order to match RTMs to PEMs, we construct a multidimensional parameter space based on shared quantities, one of which is the luminosity of the YSO’s central source. However, the PEMs track both the intrinsic luminosity of the source (i.e. purely thermal radiation from the protostar itself stemming from contraction/deuterium burning/etc.) and the total luminosity, which includes accretion. The presence of both of these components in the PEMs raises the question of how the inclusion or exclusion of accretion luminosity impacts our results, and as a consequence, which scenario is more appropriate for our purposes.

\begin{figure*}
    \centering
    \includegraphics[width=0.49\textwidth]{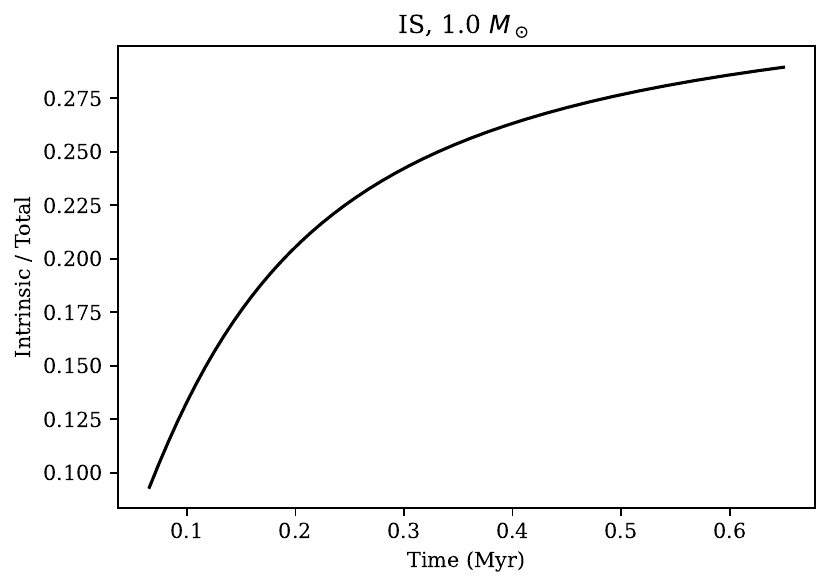}
    \includegraphics[width=0.49\textwidth]{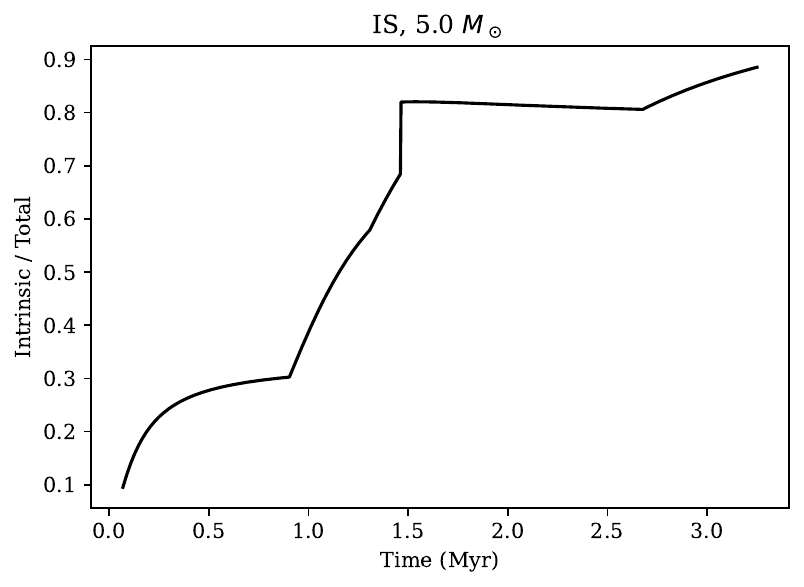}
    \caption{Intrinsic luminosity as a fraction of total luminosity for two PEMs following an isothermal-sphere accretion history. Plotted for protostars with final masses of one (\textit{left}) and five (\textit{right}) $M_\odot$.}
    \label{fig:lum_ratios}
\end{figure*}

As modeled by K12, accretion luminosity is often the dominant component of total luminosity. This is particularly true for protostars with final stellar masses less than $\sim3-4\,M_\odot$, but is also the case at early times for more massive protostars. Figure \ref{fig:lum_ratios} serves as an illustration; following an IS accretion history, the intrinsic luminosity of a 1 $M_\odot$ protostar will be subdominant for its entire accretion time, and the same will be true for roughly half the accretion time of a 5 $M_\odot$ protostar.

\begin{figure*}
    \centering
    \includegraphics[width=0.49\textwidth]{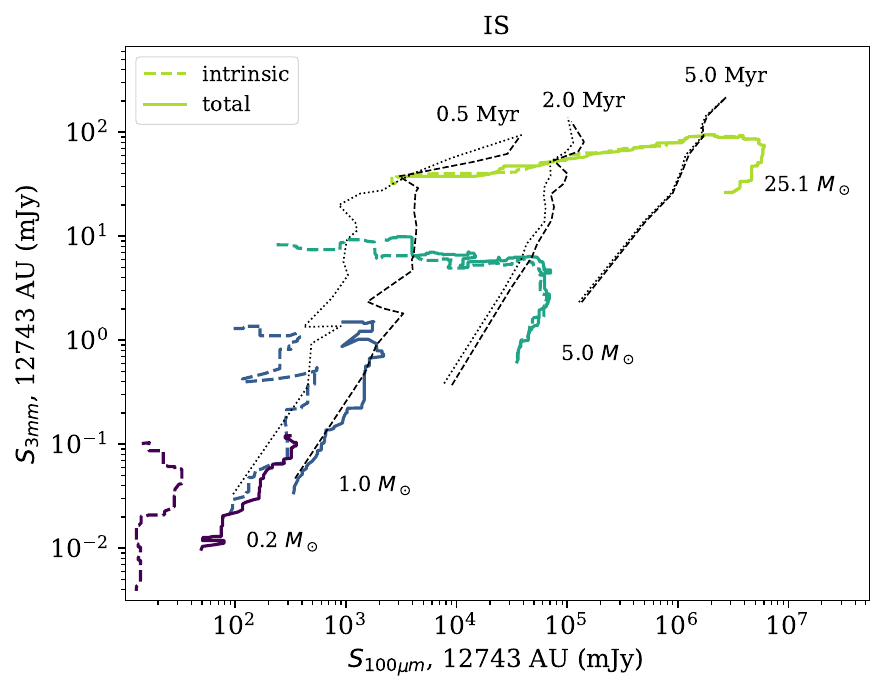}
    \includegraphics[width=0.49\textwidth]{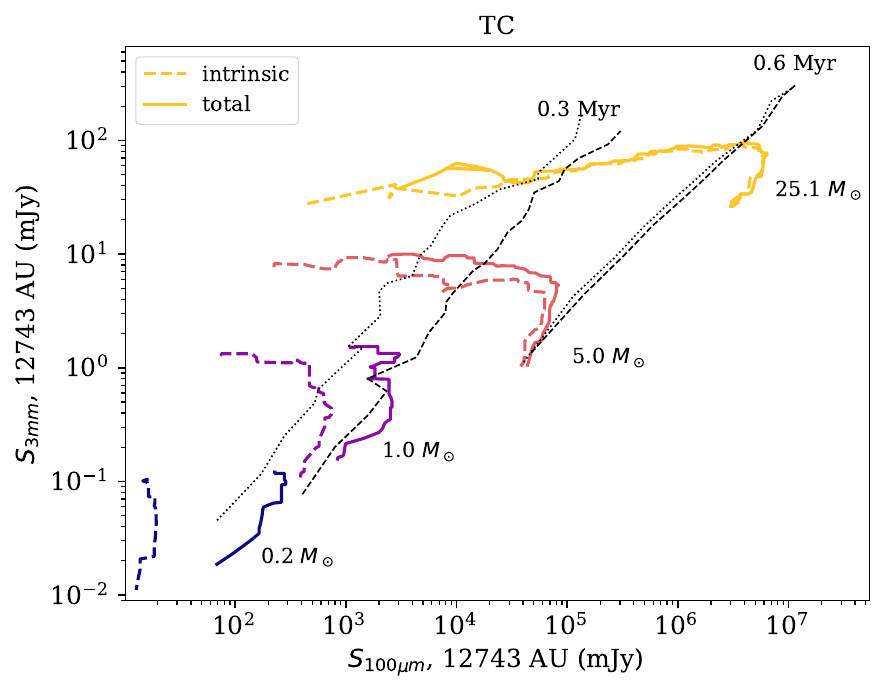}
    \includegraphics[width=0.49\textwidth]{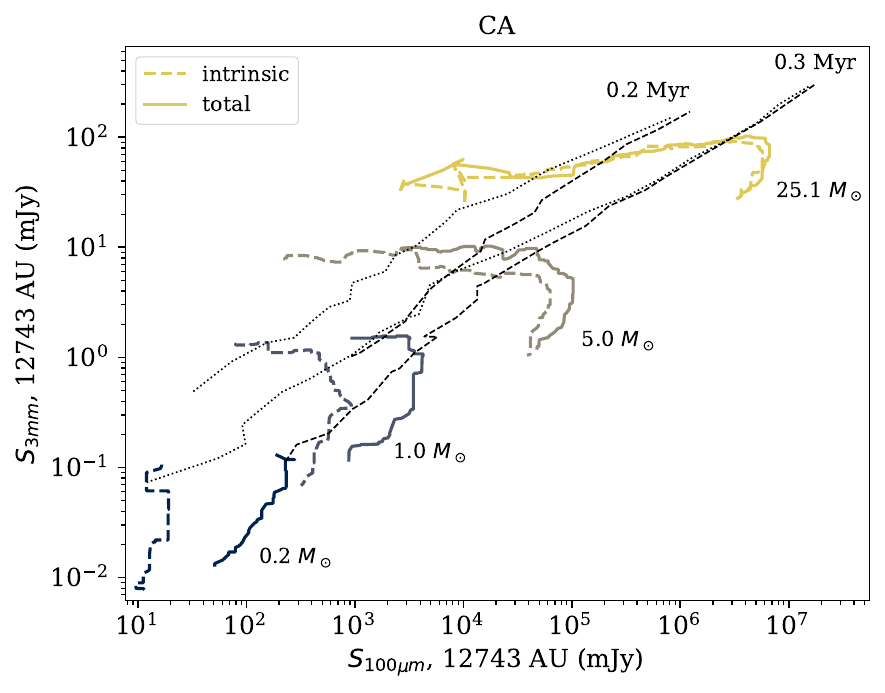}
    \caption{100-$\mu$m vs. 3-mm flux tracks for a set of modeled YSOs as in Figure \ref{fig:protohr}, but plotted using RTMs matching the intrinsic (\textit{dashed}) and total (\textit{solid}) luminosities of the base PEMs. We also plot isochrones showing the position of intrinsic (\textit{dotted}) and total (\textit{dashed}) luminosities in this flux space.}
    \label{fig:total_v_int}
\end{figure*}

The disparity between the intrinsic and total luminosity of our PEMs means that the choice of luminosity affects the RTMs that are chosen to correspond to a PEM. In turn, this affects the fluxes that we predict using those RTMs (as in \S\ref{sec:3.1}), at least for wavelengths that are dependent on the luminosity of the central source. As a way to quantify the effects of this choice, we compare flux predictions made using intrinsic and total luminosity in Figure \ref{fig:total_v_int}.

The choice of luminosity has a large impact on our predicted 100-$\mu$m fluxes. For low-mass stars, there is roughly an order-of-magnitude difference in the flux values over their entire accretion time between the ``intrinsic" and ``total" tracks, regardless of accretion history. This disparity is less present in higher-mass stars that are eventually able to outshine the luminosity from accretion, but still exists early in their accretion time when the instantaneous mass of the protostar is low.
The increase in flux due to accretion luminosity is qualitatively consistent with behavior observed in the modeling of \citet{fischer2024}, reaffirming the utility of far-IR radiation as a tracer of protostellar accretion.
The effects of accretion luminosity on the 3-mm flux, on the other hand, are fairly muted regardless of mass. On the whole, however, it is clear that because accretion makes up a nontrivial component of the total luminosity of many protostars across time and is capable of greatly affecting our flux predictions, the total luminosity should be preferred.

%% file: comp_appendix.tex
In Section \ref{sec:3.2} we provided a diagnostic for the accuracy of our flux predictions: attempting to reproduce the fluxes of models within our set by averaging over their 10 nearest neighbors, as we do when modeling YSO SEDs (\S\ref{sec:2.2}). This approach leads to good accuracy and reasonable precision. However, since the number of neighbors and the definition of ``nearest" determine the size and composition of the sample of RTMs picked to represent a YSO, we evaluate the performance of our framework as these are varied here.

Our definition of distance is Equation \ref{eq:quantdist}, which is based on the CDFs of our tracked properties (our ``quantile distance"). (See \S\ref{sec:2.3} for details.) In the course of this research, we have developed alternate definitions; here, we provide an overview.

One of the main motivations behind our definition is the large disparity between the possible values between dimensions. $T_{\star}$ is limited between $\sim$10$^3$-10$^4$ K while $L_\star$ and $M_{\rm core}$ can vary by multiple orders of magnitude, meaning that a standard Cartesian distance is likely to place unequal weight on dimensions for reasons independent from physics, which we want to avoid. One way of compensating for the level of difference is to transform every value to log-space and take the Cartesian distance there instead, which we call our ``log distance":
\begin{equation}
    D^2 = \log^{2}\left(\frac{T_{\rm model}}{T_{\rm track}}\right)+\log^{2}\left(\frac{L_{\rm model}}{L_{\rm track}}\right)+\log^{2}\left(\frac{M_{\rm model}}{M_{\rm track}}\right)
    \label{eq:distlog}
\end{equation}

From here, we can perform a normalization of sorts: dividing each value by the maximum possible range in each dimension in log-space (e.g. $\log(T_\star)\,\rightarrow\,\log(T_\star)\,/\,\log(T_{\rm max}/T_{\rm min})$) so that when distance is calculated, the offset in each dimension is weighed by its magnitude relative to the total span of values. This is our ``normalized log distance":
\begin{equation}
        D^2 = \frac{\log^2(T_{\rm model}/T_{\rm track})}{\log^2(T_{\rm max}/T_{\rm min})}
        +\frac{\log^2(L_{\rm model}/L_{\rm track})}{\log^2(L_{\rm max}/L_{\rm min})}
        +\frac{\log^2(M_{\rm model}/M_{\rm track})}{\log^2(M_{\rm max}/M_{\rm min})}
    \label{eq:distnormlog}
\end{equation}

Since our parameter space is built with long-wavelength emission in mind, we also create a definition for distance that weights the dimensions by their effect on the long-wavelength flux in order to prioritize the quantities that matter most. As in the main body of this work, we use 1 mm as the representative for long wavelengths. We make the ansatz that the 1-mm flux behaves as follows:
\begin{equation}
    S_{\rm1 mm}(T_\star,L_\star,M_{\rm core}) = A\times T_\star^{\alpha}+B\times L_\star^{\beta}+C\times M_{\rm core}^{\gamma}
    \label{eq:fluxguess}
\end{equation}
or, in other words, the flux has a power-law dependence on each quantity. To calculate the unknowns, we divide each dimension into 20 evenly log-spaced bins and fit the flux as a function of each term within bins of the other two parameters (for example, fitting $S_{\rm 1 mm} = A\times T_\star^\alpha$ within bins of $L_\star$ and $M_{\rm core}$). Doing so effectively holds the nonfit parameters as constant as possible while still maintaining good sample size (we do not consider bins with less than 50 models to avoid outliers). Once this fit is performed within each combination of bins, we find the mean of the results.

Our values for the exponents are $\alpha=0.02$, $\beta=0.28$, and $\gamma=0.91$. There is some physical basis for these values. In the limit of optically thin dust at long wavelengths, we expect the flux for a YSO with an envelope to vary roughly as:
\begin{equation}
    S_{\rm\lambda,\,long}\propto M_{\rm core}T_{\rm core} = M_{\rm core}L_{\rm core}^{1/4}
    \label{eq:optthinflux}
\end{equation}
The core temperature is highly insensitive to the source temperature, meaning that the observed flux should not depend on the source temperature significantly, which we find to be the case. Conversely, the core mass should vary roughly linearly with the 1-mm flux, which we also find to be the case. The luminosity exponent we expect from Equation \ref{eq:optthinflux} is 0.25, which is close to what we find assuming that core luminosity is proportional to $L_\star$.

With the determination of weights, we can now define our ``weighted distance":
\begin{equation}
    D^2 = \log^{2}\left(\frac{T_{\rm model}}{T_{\rm track}}\right)^\alpha+\log^{2}\left(\frac{L_{\rm model}}{L_{\rm track}}\right)^\beta+\log^{2}\left(\frac{M_{\rm model}}{M_{\rm track}}\right)^\gamma
    \label{eq:distweighted}
\end{equation}
where we apply the exponents derived through the power-law fitting to the respective dimension of our log distance. As in the log case, we can further divide each dimension by the maximum span to normalize distances. This results in our ``normalized weighted" distance:
\begin{align}
\begin{split}
        D^2 = \left(\frac{\log^2(T_{\rm model}/T_{\rm track})}{\log^2(T_{\rm max}/T_{\rm min})}\right)^{1/\alpha} 
        +&\left(\frac{\log^2(L_{\rm model}/L_{\rm track})}{\log^2(L_{\rm max}/L_{\rm min})}\right)^{1/\beta}\\
        +&\left(\frac{\log^2(M_{\rm model}/M_{\rm track})}{\log^2(M_{\rm max}/M_{\rm min})}\right)^{1/\gamma}
    \label{eq:distnormweighted}
\end{split}
\end{align}
Since we have ensured that each distance value will now be at most 1, we invert the exponents to ensure that a smaller exponent in a dimension reduces the relative contribution from that dimension instead of magnifying it.

\begin{figure*}
    \centering
    \includegraphics[width=0.49\textwidth]{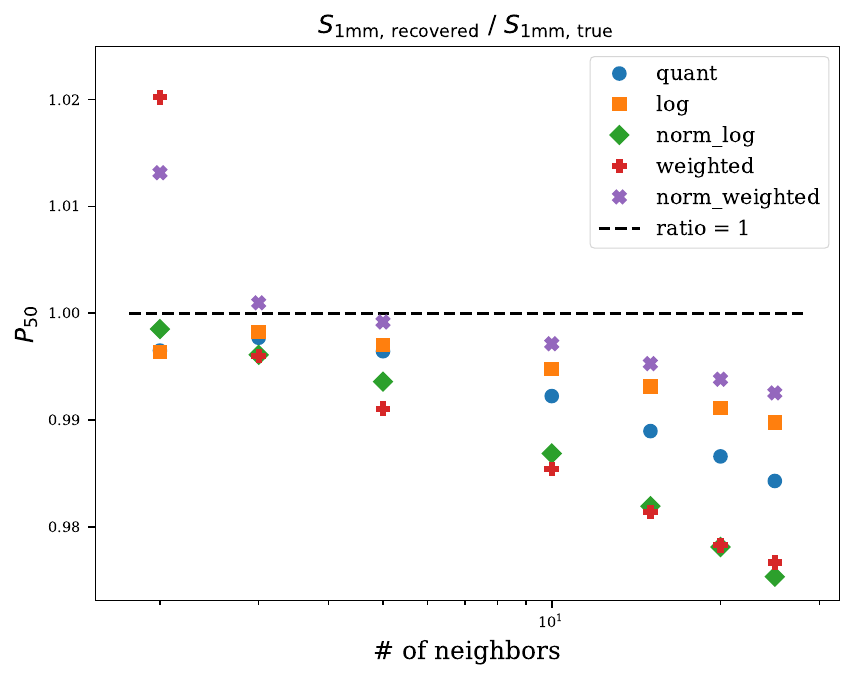}
    \includegraphics[width=0.49\textwidth]{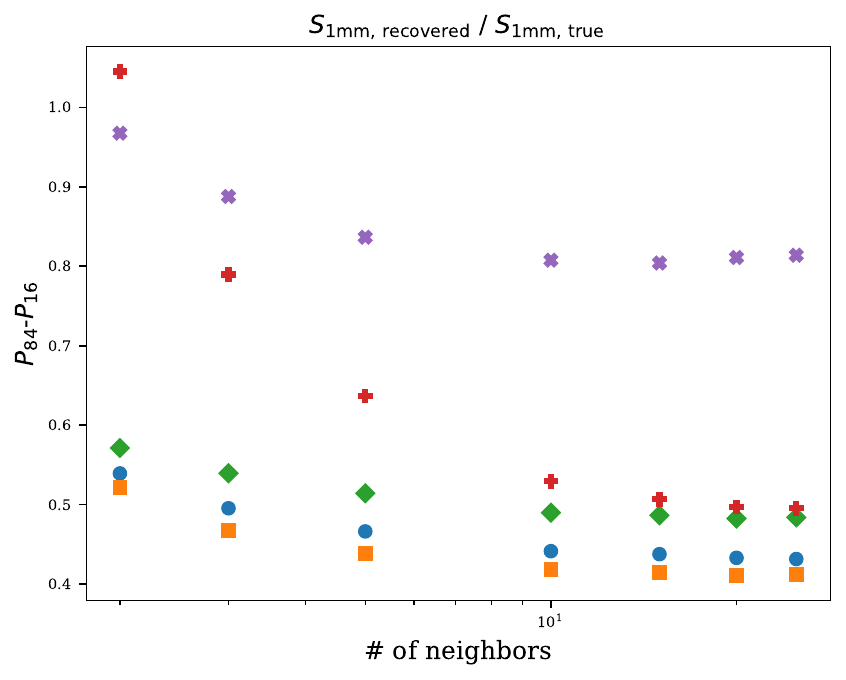}
    \includegraphics[width=0.49\textwidth]{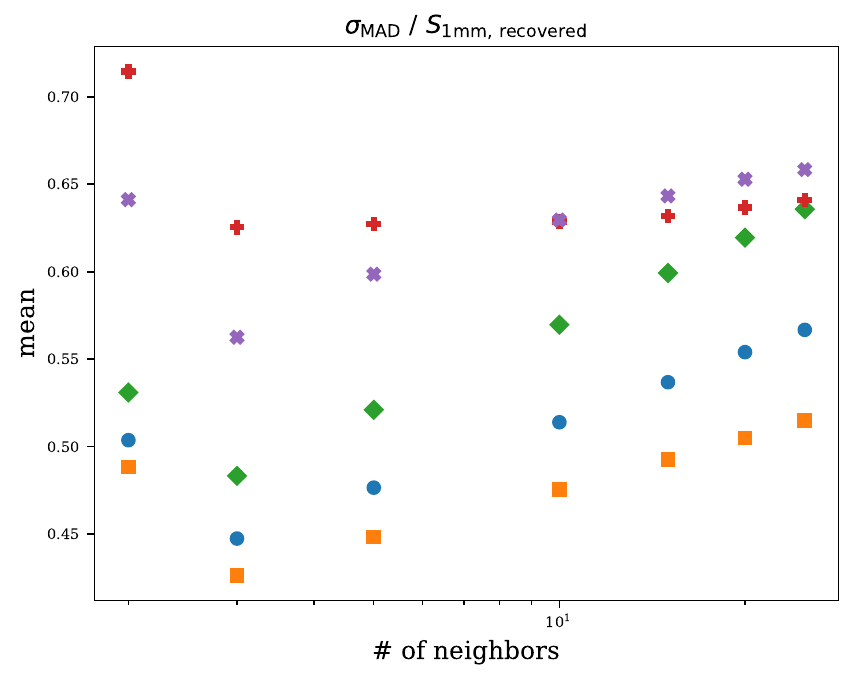}
    \includegraphics[width=0.49\textwidth]{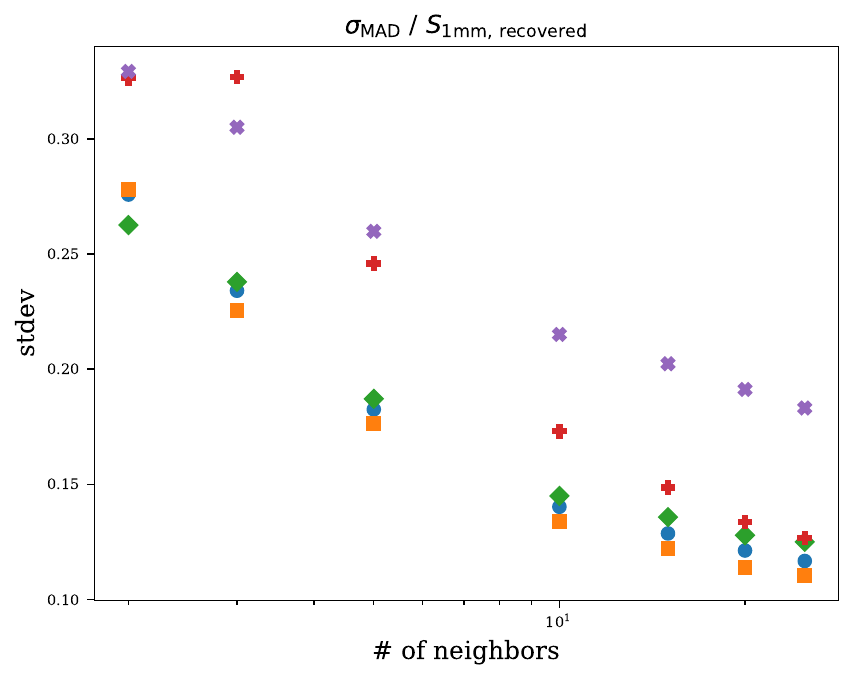}
    \caption{Comparison of the quality of 1-mm flux reproductions, as derived from plots like Figure \ref{fig:performance}, for all our definitions of distance and with varying numbers of selected models. The top row shows the 50th percentile of the distribution of flux ratios (\textit{left}) and the difference between the 84th and 16th percentiles (\textit{right}). The bottom row shows the arithmetic means (\textit{left}) and standard deviations (\textit{right}) of log-normal distributions fit to the distribution of $\sigma_{\rm MAD}$s of the RTM SEDs that are averaged over to produce recovered SEDs (scaled by the recovered value).} 
    \label{fig:distance_comp}
\end{figure*}
In Figure \ref{fig:distance_comp}, we compare the quality of 1-mm R24 model fluxes recovered using these definitions to that of ones obtained using our default method, as well as allowing the number of selected nearest-neighbor models to vary. This figure characterizes the performance of different distance definitions through percentile values from the distribution of the ratios of recovered to original model fluxes, and from log-normal distributions fit to the distribution of $\sigma_{\rm MAD}$s (scaled to standard deviation and the recovered flux value); see Figure \ref{fig:performance} and Section \ref{sec:3.2} for context.

Among the definitions, the quantile distance is consistently a good performer in both accuracy and precision. It consistently produces distributions of flux ratios with a 50th percentile close to 1 and one of the consistently lowest spreads between $P_{16}$ and $P_{84}$, regardless of the number of neighbors; that uncertainty improves slightly when adding more neighbors, but that improvement is marginal beyond 10 models. The scaled $\sigma_{\rm MAD}$s (derived from the SEDs of selected RTMs) for the quantile distance generally hover around 0.5, i.e. 50\% of the recovered value. $\sigma_{\rm MAD}$ receives a very slight boost in performance from more neighbors before the increased number begins to cause more deviation from the median. This upward trend with more neighbors provides another incentive to stop at about 10. Combined with the flux ratios, a set size of 10 neighbors represents the last point where adding a model to the set used to recover the flux will increase the precision of flux recovery while maintaining an average recovered flux value within 1\% of the true value and limiting the rise in the uncertainty of each individual predicted SED.

Interestingly, results from the log distance are slightly more precise than those from the quantile distance, both having a slightly lower spread in flux ratio and a slightly lower $\sigma_{\rm MAD}$. However, the improvement is small, and this definition of distance is not as universally applicable as the quantile distance to different scenarios (e.g. five parameters with a mix of log and linear sampling). Meanwhile, the normalized log distance and both flux-weighted definitions have uncertainties that are uniformly higher than the best performers.

%% file: plot_appendix.tex
\begin{figure}
    \centering
    \includegraphics[width=0.32\textwidth]{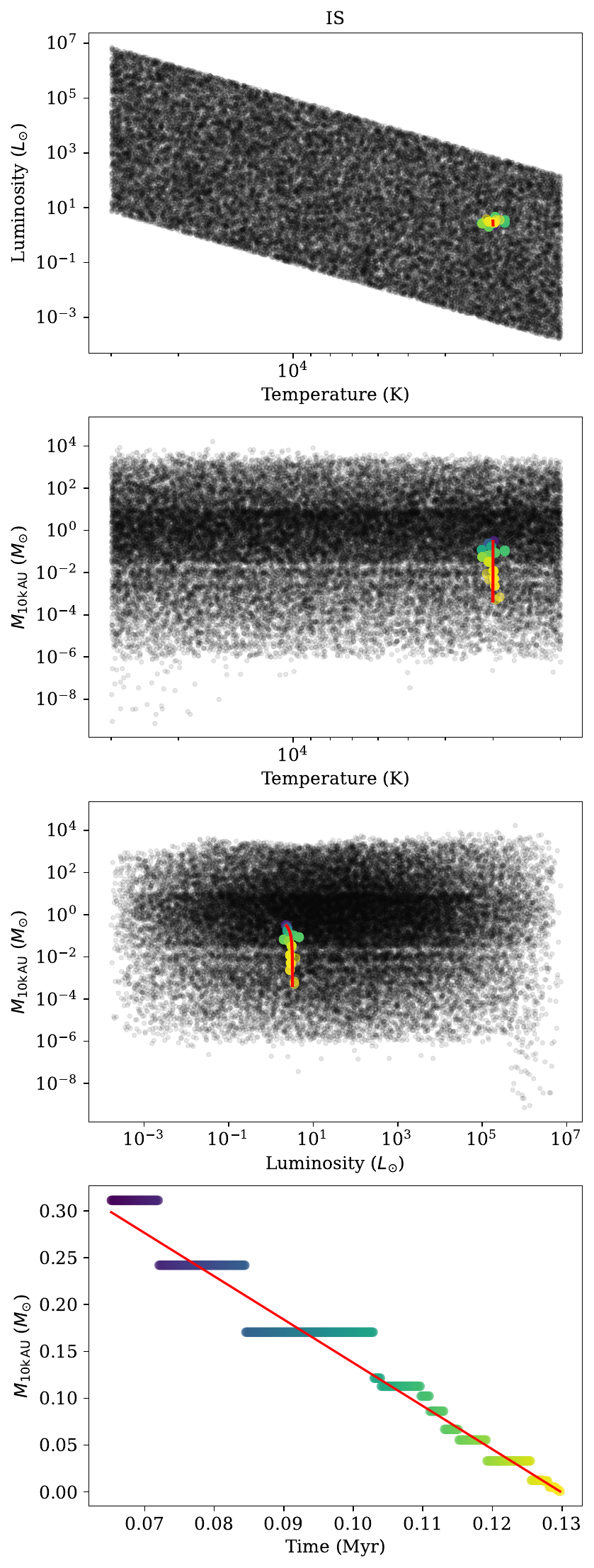}
    \includegraphics[width=0.32\textwidth]{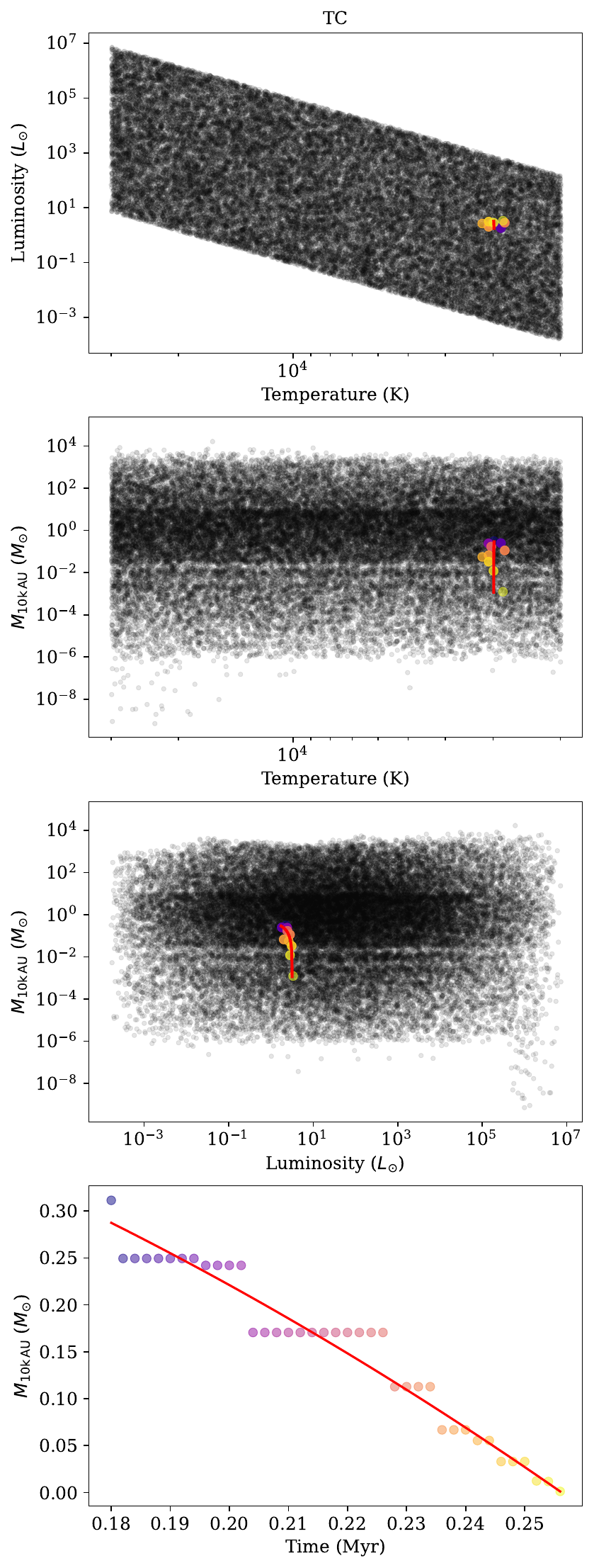}
    \includegraphics[width=0.32\textwidth]{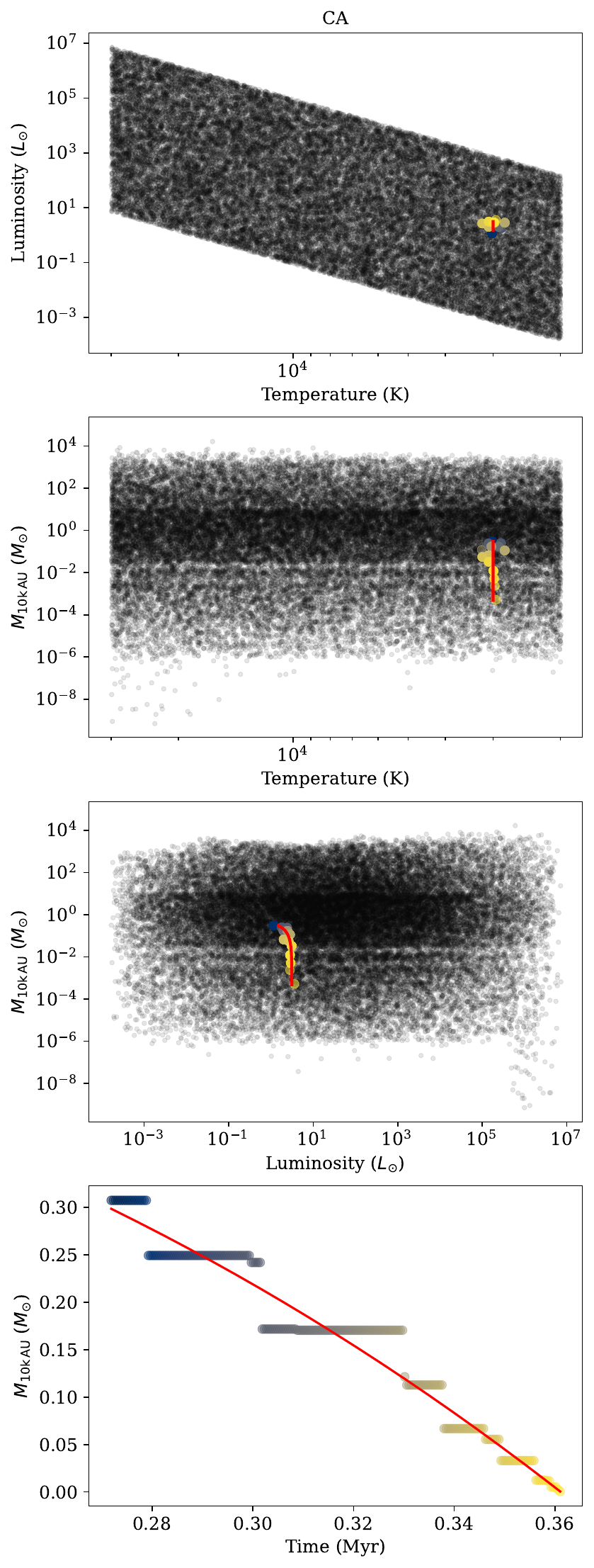}
    \caption{The same as Figure {\ref{fig:modeltrack}} for $0.2\, M_{\odot}$ stars.}
    \label{fig:fourplot1}
\end{figure}
\begin{figure}
    \centering
    \includegraphics[width=0.32\textwidth]{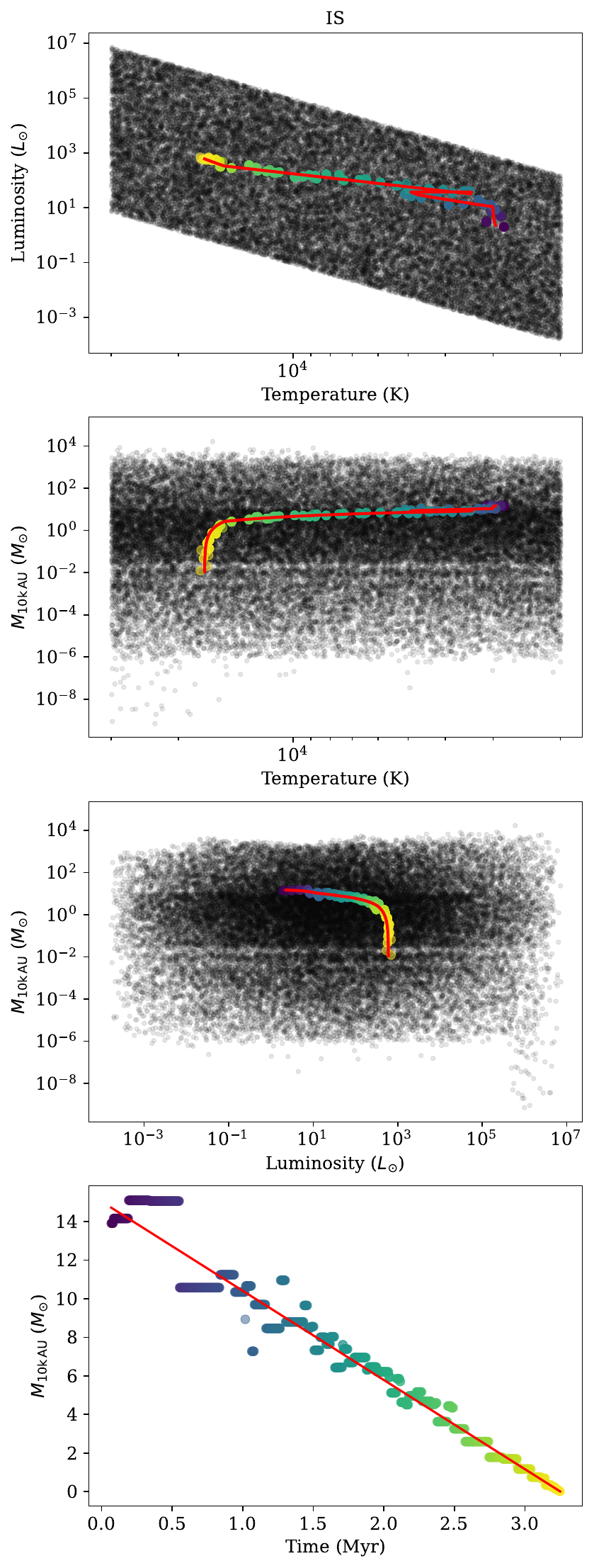}
    \includegraphics[width=0.32\textwidth]{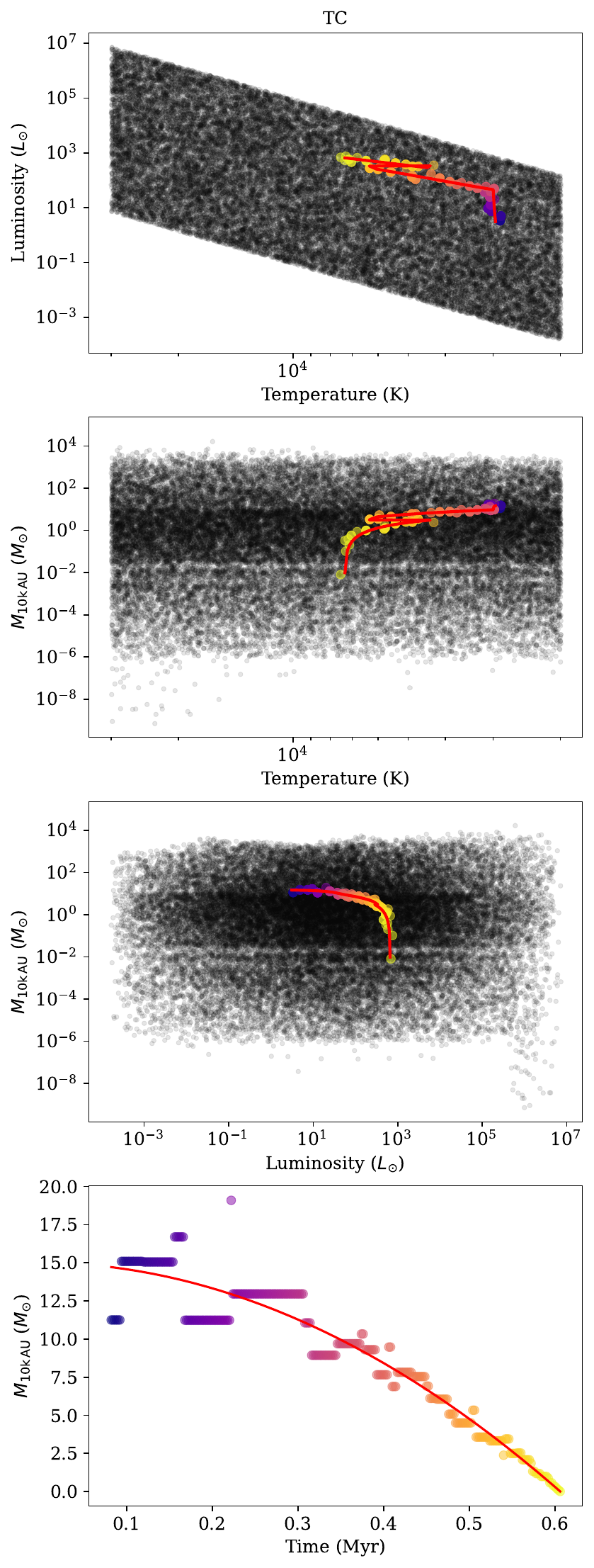}
    \includegraphics[width=0.32\textwidth]{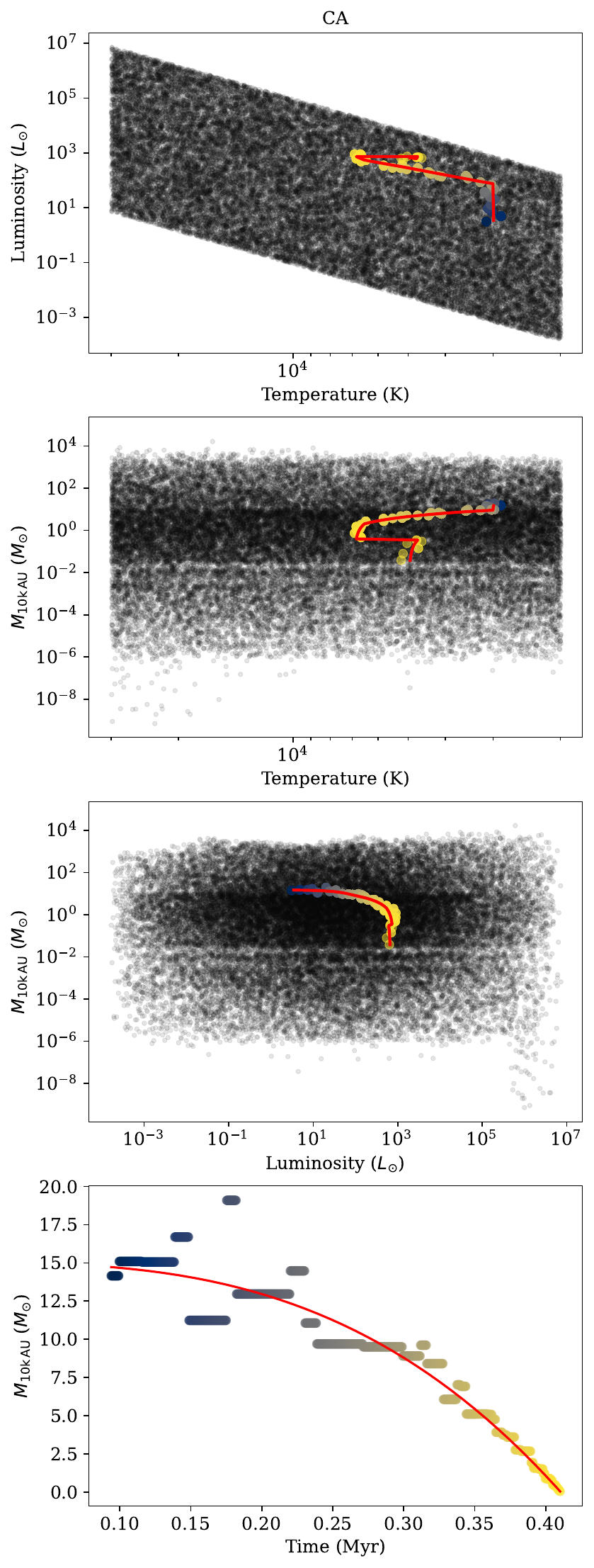}
    \caption{The same as Figure {\ref{fig:modeltrack}} for $5\, M_{\odot}$ stars.}
    \label{fig:fourplot2}
\end{figure}

Here, we show versions of Figure \ref{fig:modeltrack} for protostars with different final masses as a fuller illustration of this work's PEMs (see \S\ref{sec:2.2}). Figure \ref{fig:fourplot1} shows tracks for a low-mass star (0.2 $M_\odot$), while Figure \ref{fig:fourplot2} shows tracks for an intermediate-mass star (5 $M_\odot$).

\newpage
